\documentclass{article}

\usepackage{arxiv}

\usepackage[utf8]{inputenc} 
\usepackage[T1]{fontenc}    
\usepackage{hyperref}       
\usepackage{url}            
\usepackage{booktabs}       
\usepackage{amsfonts}       
\usepackage{nicefrac}       
\usepackage{microtype}      
\usepackage{lipsum}
\usepackage{graphicx}
\graphicspath{ {./images/} }
\usepackage{epstopdf}
\ifpdf
  \DeclareGraphicsExtensions{.eps,.pdf,.png,.jpg}
\else
  \DeclareGraphicsExtensions{.eps}
\fi

\usepackage{caption}
\usepackage{subcaption}
\usepackage{mathtools}
\usepackage{amssymb}
\usepackage{multirow}
\usepackage{siunitx}
\usepackage{algorithm}
\usepackage{algpseudocode}

\usepackage{enumitem}
\setlist[enumerate]{leftmargin=.5in}
\setlist[itemize]{leftmargin=.5in}

\newenvironment{codefont}{\fontfamily{lmtt}\selectfont}{\par}
\DeclareTextFontCommand{\codetext}{\codefont}

\title{Covariance-free Bi-fidelity Control Variates Importance Sampling for Rare Event Reliability Analysis}

\author{
 Promit Chakroborty \\
  Department of Civil and Systems Engineering\\
  Johns Hopkins University\\
  Baltimore, MD 21218 \\
  \texttt{pchakro1@jhu.edu} \\
   \And
 Somayajulu L.N. Dhulipala \\
  Idaho National Laboratory\\
  Idaho Falls, ID 83415 \\
  \texttt{Som.Dhulipala@inl.gov} \\
  \And
 Michael D. Shields \\
  Department of Civil and Systems Engineering\\
  Johns Hopkins University\\
  Baltimore, MD 21218 \\
  \texttt{michael.shields@jhu.edu} \\
}

\begin{document}

\newtheorem{proof}{Proof}
\newtheorem{theorem}{Theorem}
\newtheorem{lemma}{Lemma}
\newtheorem{corollary}{Corollary}
\newtheorem{proposition}{Proposition}
\newtheorem{definition}{Definition}
\newtheorem{remark}{Remark}
\newtheorem{stipulation}{Stipulation}




\maketitle

\begin{abstract}
Multifidelity modeling has been steadily gaining attention as a tool to address the problem of exorbitant model evaluation costs that makes the estimation of failure probabilities a significant computational challenge for complex real-world problems, particularly when failure is a rare event. To implement multifidelity modeling, estimators that efficiently combine information from multiple models/sources are necessary. In past works, the variance reduction techniques of Control Variates (CV) and Importance Sampling (IS) have been leveraged for this task. In this paper, we present the CVIS framework; a creative take on a coupled CV and IS estimator for bifidelity reliability analysis. The framework addresses some of the practical challenges of the CV method by using an estimator for the control variate mean and side-stepping the need to estimate the covariance between the original estimator and the control variate through a clever choice for the tuning constant. The task of selecting an efficient IS distribution is also considered, with a view towards maximally leveraging the bifidelity structure and maintaining expressivity. Additionally, a diagnostic is provided that indicates both the efficiency of the algorithm as well as the relative predictive quality of the models utilized. Finally, the behavior and performance of the framework is explored through analytical and numerical examples.
\end{abstract}


\section{Introduction}
\label{section:Introduction}


Understanding the performance of a system is an essential component of analysis and design within nearly all fields of engineering. Reliability Analysis is one way to do so, and its primary task is to estimate the probability of failure ($ P_F $) associated with the system under certain known or estimated input uncertainties. We represent these uncertainties by input random variables, which are collected in the random vector $ \mathbf{X} $ defined on the probability space $ \left( \Omega, \mathcal{F}, \mathcal{P} \right) $, with sample space $ \Omega $, $ \sigma $-field $ \mathcal{F} $, and probability measure $ \mathcal{P} $. Further, $ \mathbf{x} $ is defined as a specific realization of this input random vector. 

Theoretically, $ P_F $ is defined in terms of the response function $ g(\mathbf{x}) $ of the system (a scalar-valued function used to judge the performance of the system), as
\begin{equation}
\label{eqn:failure_probability_definition_response_function}
    P_F = \int_{ \left\{\mathbf{x} : g(\mathbf{x}) \leq 0 , \mathbf{x} \in \Omega \right\} } f_\mathbf{X} (\mathbf{x}) d \mathbf{x}
\end{equation}
where failure is defined to occur when $ g(\mathbf{x}) \leq 0 $, and $ f_\mathbf{X} (\mathbf{x}) $ is the joint probability density function of the random vector $ \mathbf{X} $. This integral can be simplified by defining the failure (or limit-state) indicator function $ I_F (\mathbf{x}) $, which takes a value of $ 1 $ whenever $ \mathbf{x} $ corresponds to system failure (i.e., $ g(\mathbf{x}) \leq 0 $) and a value of $ 0 $ otherwise. Thus,
\begin{equation}
\label{eqn:failure_probability_definition}
    P_F = \int_{\Omega} I_F (\mathbf{x}) f_\mathbf{X} (\mathbf{x}) d \mathbf{x} = \mathbb{E}_f \left[ I_F (\mathbf{x}) \right]
\end{equation}
where $ \mathbb{E}_f \left[ \cdot \right] $ is the expectation operator with respect to $ f_\mathbf{X} (\mathbf{x}) $.

For all except the simplest problems, this integral is analytically intractable due to a combination of dimensionality, form of the input distribution, and complexity of the failure indicator function. Instead, simulation-based methods such as Monte Carlo Simulation are used, whereby $ I_F (\mathbf{x}) $ is evaluated on a set of samples, $ \mathbf{x}_i $, generated from $ f_\mathbf{X} (\mathbf{x}) $, and these values are used to statistically estimate $ P_F $. The simplest such estimator, henceforth referred to as the crude Monte Carlo (MC) estimator, is expressed as follows
\begin{equation}
    \hat{P}_F = \frac{1}{N} \sum_{i=1}^N I_F (\mathbf{x}_i)
\end{equation}
where $ N $ is the total number of samples generated. Unfortunately, the coefficient of variation of $\hat{P}_F$, $ \operatorname{CoV} \left[ \hat{P}_F \right] $, is $ O \left( \nicefrac{1}{\sqrt{N}} \right) $ 
; thus, even a modest improvement in the confidence of the estimate requires a significant increase in the sample size $ N $. Additionally, $ \operatorname{CoV} \left[ \hat{P}_F \right] \propto \nicefrac{1}{P_F} $, and since for most real-world systems failure is a rare event, this further increases the sample size requirement for confident estimation. This issue is exacerbated by the fact that for most complex real-world systems, $ g (\mathbf{x}) $ is defined implicitly and is highly computationally expensive to evaluate. Altogether, these factors make the MC estimator infeasible in practice. Thus, the problem of improving the computational efficiency of reliability analysis has received extensive attention in the literature.

\subsection{A Brief Review of Reliability Methods}
\label{section:literature_review}

One way to improve the efficiency of reliability analysis is to use variance reduction methods, which reduce the total sample size requirement by creating estimators that have a lower variance than the MC estimator~\cite{RubinsteinKroese2016}. Two such methods, namely Control Variates (CV) and Importance Sampling (IS), underpin the method proposed in this paper. IS, in particular, has been widely used in the field of reliability analysis~\cite{Ang1992OptimalID,AUBECK1999,BUCHER1988119,DERKIUREGHIAN199837,MELCHERS19893}. Furthermore, many of the most efficient variance reduction-based reliability analysis algorithms belong to a broad class of variance reduction methods where sample points are pushed toward the failure region in a sequence of sampling steps. This general class includes methods such as subset simulation, sequential importance sampling, and tempering methods, among others~\cite{AuBeck2001,Catanach2018,PAPAIOANNOU201666,XIAO2019106248}.

Another popular solution to reduce the computational expense associated with MC is to construct fast-running surrogate models for the response function, which can be used in place of $ g (\mathbf{x}) $ (and hence $ I_F (\mathbf{x}) $) in the failure probability estimator. A wide variety of methods are used to create such surrogates for use in reliability analysis, such as Neural Networks~\cite{HURTADO2001113,PAPADRAKAKIS1996145,SUNDAR20161}, Polynomial Chaos Expansions~\cite{Choi2004,SudretPCKriging}, and Gaussian Process Regression~\cite{Bichon2008,AKMCS}. In addition, many methods utilize these surrogate models in conjunction with variance reduction techniques or other estimators designed for efficient reliability analysis~\cite{AKSS,Lelievre2018a,Razaaly2020a,sundar2019reliability,AKEE-SS}. However, the predictive accuracy and/or convergence rate of many of the surrogate modeling approaches are highly dependent on the smoothness, complexity, and/or dimensionality of the function they approximate.

An alternative way to reduce the cost of model evaluations is through multifidelity modeling, which incorporates system response information from multiple models/sources. Peherstorfer et al.~\cite{Peherstorfer2018} classify multifidelity methods into three broad categories: adaptation, filtering, and fusion. \textit{Adaptation} involves using information from the more accurate but more expensive high-fidelity (HF) model to enhance the quality of the response predictions made by the comparatively less accurate but cheaper low-fidelity (LF) models within the runtime of the algorithm. Many active-learning methods use adaptation strategies to learn surrogate model-based corrections to the LF model~\cite{LFMC, SingleLF, Zhang2022}. (In this context, a surrogate model may also be considered a LF model.) \textit{Filtering} is the idea of using LF information as a guide for where the HF model needs to be invoked, such as evaluating the HF model at sample points where the LF model passes or fails a certain criterion~\cite{LFMC,ChristenFox}, or using a large set of LF evaluations to direct IS on the HF model~\cite{PEHERSTORFER_MFIS}. Finally, \textit{fusion} combines information from all models together. Most multifidelity CV methods such as multilevel Monte Carlo (MLMC)~\cite{Giles2020,giles_2015}, multi-index Monte Carlo (MIMC)~\cite{HajiAliNobile}, and multifidelity Monte Carlo (MFMC)~\cite{Geraci_osti_1505910} fall under this category. The Approximate Control Variates (ACV) framework, proposed by Gorodetsky et al.~\cite{Gorodetsky2020}, was shown to generalize most of these CV-based methods.

Recently, some attention has also been given to the idea of further accelerating Reliability Analysis algorithms by combining the concepts of CV and IS within a multifidelity setting~\cite{MEHNI2023109014,PhamGorodetsky2022,RASHKI2018220}. 

\subsection{Goals and Contributions of the Proposed Method}
\label{section:goals_and_contributions}

The method proposed in this paper approaches the idea of combining CV and IS from a new angle. Rather than focusing purely on optimal variance reduction, the approach emphasizes practical benefits in implementation and interpretability while still achieving significant improvements in efficiency. The resultant contribution is the Control Variates - Importance Sampling (CVIS) estimator, which leverages one high-fidelity and one low-fidelity model to predict failure probabilities associated with rare events. A novel CVIS algorithm is presented, providing users with a simple and robust way to compute the CVIS estimator. In addition to significantly improving the efficiency of estimating small failure probabilities compared to crude Monte Carlo and providing comparable efficiency to existing multi-fideliity CV and IS-based methods, our framework has the following benefits:

\begin{itemize}
    \item It leverages the bifidelity structure of the models to construct an importance sampling density (ISD) that retains expressivity by not assuming an underlying form of the distribution.
    \item It bypasses practical challenges, such as the need to estimate the covariance between the models or deal with ISD normalization.
    \item A diagnostic is provided, at no additional cost, that can indicate when variance reduction is not being achieved in practice.
    \item A closed-form variance estimator is presented that incorporates all aspects of uncertainty in the failure probability estimate in an easily separable form.
\end{itemize}

\section{Preliminaries}
\label{section:Preliminaries}

In this section, we first describe the bifidelity reliability analysis problem setting and define some basic notation. Then, we briefly describe the two variance reduction techniques we primarily use in our proposed framework - Importance Sampling and Control Variates.

\subsection{Basic Notation for the Bifidelity Reliability Problem}

We consider two models; one high-fidelity (HF) model, which is considered highly accurate and complete but computationally expensive, and one low-fidelity (LF) model, which is cheaper but less accurate. The two models are characterized by the same vector of input random variables, denoted by $ \mathbf{X} $, defined on the same probability space $ \left( \Omega, \mathcal{F}, \mathcal{P} \right) $ 
We define the response function for each model and, hence, the corresponding failure indicator function. For the HF model, we define $ H (\mathbf{x}) : ~ \Omega \mapsto \mathbb R $ such that the corresponding indicator function is defined as
\begin{equation}
\label{eqn:HF_indicator}
    I_H (\mathbf{x}) = \begin{cases}
        1 & H (\mathbf{x}) \leq 0 \; \text{ i.e., \textit{failure}} \\
        0 & H (\mathbf{x}) > 0 \; \text{ i.e., \textit{safety}}
    \end{cases}
\end{equation}
Similarly, we define the LF model response function as $ L (\mathbf{x}) : ~ \Omega \mapsto \mathbb R $, and the corresponding indicator function is as
\begin{equation}
\label{eqn:LF_indicator}
    I_L (\mathbf{x}) = \begin{cases}
        1 & L (\mathbf{x}) \leq 0 \; \text{ i.e., \textit{failure}} \\
        0 & L (\mathbf{x}) > 0 \; \text{ i.e., \textit{safety}}
    \end{cases}
\end{equation}

The failure probability for the system predicted by each model is the mean value of the corresponding indicator functions (each equivalent to Eq.~\eqref{eqn:failure_probability_definition}), where $ \mathbf{X} $ follows the distribution $ f_\mathbf{X}(\mathbf{x}) $. We are specifically interested in estimating the HF failure probability. Thus
\begin{align}
    \text{The HF model failure probability: } P_F &= \mathbb{E}_f \left[ I_H (\mathbf{X}) \right] \label{eqn:HF_failure} \\
    \text{The LF model failure probability: } P_{F_L} &= \mathbb{E}_f \left[ I_L (\mathbf{X}) \right] \label{eqn:LF_failure}
\end{align}

\subsection{Importance Sampling (IS)}
\label{section:Importance_Sampling}

In Importance Sampling, variance reduction is achieved by drawing samples from an alternate density, called the Importance Sampling Density (ISD), and weighting their influence in the constructed IS estimator according to the ratio of the densities of the original distribution to the ISD. The ISD is constructed such that most of its density lies in the ``important" region of the function of interest. In the specific case of reliability analysis, where we aim to estimate $ P_F $ (Eq.~\eqref{eqn:HF_failure}), a ``good" ISD would have most of its density covering the region where $ I_H (\mathbf{x}) = 1 $. If the ISD is $ q_\mathbf{X}(\mathbf{x}) $, then IS redefines $ P_F $ as
\begin{equation}
\label{eqn:IS_framework_definition}
    P_F = \mathbb{E}_q \left[ I_H (\mathbf{X}) \frac{f_\mathbf{X} (\mathbf{X})}{q_\mathbf{X} (\mathbf{X})} \right]
\end{equation}
where now $ \mathbf{X} \sim q_\mathbf{X}(\mathbf{x}) $ and so we take the expectation with respect to the ISD. The resultant estimator is given by
\begin{equation}
\label{eqn:IS_estimator_general}
    \hat{P}_{F_{IS}} = \frac{1}{N} \sum_{i=1}^N \left[ I_H (\mathbf{x}_i) \frac{f_\mathbf{X} (\mathbf{x}_i)}{q_\mathbf{X} (\mathbf{x}_i)} \right]
\end{equation}
where $ N $ is the total number of samples, and each $ \mathbf{x}_i $ is drawn from $ q_\mathbf{X}(\mathbf{x}) $. 
Note that 
the IS estimator in Eq.~\eqref{eqn:IS_estimator_general} is only unbiased when $ \operatorname{supp} (q_\mathbf{X} (\mathbf{x})) \supseteq \operatorname{supp} (I_H (\mathbf{x}) f_\mathbf{X} (\mathbf{x})) $ \cite{Owen2013, RubinsteinKroese2016}, where $ \operatorname{supp} (q_\mathbf{X}(\mathbf{x})) = \left\{ \mathbf{x} \in \Omega : q_\mathbf{X}(\mathbf{x}) > 0 \right\} $ denotes the support of the distribution $ q_\mathbf{X} (\mathbf{x}) $.

It can be shown that the variance of the resultant estimator $ \hat{P}_{F_{IS}} $ is minimum when the ISD takes the following optimal form:
\begin{equation}
\label{eqn:optimal_ISD_form}
    q^*_\mathbf{X} (\mathbf{x}) \propto I_H (\mathbf{x}) f_\mathbf{X} (\mathbf{x})
\end{equation}

Tabandeh et al. \cite{Tabandeh2022} classify IS for reliability analysis into two broad groups: (a) Density approximation methods, which approximate $ q^*_\mathbf{X} (\mathbf{x}) $ using a specific member of a chosen family of distributions, and (b) Limit-state approximation methods, which utilize an approximation of the limit-state function to substitute for $I_H (\mathbf{x})$ in Eq.~\eqref{eqn:optimal_ISD_form}.

\subsection{Control Variates (CV)}
\label{section:Control_Variates}

Classical Control Variates is a technique used to improve the accuracy of statistical estimates of a primary quantity of interest by making use of its correlation with some secondary quantity, called the control variate. 
For mean value estimation with CV, a set of samples (say $ \overline{\mathbf{z}} $) is generated from the input distribution of the problem (say $ f_Z(z) $). Then, both the primary quantity of interest (say $ U $) and the control variate (say $ V $) are evaluated for each sample. By doing so, we can achieve some estimate of the mean of both $ U $ and $ V $, as $ \hat{U}(\overline{\mathbf{z}}) $ and $ \hat{V}(\overline{\mathbf{z}}) $, respectively. Additionally, in classical CV, the true mean of the control variate is also assumed to be known (say $ \mu_V $). Then, the classical CV estimate of the mean of $ U $ is constructed as
\begin{equation}
    \label{eqn:general_classical_CV}
    \hat{\mu}_{U, CV}(\overline{\mathbf{z}}) = \hat{U}(\overline{\mathbf{z}}) + \alpha \left( \hat{V}(\overline{\mathbf{z}}) - \mu_V \right)
\end{equation}
where $ \alpha $ is a scalar tuning constant. Depending on the correlation between $ \hat{U}(\overline{\mathbf{z}}) $ and $ \hat{V}(\overline{\mathbf{z}}) $ and the chosen value of $ \alpha $, the variance of $ \hat{\mu}_{U, CV}(\overline{\mathbf{z}}) $ can be much lower than the variance of $ \hat{U}(\overline{\mathbf{z}}) $.

For our case of bifidelity reliability analysis, the quantity of interest is $ I_H (\mathbf{x}) $ (Eq.~\eqref{eqn:HF_indicator}), and the natural choice of the control variate is $ I_L (\mathbf{x}) $ (Eq.~\eqref{eqn:LF_indicator}). Thus, for bifidelity reliability where $ \mathbb{E} [I_L (\mathbf{X})] = P_{F_L} $, the classical CV estimator takes the form
\begin{equation}
\label{eqn:CV_estimator_classical}
    \hat{P}_{F_{CV}} = \hat{Q} + \alpha \left( \hat{Q}_L - P_{F_L} \right)
\end{equation}
where $ \hat{Q} = \sum_{i=1}^N \left( \nicefrac{I_H (\mathbf{x}_i)}{N} \right) $ and $ \hat{Q}_L = \sum_{i=1}^N \left( \nicefrac{I_L (\mathbf{x}_i)}{N} \right) $ are typically the MC\footnote{In general $ \hat{Q} $ and $ \hat{Q}_L $ can be any estimators of $ \mathbb{E} [I_H (\mathbf{X})] $ and $ \mathbb{E} [I_L (\mathbf{X})] $} estimators of $ \mathbb{E} [I_H (\mathbf{X})] $ and $ \mathbb{E} [I_L (\mathbf{X})] $ respectively, evaluated on the same set of samples $ \left\{ \mathbf{x}_i \right\} $ drawn from the input distribution $ f_\mathbf{X} (\mathbf{x}) $.\footnote{The explicit dependence of $ \hat{Q} $ and $ \hat{Q}_L $ on the samples $ \left\{ \mathbf{x}_i \right\} $ is dropped for simplicity of notation.} $ N $ is the total number of samples generated. The challenge with the classical CV formulation is that $ \mathbb{E} [I_L (\mathbf{X})] = P_{F_L} $ is rarely known exactly.

The Approximate Control Variates (ACV) framework was introduced by Gorodetsky et al.~\cite{Gorodetsky2020} for the more practical situation when the control variate mean is unknown. In this case, it must also be estimated, and thus the ACV estimator becomes
\begin{equation}
    \label{eqn:ACV_estimator_general}
    \hat{P}_{F_{ACV}} = \hat{Q} + \alpha \left( \hat{Q}_L - \hat{P}_{F_L} \right) = \hat{Q} + \alpha \Delta_L
\end{equation}
where $ \Delta_L = \left( \hat{Q}_L - \hat{P}_{F_L} \right) $, and $ \hat{P}_{F_L} $ is some estimator of the LF failure probability that is different from $ \hat{Q}_L $, in either form or sample set. Note that this makes $ \Delta_L $ a random variable as well.

They further show that the variance of this estimator is minimized for the following optimal choice of $ \alpha $, given by
\begin{equation}
\label{eqn:ACV_optimal_alpha}
    \alpha^* = - \frac{ \operatorname{\mathbb{C}ov} \left[ \Delta_L, \hat{Q} \right]}{ \operatorname{\mathbb{V}ar} \left[ \Delta_L \right]}
\end{equation}
The terms in Eq.~\eqref{eqn:ACV_optimal_alpha} are rarely known a priori. The most natural solution to this problem is to estimate these terms as well, as explored by Pham \& Gorodetsky~\cite{PhamGorodetsky2022}. However, their method requires the generation of an ensemble of the estimators $ \hat{Q} $, $ \hat{Q}_L $, and $ \hat{P}_{F_L} $, i.e., they estimate sample-statistics of the $ \hat{Q} $, $ \hat{Q}_L $, and $ \hat{P}_{F_L} $ estimators by generating multiple realizations of each. This requires considerable computational effort or the use of multiple small batches of samples for each realization of the estimators. Thus, we see that implementing a CV estimator requires knowledge of the covariance between the estimators of the statistics of the HF and LF models, which is almost certainly unavailable exactly and is very tricky to estimate.

\section{Proposed Methodology: The Control Variates - Importance Sampling (CVIS) Estimator}
\label{section:methodology}


In the proposed approach, we utilize the ACV formulation with an alternate choice of $ \alpha $ that circumvents the need to know or estimate the inter-model covariances. Rearranging the terms from Eq.~\eqref{eqn:ACV_estimator_general} (and flipping the sign of $ \alpha $), the general ACV estimator can be expressed
\begin{equation}
\label{eqn:rearranged_acv_estimator}
    \hat{P}_{F_{CV}} = \alpha \hat{P}_{F_L} + \left[ \hat{Q} - \alpha \hat{Q}_L \right]
\end{equation}
Henceforth, we use the subscript ``CV'' to denote all variations of the ACV estimator (the generic version of which was defined as $ \hat{P}_{F_{ACV}} $ in Eq.~\eqref{eqn:ACV_estimator_general}). This is done both for simplicity of notation and to acknowledge that it is the ACV formulation that is used in practice, as the classical CV formulation is almost never applicable.

This estimator behaves differently depending on the estimators used for $ \hat{Q} $, $ \hat{Q}_L $, and $ \hat{P}_{F_L} $. We define two versions of this estimator that will be useful to us (particularly in Section~\ref{section:alpha_as_diagnostic}) based on different formulations of $ \hat{Q} $ and $ \hat{Q}_L $; in both cases, $ \hat{P}_{F_L} $ can be chosen freely. The first case (the Control Variates Monte Carlo, CVMC estimator) uses CV as the lone variance reduction method; MC estimators are used for $ \hat{Q} $ and $ \hat{Q}_L $, i.e., $ \hat{Q} = \hat{Q}_{MC} = \sum_{i=1}^N \left( \nicefrac{I_H (\mathbf{x}_i)}{N} \right) $ and $ \hat{Q}_L = \hat{Q}_{L_{MC}} = \sum_{i=1}^N \left( \nicefrac{I_L (\mathbf{x}_i)}{N} \right) $, with each $ \mathbf{x}_i $ drawn from $ f_{\mathbf{X}} (\mathbf{x}) $, leading to
\begin{equation}
\label{eqn:acv_with_cmc}
    \hat{P}_{F_{CVMC}} = \alpha \hat{P}_{F_L} + \left[ \hat{Q}_{MC} - \alpha \hat{Q}_{L_{MC}} \right]
\end{equation}
The second case combines IS with CV; IS estimators are used for $ \hat{Q} $ and $ \hat{Q}_L $, i.e., $ \hat{Q} = \hat{Q}_{IS} = \sum_{i=1}^N \left( \nicefrac{I_H (\mathbf{x}_i) f_{\mathbf{X}} (\mathbf{x}_i)}{N q_{\mathbf{X}} (\mathbf{x}_i)} \right) $ and $ \hat{Q}_L = \hat{Q}_{L_{IS}} = \sum_{i=1}^N \left( \nicefrac{I_L (\mathbf{x}_i) f_{\mathbf{X}} (\mathbf{x}_i)}{N q_{\mathbf{X}} (\mathbf{x}_i)} \right) $, with each $ \mathbf{x}_i $ drawn from some ISD $ q_{\mathbf{X}} (\mathbf{x}) $, giving us the most general version of the CVIS estimator as
\begin{equation}
\label{eqn:acv_with_is}
    \hat{P}_{F_{CVIS}} = \alpha \hat{P}_{F_L} + \left[ \hat{Q}_{IS} - \alpha \hat{Q}_{L_{IS}} \right]
\end{equation}


Next, instead of computing the optimal value $ \alpha^* $ as in Eq.~\eqref{eqn:ACV_optimal_alpha}, we choose $ \hat{\alpha} $ as the value that makes the second term in Eq.~\eqref{eqn:rearranged_acv_estimator} zero. Therefore,
\begin{equation}
\label{eqn:proposed_alpha_definition}
    \hat{\alpha} = \frac{\hat{Q}}{\hat{Q}_L}
\end{equation}
and our proposed ``simple'' (as opposed to optimal) ACV (SACV) estimator takes the form
\begin{equation}
\label{eqn:S-ACV_estimator}
    \hat{P}_{F_{SACV}} = \hat{\alpha} \hat{P}_{F_{L}}
\end{equation}
Note that we use the notation $ \hat{\alpha} $ to signify that our choice of $ \alpha $ is itself an estimate as it relies on estimated values; therefore, it is a random variable and not a constant. If the estimators on the right-hand-side of Eq.~\eqref{eqn:proposed_alpha_definition} were replaced with their limiting values (as the number of samples goes to infinity), we would get the limiting value of this choice of $ \alpha $ as
\begin{equation}
\label{eqn:limiting_or_true_value_of_cvis_alpha}
    \alpha^\dagger = \frac{P_F}{P_{F_L}}
\end{equation}
Substituting Eq.~\eqref{eqn:proposed_alpha_definition} into Eq.~\eqref{eqn:acv_with_cmc} gives
\begin{equation}
    \hat{P}_{F_{SCVMC}} = \hat{P}_{F_{L}} \frac{\hat{Q}_{MC}}{\hat{Q}_{L_{MC}}}
\end{equation}
and substituting it into Eq.~\eqref{eqn:acv_with_is} gives
\begin{equation}
    \hat{P}_{F_{SCVIS}} = \hat{P}_{F_{L}} \frac{\hat{Q}_{IS}}{\hat{Q}_{L_{IS}}}
\end{equation}

Our proposed CVIS framework is a special case of the general $ \hat{P}_{F_{SCVIS}} $ estimator when a particular ISD is used as described below. Thus, the proposed CVIS estimator, simply referred to as the CVIS estimator henceforth, can be represented as
\begin{gather}
    \Tilde{P}_{F_{CVIS}} = \Tilde{P}_F = \Tilde{\alpha} \Tilde{P}_{F_L} \label{eqn:CVIS_estimator} \\
    \Tilde{\alpha} = \frac{\Tilde{Q}}{\Tilde{Q}_L} \label{eqn:CVIS_alpha}
\end{gather}
Note that here, and for the remainder of the document, we notationally replace the $\hat{\cdot}$ with a $\tilde{\cdot}$ to distinguish the proposed estimators from those of previous works. Additionally, $ \Tilde{\alpha} $ is henceforth referred to as the CVIS constant.



The CVIS framework uses IS for $ \tilde{Q} $ and $ \tilde{Q}_{L} $\footnote{As per Section~\ref{section:Control_Variates}, both must use the same set of samples.} to focus the samples around the failure region and improve the overall efficiency of the method. This is because, although the SCVMC formulation does result in some theoretical variance reduction, the practical issue of generating a statistically sufficient number of failure samples in a rare-event simulation problem persists, as mentioned in Section~\ref{section:Importance_Sampling}. Any efficient, unbiased estimator of $ P_{F_L} $ is acceptable for $ \tilde{P}_{F_{L}} $.

The specific choice of ISD exploits the bifidelity structure once again by applying the limit-state approximation concept (as defined in Section~\ref{section:Importance_Sampling}). 
This is done by replacing $ I_H (\mathbf{x}) $ in Eq.~\eqref{eqn:optimal_ISD_form} with a logistic approximation of $ I_L (\mathbf{x}) $ given by
\begin{equation}
\label{eqn:logistic_form}
    S_L (\mathbf{x}, \beta) = \frac{1}{1+\exp{ \left( \beta L(\mathbf{x}) \right) }}
\end{equation}
where $ \beta $ is a tuning parameter such that $ \lim_{\beta \to 0} S_L (\mathbf{x}, \beta) = 0.5 $ and $ \lim_{\beta \to \infty} S_L (\mathbf{x}, \beta) \approx I_L (\mathbf{x}) $ $ \forall ~\mathbf{x} \in \Omega $. 
The resulting ISD is
\begin{equation}
\label{eqn:proposed_ISD}
    \mathfrak{q}_{\mathbf{X}} (\mathbf{x}, \beta) = \frac{S_L (\mathbf{x}, \beta) f_{\mathbf{X}} (\mathbf{x})}{C_S}
\end{equation}
where $ C_S = \int_{\Omega} S_L (\mathbf{x}, \beta) f_{\mathbf{X}} (\mathbf{x}) d\mathbf{x} $ 
is the normalizing constant. Finally, from Eq.~\eqref{eqn:IS_estimator_general} and after some simplifications, for a set of samples $ \left\{ \mathbf{x}_i \right\} $ drawn from $ \mathfrak{q}_{\mathbf{X}} (\mathbf{x}, \beta) $, we get
\begin{align}
    \tilde{Q} &= \frac{C_S}{N} \sum_{i=1}^N \left[ \frac{I_H (\mathbf{x}_i)}{S_L (\mathbf{x}_i, \beta)} \right] \label{eqn:HF_IS_estimator_for_CVIS} \\
    \tilde{Q}_{L} &= \frac{C_S}{N} \sum_{i=1}^N \left[ \frac{I_L (\mathbf{x}_i)}{S_L (\mathbf{x}_i, \beta)} \right] \label{eqn:LF_IS_estimator_for_CVIS}
\end{align}
Note that due to the form of $ \tilde{\alpha}$ (Eq.~\eqref{eqn:CVIS_alpha}), the normalizing constant $ C_S $ cancels out. Thus, it is of practical use to define $ \tilde{\mathcal{Q}} = \nicefrac{\tilde{Q}}{C_S} $ and $ \tilde{\mathcal{Q}}_{L} = \nicefrac{\tilde{Q}_{L}}{C_S} $, such that
\begin{equation}
\label{eqn:proposed_alpha_definition_alternate}
    \tilde{\alpha} = \frac{\tilde{\mathcal{Q}}}{\tilde{\mathcal{Q}}_{L}}
\end{equation}

The assembled collection of Eqs.~\eqref{eqn:CVIS_estimator},~\eqref{eqn:CVIS_alpha},~\eqref{eqn:HF_IS_estimator_for_CVIS}, and~\eqref{eqn:LF_IS_estimator_for_CVIS} together form the proposed CVIS estimator. In the special case where the HF failure region is a subset of the LF failure region, the CVIS estimator reduces to a simpler form, presented in Appendix~\ref{appendix:nested_failure}. 


\subsection{Stability and Diagnostic}
\label{section:stability_and_diagnostic}

In this section, we first present a stipulation that the chosen pair of HF and LF models must satisfy in order for the proposed IS estimators $ \tilde{Q} $ and $ \tilde{Q}_{L} $ to be stable and well-defined. We also discuss how the stipulation allows for the definition of a ``well-tuned'' ISD. Then, we provide a few guarantees for the performance of the proposed method and present a diagnostic that is available for no additional model evaluations. Finally, we make some observations about the relative influence of the CV formulation and the chosen IS estimators on the variance reduction achieved by the CVIS methodology.

\subsubsection{Model Relationship Requirements}
\label{section:model_relationship_requirements}

The CVIS estimator, as presented 
above, is not stable for all possible pairs of HF and LF models. This is primarily because of how IS is implemented. For $ \mathfrak{q}_{\mathbf{X}} (\mathbf{x}, \beta) $ to function as a ``good'' ISD, $ S_L (\mathbf{x}, \beta) $ has to approximate $ I_H (\mathbf{x}) $ reasonably well; however, if the LF model itself is poor at predicting failure, then this is not possible.

Let us define the failure domain $ \Omega_{\mathcal{F}} = \left\{ \mathbf{x} \in \Omega : I_H (\mathbf{x}) = 1 \right\} $ and its complement $ \Omega_{\mathcal{F}}^C = \Omega \setminus \Omega_{\mathcal{F}} $. Per Section~\ref{section:Importance_Sampling}, the CVIS estimator is only unbiased if $ \mathfrak{q}_{\mathbf{X}} (\mathbf{x}, \beta) > 0 ~~ \forall ~ \mathbf{x} \in \Omega_{\mathcal{F}} $ (under the assumption that $ f_{\mathbf{X}} (\mathbf{x}) > 0 ~~ \forall ~ \mathbf{x} \in \Omega_{\mathcal{F}} $, which is not restrictive as usually $ \operatorname{supp} \left( f_{\mathbf{X}} (\mathbf{x}) \right) = \Omega $). While this is always satisfied in theory, since $ S_L (\mathbf{x}, \beta) > 0 ~~ \forall ~ \mathbf{x}$, in practice $ \mathfrak{q}_{\mathbf{X}} (\mathbf{x}, \beta) \approx 0 $ far from the LF failure region (except when $ \beta $ is very small, in which case $ \mathfrak{q}_{\mathbf{X}} (\mathbf{x}, \beta) \approx f_{\mathbf{X}} (\mathbf{x}) $ and there is effectively no importance sampling).


To ensure the estimator is unbiased in practice, while still allowing for appreciable variance reduction through IS, we make the following stipulation.

\begin{stipulation}[The Model Relationship Requirement]
\label{stipulation:fundamental_model_quality_assumption}
    There must exist a sufficiently small value $ \delta_L > 0 $ such that the LF response function $L(\mathbf{x})$ satisfies:
    \begin{equation}
        L (\mathbf{x}) < \delta_L ~~ \forall ~ \mathbf{x} \in \Omega_{\mathcal{F}}
    \end{equation}
\end{stipulation}

\begin{remark}
\label{remark:smallest_contour_of_LF_that_covers_HF_failure_region}
    Additionally, we define $ \delta_{\mathcal{L}} $ as the smallest value of $ \delta_L $ that satisfies Stipulation~\ref{stipulation:fundamental_model_quality_assumption}, i.e., $ L (\mathbf{x}) < \delta_{\mathcal{L}} ~~ \forall ~ \mathbf{x} \in \Omega_{\mathcal{F}} $, such that for any $ \delta'_L < \delta_{\mathcal{L}} $, $ \exists ~ \mathbf{x} \in \Omega_{\mathcal{F}} $ with $ L (\mathbf{x}) \geq \delta'_L $. (In the case where the HF failure region is a subset of the LF failure region, we define $ \delta_{\mathcal{L}} \to 0^+ $, i.e., a positive number infinitesimally larger than $ 0 $.) We also define $ \Omega_{\mathcal{L}} = \left\{ \mathbf{x} \in \Omega : L (\mathbf{x}) < \delta_{\mathcal{L}} \right\} $ and its complement $ \Omega_{\mathcal{L}}^C = \Omega \setminus \Omega_{\mathcal{L}} $.
\end{remark}

\begin{figure}[!htbp]
\centering
\begin{subfigure}{.45\textwidth}
  \centering
  \includegraphics[width=\linewidth]{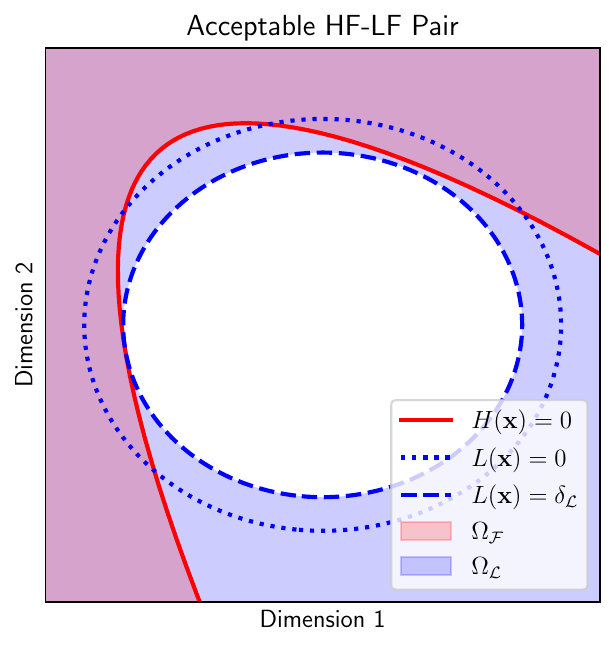}
  \caption{}
  \label{fig:Model_set_visualization_acceptable}
\end{subfigure}%
\begin{subfigure}{.45\textwidth}
  \centering
  \includegraphics[width=\linewidth]{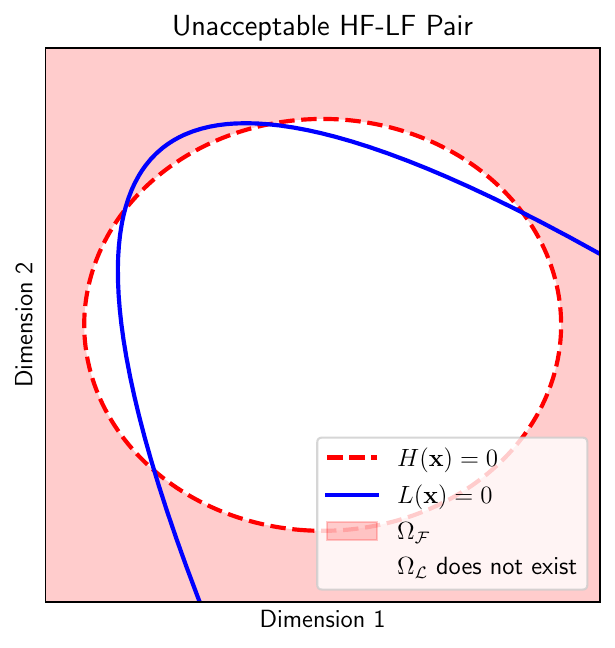}
  \caption{}
  \label{fig:Model_set_visualization_unacceptable}
\end{subfigure}
\caption{A visualization of relative failure limit surfaces for a representative pair of HF and LF models when (a) the pair follows Stipulation~\ref{stipulation:fundamental_model_quality_assumption} and (b) the pair does not follow Stipulation~\ref{stipulation:fundamental_model_quality_assumption}. Additionally, the sets $ \Omega_{\mathcal{F}} $ the HF failure region) and $ \Omega_{\mathcal{L}} $ (the smallest LF contour that covers $ \Omega_{\mathcal{F}} $, as defined in Remark~\ref{remark:smallest_contour_of_LF_that_covers_HF_failure_region}) have also been visualized. In case (b) Stipulation~\ref{stipulation:fundamental_model_quality_assumption} is not followed, i.e., $ \nexists \; \delta_{\mathcal{L}} $ such that $ L (\mathbf{x}) < \delta_{\mathcal{L}} ~ \forall \; \mathbf{x} \in \Omega_{\mathcal{F}} $. Hence, $ \Omega_{\mathcal{L}} $ cannot be constructed.}
\label{fig:Model_set_visualization}
\end{figure}

Informally, Stipulation~\ref{stipulation:fundamental_model_quality_assumption} states that the LF limit-state function must either indicate failure or be \emph{``close to''} indicating failure wherever the HF limit-state indicator function does. And it further follows from Remark~\ref{remark:smallest_contour_of_LF_that_covers_HF_failure_region} that Stipulation~\ref{stipulation:fundamental_model_quality_assumption} can be interpreted to mean that $ \Omega_{\mathcal{F}} \subseteq \Omega_{\mathcal{L}}$ must be satisfied. An illustration of this concept is provided in  Figure~\ref{fig:Model_set_visualization}, which shows an acceptable HF-LF pair and an unacceptable HF-LF pair. The pair in Figure~\ref{fig:Model_set_visualization_acceptable} satisfies the requirement because the LF model predicts failure in the region where it does not accurately approximate the HF model. The pair in Figure~\ref{fig:Model_set_visualization_unacceptable}, on the other hand, is not able to predict failures that occur in the lower right corner of the domain.

Stipulation~\ref{stipulation:fundamental_model_quality_assumption} excludes cases where the LF model completely misses certain failure regions, but on its own it does not guarantee that an ISD of the form in Eq.~\eqref{eqn:proposed_ISD} leads to practically well-behaved estimators. We must further ensure that our chosen formulation is able to utilize the inter-model relationship enforced by the stipulation. To this end, we provide a theoretical framework in Appendix~\ref{appendix:ISD_tuning_insights} that allows us to define an ``optimally-tuned'' ISD under our chosen form (Eq.~\eqref{eqn:proposed_ISD}), which amounts to choosing a value $ \beta^* $ for the ISD tuning parameter that achieves the best possible concentration of the ISD around the failure region. Note that $ \mathfrak{q}_{\mathbf{X}} \left( \mathbf{x}, \beta^* \right) $ 
is not the same as the optimal ISD from Eq.~\eqref{eqn:optimal_ISD_form}, because $ \Omega_{\mathcal{L}} \neq \Omega_{\mathcal{F}} $ in general, and $ \int_{\Omega_{\mathcal{L}}^C} \mathfrak{q}_{\mathbf{X}} (\mathbf{x}, \beta^*) d\mathbf{x} \neq 0 $. Instead, the ``optimally-tuned'' ISD is the best approximation of the optimal ISD we can construct given Stipulation~\ref{stipulation:fundamental_model_quality_assumption} and the chosen form of $ \mathfrak{q}_{\mathbf{X}} (\mathbf{x}, \beta) $. The problem of estimating $ \beta^* $ 
is addressed in Section~\ref{section:CVIS_algorithm_and_implementation}.

Although the proposed method constructs a reasonably efficient ISD, due to the form chosen for $ \mathfrak{q}_{\mathbf{X}} (\mathbf{x}, \beta) $ (Eq.~\eqref{eqn:proposed_ISD}), even the best possible ISD that can be constructed within the CVIS framework (the ``optimally-tuned'' ISD defined above) necessarily places some appreciable density outside the HF failure region ($ \Omega_{\mathcal{L}} \setminus \Omega_{\mathcal{F}} $ in particular). Depending on how closely or poorly the LF model approximates the HF model (i.e., how small $ \delta_{\mathcal{L}} $ is), the necessary density outside the HF failure region might lead to a poorer ISD than an ISD construction scheme that directly targets the HF failure region. However, such schemes either assume a parametric form for the distribution or construct a surrogate for the HF model to use for the ISD (as per Tabandeh et al. \cite{Tabandeh2022}). Compared to the former case, our method is often easier to tune since it requires the selection of only one parameter, and has better expressivity due to the lack of an assumed distribution form. Compared to the latter case, our method is cheaper, as it does not require any HF model evaluations for training data. 





\subsubsection{Diagnosing Variance Reduction}
\label{section:alpha_as_diagnostic}

Since neither CV nor IS guarantees variance reduction for all $ \alpha $ values or ISDs, 
we would like to know whether a specific choice of $ \tilde{\alpha} $ and $ \mathfrak{q}_{\mathbf{X}} (\mathbf{x}, \beta) $ in CVIS achieves variance reduction. To this end, we derive here 
a performance diagnostic $ \kappa $ (defined in Theorem~\ref{thm:CVIS_alpha_variance_reduction_diagnostic}) that requires no additional HF or LF model evaluations. Let us begin with the following theorem, proven in Appendix~\ref{appendix:proof_CVIS_variance_less_than_CV_variance}.

\begin{theorem}
\label{thm:CVIS_variance_less_than_CV_variance}
Consider the estimators $ \hat{P}_{F_{CVMC}} $ (Eq.~\eqref{eqn:acv_with_cmc}) and $ \hat{P}_{F_{CVIS}} $ (Eq.~\eqref{eqn:acv_with_is}).
    If the ISD used for $ \hat{P}_{F_{CVIS}} $ is $ \mathfrak{q}_{\mathbf{X}} (\mathbf{x}, \beta) $, the same LF failure estimator $ \hat{P}_{F_L} $ is used for both cases such that it is independent of $ \hat{Q}_{MC} $ \& $ \hat{Q}_{L_{MC}} $ and $ \hat{Q}_{IS} $ \& $ \hat{Q}_{L_{IS}} $, and $ N $ statistically independent samples are used for $ \hat{Q}_{MC} $ \& $ \hat{Q}_{L_{MC}} $ and $ \hat{Q}_{IS} $ \& $ \hat{Q}_{L_{IS}} $, generated from their respective distributions (with $ N $ being the same in all cases). Then, for any arbitrary fixed value of $ \alpha $
    \begin{equation}
    \label{eqn:CVIS_variance_less_than_CV_variance}
        \operatorname{\mathbb{V}ar} \left[ \hat{P}_{F_{CVIS}} \right] \leq \operatorname{\mathbb{V}ar} \left[ \hat{P}_{F_{CVMC}} \right]
    \end{equation}
\end{theorem}


This theorem broadly states that, under proper conditions, the $ P_F $ estimate that combines IS (using the CVIS ISD) with CV will have variance that is less than or equal to the $P_F$ estimate using CV alone. Intuitively, this makes sense. 
Consequently, 
any value of $ \alpha $ that leads to variance reduction in $ \hat{P}_{F_{CVMC}} $ 
will also lead to variance reduction in $ \hat{P}_{F_{CVIS}} $, which is the CVIS estimator $ \Tilde{P}_F $ when $ \alpha = \tilde{\alpha} $, and $ \hat{P}_{F_L} = \tilde{P}_{F_{L}} $. (Note that $ \hat{Q}_{IS} = \tilde{Q} $ and $ \hat{Q}_{L_{IS}} = \tilde{Q}_{L} $ since $ \mathfrak{q}_{\mathbf{X}} (\mathbf{x}, \beta) $ is used as the ISD for $ \hat{P}_{F_{CVIS}} $) 

We now state two theorems that help us determine values of $ \alpha $ that yield variance reduction in $ \hat{P}_{F_{CVMC}} $. They are proven in Appendices~\ref{appendix:proof_optimal_alpha_in_model_limit} and~\ref{appendix:proof_variance_reduction_diagnostic}, respectively.

\begin{theorem}
\label{thm:model_limit_CVIS_alpha_optimal}
    Define $ \Omega_{\Delta} \coloneqq \left\{ \mathbf{x} \in \Omega : I_H (\mathbf{x}) \neq I_L (\mathbf{x}) \right\} $, and 
    let $ \alpha^* $ be the value of $ \alpha $ that leads to the maximum variance reduction allowed by pure CV for $ \hat{P}_{F_{CVMC}} $ (see Section~\ref{section:Control_Variates}). If
    
    \begin{enumerate}
        \item $ \hat{Q}_{MC} $ and $ \hat{Q}_{L_{MC}} $ are estimated from the same set of $ N $ statistically independent samples generated according to $ f_{\mathbf{X}} (\mathbf{x}) $,
        \item The target failure probability is small (i.e., $ P_F \ll 1 $), and
        \item $ \operatorname{\mathbb{V}ar} \left[ \hat{P}_{F_L} \right] + \operatorname{\mathbb{V}ar} \left[ \hat{Q}_{L_{MC}} \right] \approx \operatorname{\mathbb{V}ar} \left[ \hat{Q}_{L_{MC}} \right] $, 
    \end{enumerate}
    Then $ \tilde{\alpha} $ (Eq.~\eqref{eqn:CVIS_alpha}) estimates $ \alpha^* $ as $ \Omega_{\Delta} \to \varnothing $, i.e.,
    \begin{equation}
        \tilde{\alpha} = \lim_{\Omega_{\Delta} \to \varnothing} \hat{\alpha}^*
    \end{equation}
\end{theorem}

\begin{remark}
    Condition 3 from Theorem~\ref{thm:model_limit_CVIS_alpha_optimal} is satisfied in practice for two reasons. First, the same small set of samples are used to compute $ \hat{Q}_{L_{MC}} $ and $ \hat{Q}_{MC} $. This set of samples is limited by the cost of HF model evaluations. Meanwhile, the number of samples used to estimate $ \hat{P}_{F_L} $ is limited only by the cost of the LF model, and thus a much larger number of samples can be used. 
    Secondly, $ \hat{Q}_{L_{MC}} $ is a MC estimate, while $ \hat{P}_{F_L} $ is chosen to be a variance-reduced failure estimator, which further reduces $\operatorname{\mathbb{V}ar} \left[ \hat{P}_{F_L}\right]$.
\end{remark}


\begin{theorem}
\label{thm:CVIS_alpha_variance_reduction_diagnostic}
    Define $ P_{HL} \coloneqq \mathbb{E}_f \left[ I_H (\mathbf{X}) I_L (\mathbf{X}) \right] $ as the probability that both the LF and the HF model indicate failure. Then, under the following conditions:
    \begin{enumerate}
        \item $ \hat{Q}_{MC} $ and $ \hat{Q}_{L_{MC}} $ are estimated from the same set of $ N $ statistically independent samples generated according to $ f_{\mathbf{X}} (\mathbf{x}) $,
        \item The target failure probability is small (i.e., $ P_F \ll 1 $)
        \item $ \operatorname{\mathbb{V}ar} \left[ \hat{P}_{F_L} \right] + \operatorname{\mathbb{V}ar} \left[ \hat{Q}_{L_{MC}} \right] \approx \operatorname{\mathbb{V}ar} \left[ \hat{Q}_{L_{MC}} \right] $, and
        \item $ \alpha = \alpha^\dagger $ (i.e., the underlying value that is estimated by $ \tilde{\alpha} $ (Eq.~\eqref{eqn:limiting_or_true_value_of_cvis_alpha}).)
    \end{enumerate}
    we have that $ \operatorname{\mathbb{V}ar} \left[ \hat{P}_{F_{CVMC}} \right] \leq \operatorname{\mathbb{V}ar} \left[ \hat{Q}_{MC} \right] $ (i.e., variance reduction occurs in $ \hat{P}_{F_{CVMC}} $), if 
    \begin{equation}
    \label{eqn:kappa_definition_and_variance_reduction_diagnostic}
        \kappa = \frac{P_{HL}}{P_F} \geq \frac{1}{2}
    \end{equation}
\end{theorem}
Theorem~\ref{thm:CVIS_alpha_variance_reduction_diagnostic} can be interpreted to mean that variance reduction will be achieved if the intersection of the LF and HF failure domains account for at least half of the true failure probability.

\begin{corollary}
    \label{corollary:CVIS_alpha_diagnostic_terms_of_correlation}
    Under the conditions 1-4 of theorem~\ref{thm:CVIS_alpha_variance_reduction_diagnostic}, $ \operatorname{\mathbb{V}ar} \left[ \hat{P}_{F_{CVMC}} \right] \leq \operatorname{\mathbb{V}ar} \left[ \hat{Q} \right] \Rightarrow \tilde{\alpha} \leq 4 \rho_{HL}^2 $, where $ \rho_{HL} $ is the correlation between $ I_H (\mathbf{x}) $ and $ I_L (\mathbf{x}) $, i.e., 
    \begin{equation}
        \rho_{HL}^2 = \frac{ \left( P_{HL} - P_F P_{F_L} \right)^2}{P_F P_{F_L} \left( 1 - P_{F} \right) \left( 1 - P_{F_L} \right)}
    \end{equation}
\end{corollary}

According to Theorem~\ref{thm:CVIS_alpha_variance_reduction_diagnostic}, variance reduction will be achieved if $\kappa\ge \dfrac{1}{2}$ or, by Corollary~\ref{corollary:CVIS_alpha_diagnostic_terms_of_correlation} if $\tilde{\alpha} \leq 4 \rho_{HL}^2$. 
The practical estimation of $\kappa$
will be discussed in Section~\ref{section:CVIS_algorithm_and_implementation}. Additionally, Theorem~\ref{thm:model_limit_CVIS_alpha_optimal} shows that $ \alpha^{\dagger} $ asymptotically converges to the optimal $ \alpha^* $ 
as the LF model predictive quality improves. This fact, along with various practical benefits discussed in Section~\ref{section:CVIS_algorithm_and_implementation}, provides strong justification for our formulation of the CVIS estimator.

\begin{remark}
    One useful interpretation of Theorem~\ref{thm:CVIS_alpha_variance_reduction_diagnostic} is that under the stipulated conditions, if CV leads to variance reduction compared to pure Monte Carlo simulation, then the ``size'' of the set where both HF and LF models indicate failure must be at least half of the ``size'' of the set where HF indicates failure. If we consider the ``size'' of a set to be the probability measure denoted by $ P \left( \cdot \right) $, we see that
    \begin{multline}
        \mathcal{P} \left( \left\{ \mathbf{x} \in \Omega : I_H (\mathbf{x}) = 1 \right\} \cap \left\{ \mathbf{x} \in \Omega : I_L (\mathbf{x}) = 1 \right\} \right) \geq \frac{1}{2} \mathcal{P} \left( \left\{ \mathbf{x} \in \Omega : I_H (\mathbf{x}) = 1 \right\} \right) \\
        \Rightarrow \mathcal{P} \left( \left\{ \mathbf{x} \in \Omega : I_H (\mathbf{x}) = 1 , I_L (\mathbf{x}) = 1 \right\} \right) \geq \mathcal{P} \left( \left\{ \mathbf{x} \in \Omega : I_H (\mathbf{x}) = 1 , I_L (\mathbf{x}) = 0 \right\} \right)
    \end{multline}
    This gives us further intuition into the model quality necessary for our framework, and, in theory, allows us to assert whether a value of $ \delta_{\mathcal{L}} $ as defined by stipulation~\ref{stipulation:fundamental_model_quality_assumption} is small enough. Simply stated, to achieve variance reduction the LF model must correctly predict failure more often than it incorrectly predicts safety.
\end{remark}

\subsubsection{Practical Variance Reduction: CV vs.\ IS}
\label{section:CV_vs_IS}

An interesting observation can be made about the relative usefulness of IS and CV in achieving variance reduction in the CVIS framework. Notice that as $ S_L \left( \mathbf{x}, \beta \right) \to I_L (\mathbf{x}) $, $ \operatorname{\mathbb{V}ar}_q \left[ I_L (\mathbf{X}) \right] \to 0 \Rightarrow \operatorname{\mathbb{V}ar}_q \left[ \nicefrac{I_L (\mathbf{X})}{S_L (\mathbf{X}, \beta)} \right] \to 0 $. Consequently, $ \operatorname{\mathbb{C}ov}_q \left[ \nicefrac{I_H (\mathbf{X})}{S_L (\mathbf{X}, \beta)}, \nicefrac{I_L (\mathbf{X})}{S_L (\mathbf{X}, \beta)} \right] \to 0 $ by definition.

Therefore, as IS gets more efficient, there is less covariance between the models for the CV formulation to leverage for further variance reduction (and less overall variance to be reduced). Additionally, 
using CV with IS for the component reliability estimators is, of course, significantly more effective than using CV with only MC estimators because classical MC estimators are notorious poor at estimating small failure probabilities. 


Thus, in practice, CV should not be viewed as the primary variance reduction mechanism for reliability. Rather, we recommend using CV as a supplementary tool to improve robustness, which reduces residual variance from suboptimality in IS. 
In this view, it becomes less relevant to estimate the optimal tuning constant $ \alpha^* $, as using any other $ \alpha $ value only leads to a minute difference in estimator efficiency, provided it does not increase the variance when applied. This provides clear motivation for sub-optimal formulations of $ \alpha $, such as ours, that provide alternative benefits.

In the particular case of the proposed CVIS framework, using $ \Tilde{\alpha} $ (Eq.~\eqref{eqn:CVIS_alpha}) simplifies the resulting estimator, which allows us to provide a closed-form variance estimator that incorporates the uncertainty in the estimation of $ \alpha $, unlike most other algorithms that treat the estimated optimal $ \alpha $ value as a constant~\cite{Gorodetsky2020}. Our framework also only requires the form of the ISD up to an integration constant, which is a significant benefit. Of course, the primary benefit remains that we do not need to estimate the covariance between $ I_H (\mathbf{x}) $ and $ I_L (\mathbf{x}) $ to apply the estimator.

\subsection{Statistical Properties of the Proposed Estimator}
\label{section:statistics_of_estimator}

We present here statistical properties of the estimators $ \Tilde{P}_{F} $ and $ \Tilde{\alpha} $, as defined in Eqs.~\eqref{eqn:CVIS_estimator} and~\eqref{eqn:CVIS_alpha} respectively.

\begin{theorem}
\label{thm:alpha_variance}
    For sufficiently large $ N $ (as required in Eqs.~\eqref{eqn:HF_IS_estimator_for_CVIS} and~\eqref{eqn:LF_IS_estimator_for_CVIS}),
    \begin{equation}
    \label{eqn:alpha_distribution}
        \ln{\left( \Tilde{\alpha} \right)} \sim \mathcal{N} \left( \mu_{\alpha}, \sigma_{\alpha}^2 \right)
        \Rightarrow \Tilde{\alpha} \sim \text{Lognormal} \left( \mu_{\alpha}, \sigma_{\alpha}^2 \right)
    \end{equation}
    where
    \begin{gather}
        \mu_{\alpha} = \ln{\left( \frac{\mathbb{E} \left[ \Tilde{\mathcal{Q}} \right]}{\mathbb{E} \left[ \Tilde{\mathcal{Q}}_{L} \right]} \right)} \label{eqn:alpha_distribution_mean_definition} \; \text{ and } \;
        \sigma_{\alpha}^2 =  \frac{\operatorname{\mathbb{V}ar} \left[ \Tilde{\mathcal{Q}} \right]}{\left( \mathbb{E} \left[ \Tilde{\mathcal{Q}} \right] \right)^2} + \frac{\operatorname{\mathbb{V}ar} \left[ \Tilde{\mathcal{Q}}_{L} \right]}{\left( \mathbb{E} \left[ \Tilde{\mathcal{Q}}_{L} \right] \right)^2} 
    \end{gather}
\end{theorem}

The proof is presented in Appendix~\ref{appendix:proof_alpha_variance}. Using known results for the Lognormal distribution, we get
\begin{gather}
    \mathbb{E} \left[ \Tilde{\alpha} \right] = \exp{ \left( \mu_{\alpha} + \frac{\sigma_{\alpha}^2}{2} \right)} \label{eqn:alpha_expected_value} \\
    \operatorname{\mathbb{V}ar} \left[ \Tilde{\alpha} \right] = \left\{ \exp{ \left( \sigma_{\alpha}^2 \right)} - 1 \right\} \exp{ \left( 2 \mu_{\alpha} + \sigma_{\alpha}^2 \right) } \label{eqn:alpha_variance}
\end{gather}

Since $ \Tilde{\mathcal{Q}} $ and $ \Tilde{\mathcal{Q}}_{L} $ are unbiased and consistent estimators by construction,
$ \sigma_{\alpha} \to 0 $ as $ N \to \infty $, implying that $ \Tilde{\alpha} $ is asymptotically unbiased and consistent.

If we once again consider that $ \Tilde{P}_{F_{L}} $ is independent of $ \Tilde{Q} $ and $ \Tilde{Q}_{L} $ (or $ \Tilde{\mathcal{Q}} $ and $ \Tilde{\mathcal{Q}}_{L} $), then $ \Tilde{P}_{F_{L}} $ is independent of $ \Tilde{\alpha} $, and we can write the variance of the proposed estimator $ \Tilde{P}_{F} $ as
\begin{equation}
    \label{eqn:variance_of_proposed_CVIS_estimator}
    \operatorname{\mathbb{V}ar} \left[ \Tilde{P}_{F} \right] = \left( \mathbb{E} \left[ \Tilde{\alpha} \right] \right)^2 \operatorname{\mathbb{V}ar} \left[ \Tilde{P}_{F_{L}} \right] + \left( \mathbb{E} \left[ \Tilde{P}_{F_{L}} \right] \right)^2 \operatorname{\mathbb{V}ar} \left[ \Tilde{\alpha} \right] + \operatorname{\mathbb{V}ar} \left[ \Tilde{\alpha} \right] \operatorname{\mathbb{V}ar} \left[ \Tilde{P}_{F_{L}} \right]
\end{equation}
which follows directly from the definition of the variance for the product of two independent random variables. Consequently, the coefficient of variation of $ \Tilde{P}_{F} $ is
\begin{equation}
    \label{eqn:cov_of_proposed_CVIS_estimator}
    \operatorname{CoV} \left[ \Tilde{P}_{F} \right] = \sqrt{ \left( \operatorname{CoV} \left[ \Tilde{\alpha} \right] \right)^2 + \left( \operatorname{CoV} \left[ \Tilde{P}_{F_{L}} \right] \right)^2 + \left( \operatorname{CoV} \left[ \Tilde{\alpha} \right] \operatorname{CoV} \left[ \Tilde{P}_{F_{L}} \right] \right)^2 }
\end{equation}

When the LF model is significantly cheaper than the HF model (such as for data-driven surrogates), it is easy to reduce the coefficient of variation of $ \Tilde{P}_{F_L} $ to near zero such that: 
\begin{equation}
    \label{eqn:cov_of_CVIS_simplified_cheap_LF}
    \operatorname{CoV} \left[ \Tilde{P}_{F} \right] \approx \operatorname{CoV} \left[ \Tilde{\alpha} \right].
\end{equation}
In these cases, the coefficient of variation of $ \Tilde{P}_F $ is limited by the accuracy of $ \Tilde{\alpha} $ as below. Since $ \Tilde{\alpha} $ always requires the highly expensive HF model evaluations, its coefficient of variation is restricted by the total computational budget even if the estimation of $ \Tilde{P}_{F_L} $ is of negligible expense.

\subsection{Sample Allocation for CVIS}
\label{section:sample_allocation}

In most CV applications, 
an important consideration is the optimal sample allocation, i.e., the number of samples that are assigned to each individual model to achieve a CV estimate with the minimum variance for a given computational budget. The general problem was formulated by Gorodetsky et al.~\cite{Gorodetsky2020} in the following manner. Let the total computational budget be given by $ C $, and let $ w $ and $ w_L $ be the cost of a single evaluation of the HF and LF models, respectively. Further, let us represent the number of HF evaluations by $ N $ and the number of LF evaluations for every HF evaluation as $ r $, such that the total number of LF evaluations is given by $ rN $ ($ N \geq 1 $ and $ r \geq 1 $). Then, the optimal sample allocation refers to finding the optimal values $ N^* $ and $ r^* $, defined as
\begin{align}
\begin{split}
    \left( N^*, r^* \right) &= \arg \min_{N, r} \mathbb{V}ar \left[ \hat{P}_{F_{CV}} \right] \label{eqn:optimal_sample_allocation_acv} \\
    &\text{subject to } N \left( w + r w_L \right) \leq C
\end{split}
\end{align}
where $ \hat{P}_{F_{CV}} $ is as defined in Eq.~\eqref{eqn:rearranged_acv_estimator}. Unfortunately, $ \mathbb{V}ar \left[ \hat{P}_{F_{CV}} \right] $ depends not only on $ N $ and $ r $, but also on the mean and variance of $ \hat{Q} $, $ \hat{Q}_L $ and $ \hat{P}_{F_L} $, as well as the covariance between $ \hat{Q} $ and $ \hat{Q}_L $. Usually, coarse estimates of these statistics are computed using pilot samples\footnote{Pilot samples are samples generated during a pilot run.} and are used to find approximately optimal values of $ N^* $ and $ r^* $. It is accepted that a misestimation of the statistics does lead to some degradation of performance compared to the theoretical optimal; however, a near-optimal sample allocation still leads to better performance compared to a blindly chosen sample allocation.

In the above context, a pilot run refers to the generation of a small number of samples (as compared to the number of samples generated for the final estimate $ \hat{P}_{F_{CV}} $) 
from the input distribution to compute preliminary estimates for the problem. In general problems, these estimates are serviceable despite having moderate-to-high variance, depending on the number of pilot samples. Unfortunately, in failure probability estimation, this requires estimating the statistics of an indicator function having an exceedingly rare probability of taking a non-zero value. In other words, due to the discrete binary nature of the failure indicator function, any pilot sample set that does not sample the failure domain a statistically significant number of times is of practically no use. 
Although the pilot estimators may be theoretically unbiased, small-sample estimates of small failure probabilities are notoriously unreliable; a small set of samples rarely explores the failure domain yielding a failure probability estimate of zero, which is of no use for the CV estimate. On the other hand, if the small sample estimate does, by chance, sample from the failure domain, then $ \hat{Q}$, $\hat{Q}_L$, or $\hat{P}_{F_L}$ will likely be greatly overestimated and, again, will not be useful for the CV estimate.
Therefore, in such a context, a pilot run loses meaning, as even coarse estimates require very large sample sizes.

Instead of pilot runs, most rare-event reliability algorithms call for an \textit{exploration run}, which is used to find the failure domain and thereby enable the generation of more informative samples for the final estimator $ \hat{P}_F $. This entails generating an appreciable number of samples (as compared to the number of samples generated for the final estimate) such that the input space $ \Omega $ is sufficiently explored. The exploration run must be carried out until reasonable confidence in having found all important failure regions is achieved. Often, the number of samples required for such an exploration run is decided adaptively until some confidence criterion is met. Clearly, it is infeasible to do such a start-up run using the HF model directly. In a multifidelity setting such as the proposed CVIS framework, the LF model can be used to carry out the exploration run instead. In addition to identifying the failure domain, the exploration run also typically results in robust estimates of the LF model statistics (e.g., $ \hat{P}_{F_L} $). This estimate is then subsequently used in the CV problem.

In the proposed CVIS framework, we estimate $ \Tilde{P}_{F_L} $ from Eq.~\eqref{eqn:CVIS_estimator} using a suitable reliability method (e.g.\ subset simulation) on the LF model as an exploration run. We then must 
accurately estimate $ \Tilde{\alpha} $ in order to estimate $ \Tilde{P}_F $. 
This calls for an equal number of HF and LF model evaluations, since both models must be evaluated on the same set of samples. Thus, in our framework, the standard CV notion of optimal sample allocation does not hold.
Instead, given a total budget $C$, a natural sample allocation scheme follows:
\begin{itemize}
    \item First, conduct an exploration run using the LF model until all important regions are identified, and estimate $ \Tilde{P}_{F_L} $ as a byproduct. Let this budget be denoted $ C_E = w_L N_E$ where the number of LF model evaluations required for exploration is $ N_E $.
    \item Next, generate a number of samples $ N_Q $ (to estimate $ \Tilde{Q} $ and $ \Tilde{Q}_L $) using IS on which both the HF and LF models are evaluated. Choose $ N_Q $, having budget $ C_Q = N_Q (w + w_L) $, 
    such that $  \operatorname{CoV}\left[ \Tilde{\alpha} \right] \approx  \operatorname{CoV} \left[ \Tilde{P}_{F_L} \right] $ (which is optimal according to Eq.~\eqref{eqn:variance_of_proposed_CVIS_estimator} \footnote{If the coefficient of variation of one of the component estimators is much larger than the other, then the larger term dominates $ \operatorname{CoV} \left[ \Tilde{P}_F \right] $.}), or until the total budget $C$ is exhausted (i.e. $C_Q=C-C_E$). 
    \item If $ C_E + C_Q < C $, denote the surplus budget as $ C_S = C - C_E - C_Q $. This surplus can now be optimally allocated according to the following procedure:
    \begin{equation}
        \begin{split}
            \left( N_{S_Q}^*, N_{S_L}^* \right) &= \arg \min_{N_{S_Q}, N_{S_L}} \mathbb{V}ar \left[ \Tilde{P}_F \right] \label{eqn:optimal_surplus_sample_allocation_cvis} \\
    &\text{subject to } (w + w_L) N_{S_Q} + w_L N_{S_L} = C_S
        \end{split}
    \end{equation}
    
    Here $ N_{S_Q} $ is defined as the additional number of samples that can be generated to refine $ \Tilde{Q} $ and $ \Tilde{Q}_L $, and $ N_{S_L} $ is the additional number of samples that can be generated to refine $ \Tilde{P}_{F_L} $. Consequently, $ \mathbb{V}ar \left[ \Tilde{P}_F \right] $ (Eq.~\eqref{eqn:variance_of_proposed_CVIS_estimator}) corresponds to $ N_Q + N_{S_Q} $ samples being used to evaluate $ \Tilde{Q} $ and $ \Tilde{Q}_L $, and $ N_E + N_{S_L} $ samples being used to evaluate $ \Tilde{P}_{F_L} $. The optimization in Eq.~\eqref{eqn:optimal_surplus_sample_allocation_cvis} can, of course, be solved because robust estimates of all the necessary statistics are available from steps 1 and 2.
\end{itemize}

At the end of this procedure, the HF model is evaluated a total of $ N_Q + N_{S_Q}^* $ times, while the LF model is evaluated $ N_E + N_Q + N_{S_Q}^* + N_{S_L}^* $ times.

    

\section{The CVIS Algorithm}
\label{section:CVIS_algorithm_and_implementation}

This section presents the proposed CVIS algorithm, which provides a practical way to use the CVIS estimator for bifidelity Reliability Analysis problems. We first outline the steps of the proposed methods in Algorithm~\ref{Algo:CVIS_procedure}. We then make recommendations for specific methods to use in each step, as appropriate. 

\begin{algorithm}[h!t]
\caption{CVIS Procedure}
\label{Algo:CVIS_procedure}
\begin{codefont}
\begin{algorithmic}[1]
\Require $ I_H (\mathbf{x}) $, $ I_L (\mathbf{x}) $, $ \mathfrak{q}_{\mathbf{X}} (\mathbf{x}, \beta) $, $ \tau$ \Comment{$ \tau $ is a pre-specified threshold.}
\State Estimate $ \Tilde{P}_{F_L} $ such that $ \operatorname{CoV} \left[ \Tilde{P}_{F_L} \right] \leq \tau $ 
\State Determine $ \beta^* $ (Eq.~\eqref{eqn:practical_choice_for_beta_star})
\State Draw $ \mathbf{x}_1, \mathbf{x}_2, \dots, \mathbf{x}_N \sim \mathfrak{q}_{\mathbf{X}} (\mathbf{x}, \beta^*) $
\State $ \Tilde{Q} = \sum_{i=1}^N I_H (\mathbf{x}_i) / N $ (see Section~\ref{section:methodology})
\State $ \Tilde{Q}_L = \sum_{i=1}^N I_L (\mathbf{x}_i) / N $ (see Section~\ref{section:methodology})
\State Estimate $ \Tilde{\alpha} = \Tilde{Q} / \Tilde{Q}_L $ (Eq.~\eqref{eqn:proposed_alpha_definition_alternate})
\State Estimate $ \Tilde{P}_F = \Tilde{\alpha} \Tilde{P}_{F_L} $ (Eq.~\eqref{eqn:CVIS_estimator})
\State Compute $ \kappa $ (see Theorem~\ref{thm:CVIS_alpha_variance_reduction_diagnostic}) \Comment{If a diagnostic is desired}
\State Compute $ \mu_{\alpha} $ and $ \sigma_{\alpha} $ (Eq.~\eqref{eqn:alpha_distribution_mean_definition})
\State Compute $ \operatorname{CoV} \left[ \Tilde{\alpha} \right] = \sqrt{ \left[ \exp (\sigma_{\alpha}^2) - 1 \right] \exp \left( \mu_{\alpha} + \left( \sigma_{\alpha}^2 / 2 \right) \right) } $ (Eqs.~\eqref{eqn:alpha_expected_value},~\eqref{eqn:alpha_variance})
\State Compute $ \operatorname{CoV} \left[ \Tilde{P}_{F} \right] $ (Eq.~\eqref{eqn:cov_of_proposed_CVIS_estimator})
\If{$ \operatorname{CoV} \left[ \Tilde{P}_{F} \right] > \tau $ and computational budget is not exhausted}
\State Refine $ \Tilde{P}_F $ in accordance with Section~\ref{section:sample_allocation}
\Else
\State Stop procedure and use $ \Tilde{P}_F $ and $ \operatorname{CoV} \left[ \Tilde{P}_{F} \right] $
\EndIf
\end{algorithmic}
\end{codefont}
\end{algorithm}

In summary, the procedure consists of first estimating $ \Tilde{P}_{F_L} $ to an acceptable, pre-decided level of accuracy $ \tau $. Next, the ISD of our chosen form is tuned, i.e., a value of $ \beta $ is selected using the samples generated while estimating $ \Tilde{P}_{F_L} $. Samples are subsequently generated from the ISD to estimate $ \Tilde{\alpha} $ such that $ \operatorname{CoV} \left[ \tilde{\alpha} \right] \approx \operatorname{CoV} \left[ \Tilde{P}_{F_L} \right] $, or the computational budget is exhausted. If any computational budget remains, then further samples can be allocated according to the optimization scheme in Eq.~\eqref{eqn:optimal_surplus_sample_allocation_cvis} (See Section~\ref{section:sample_allocation} for more details). At the outset of this procedure, we assume the threshold $ \tau $ is sufficiently small that all estimators are asymptotically stable.

To estimate $ \Tilde{P}_{F_{L}} $, we recommend using Subset Simulation (SuS)~\cite{AuBeck2001}, which is the state-of-the-art for single fidelity rare-event failure probability estimation. The method works by constructing a sequence of nested subsets $ \mathcal{S}_i = \left\{ \mathbf{x} \in \Omega: g (\mathbf{x}) \leq b_i \right\} $, $ i = 0, 1, \dots M_{S} $,
such that $\mathcal{S}_{i}\subseteq \mathcal{S}_{i-1}$, that converges to the failure region $ \Omega_{\mathcal{F}} = \mathcal{S}_{M_{S}} = \left\{ \mathbf{x} \in \Omega: g (\mathbf{x}) \leq b_{M_{S}} = 0 \right\} $.
Samples are then drawn from the conditional probability distributions $ f_{\mathbf{X}} (\mathbf{x} \ | \ \mathbf{x} \in \mathcal{S}_{i-1}) $ to estimate a sequence of conditional probabilities $P_{i|i-1}$ from which the probability of failure is estimated as $P_F=P_0\prod_{i=1}^{M_S}P_{i|i-1}$ (where $ P_0 = 1 $ as $ b_0 = \infty \Rightarrow \mathcal{S}_0 = \Omega $ by definition). 
This allows us to efficiently compute $ \Tilde{P}_{F_{L}} $ while keeping $ \operatorname{CoV} \left[ \Tilde{P}_{F_{L}} \right] $ low by simply setting $ g (\mathbf{x}) = L (\mathbf{x}) $. 
Next, we use the response function threshold value $ b_{M_{S}-1} $ and the associated probability $ P_{M_{S}|(M_{S}-1)} $ to estimate $ \beta^* $ directly (without having to compute $ \delta_{\mathcal{L}} $) as:
\begin{equation}
\label{eqn:practical_choice_for_beta_star}
    \beta^* = \frac{2}{b_{M_{S}-1}} \ln \left( \frac{2}{P_{M_{S}|(M_{S}-1)}} - 1 \right) 
\end{equation}
A brief justification for Eq.~\eqref{eqn:practical_choice_for_beta_star} is given in Appendix~\ref{appendix:LF_Pf_estimator_and_ISD_construction}. Although this is a heuristic approach, we have found this estimate to work well in practice. All the examples studied in Section~\ref{section:results_experiments_discussions} use this procedure to define the ISD.

Since our ISD is constructed to retain the full complexity of the LF model, it can, in general, be non-elliptic, multimodal, poorly scaled, or even degenerate. To sample from such complex distributions, as is necessary for evaluating $ \Tilde{\mathcal{Q}} $ and $ \Tilde{\mathcal{Q}}_{L} $, we use Differential Evolution Markov Chain (DE-MC)~\cite{terBraak,terBraakVrugt}, which is a Markov Chain Monte Carlo (MCMC) sampling method adapted from genetic algorithms for numerical optimization.
See Appendix~\ref{appendix:Sampling_from_k} for a brief discussion about initializing DE-MC in this context. However, any similarly robust sampling scheme can be applied.
Subsequently, to evaluate the estimator variance of $ \Tilde{\mathcal{Q}} $ and $ \Tilde{\mathcal{Q}}_{L} $, 
we recommend using the Replicated Batch Means (RBM) estimator~\cite{ArgonAndradottirRBM,gupta2020globallycenteredbatchmeans}, which is specially constructed for MCMC methods that use multiple parallel independent Markov Chains. (Note that a specialized estimator is necessary, as the naive estimator of the Markov Chain asymptotic variance is not a consistent estimator. Further details are provided in Appendix~\ref{appendix:estimating_Q_or_alpha_stats}) 

Finally, to estimate the diagnostic $ \kappa $, we can use the following estimator,
\begin{gather}
    \Tilde{\kappa} = \frac{\tilde{\mathcal{Q}}_{HL}}{\Tilde{\mathcal{Q}}} \label{eqn:kappa_estimator} \\
    \Tilde{\mathcal{Q}}_{HL} = \frac{1}{N} \sum_{i=1}^N \left[ \frac{I_H (\mathbf{x}_i) I_L (\mathbf{x}_i)}{S_L (\mathbf{x}_i, \beta^*)} \right] \label{eqn:model_intersection_IS_estimators}
\end{gather}
where the same $ \mathbf{x}_i $, $ i = 1, 2, \dots, N $ are used for $ \Tilde{\mathcal{Q}}_{HL} $ as were used for $ \Tilde{\mathcal{Q}} $ and $ \Tilde{\mathcal{Q}}_{L} $. Hence, this estimator requires no additional model calls beyond those necessary for estimating $ \Tilde{\mathcal{Q}} $ and $ \Tilde{\mathcal{Q}}_{L} $.

\section{Numerical Examples, Results, and Discussion}
\label{section:results_experiments_discussions}

The performance of the CVIS algorithm is presented here, along with relevant insights, through two analytical case studies and one numerical problem. For the analytical examples, CVIS is compared to the \textit{Multifidelity Importance Sampling} (MFIS) method by Peherstorfer et al.~\cite{PEHERSTORFER_MFIS} and the \textit{Ensemble Approximate Control Variates} (E-ACV) method by Pham \& Gorodetsky~\cite{PhamGorodetsky2022}, with particular emphasis on model call requirements and the total uncertainty in the failure prediction. These methods 
are the best analogs for the CVIS method in the current literature. However, a few adaptations have been made in the application of these algorithms 
in order to ensure consistency with the proposed CVIS method.
These changes are discussed first in Section~\ref{section:Implementation_details}. The subsequent sections present the three examples.

\subsection{Implementation Details for MFIS and E-ACV}
\label{section:Implementation_details}

The following adjustments are made in the MFIS and E-ACV methodologies to ensure proper comparison to CVIS:

\begin{itemize}
    \item Instead of using Gaussian Mixture Models, as originally suggested for both the MFIS and E-ACV methods, the ISD is constructed according to the procedure proposed for the CVIS algorithm (as described in Section~\ref{section:CVIS_algorithm_and_implementation}). This allows us to use exactly the same sampling density for all three methods.
    \item Our choice of ISD construction provides only the unnormalized ISD. This does not affect CVIS (due to the use of the unnormalized estimators $ \Tilde{\mathcal{Q}} $ and $ \Tilde{\mathcal{Q}}_L $ (Eq.~\eqref{eqn:proposed_alpha_definition_alternate})). However, this issue must be addressed for MFIS and E-ACV. Self-normalized Importance Sampling, which is designed for IS when one or both of the input distribution or ISD are unnormalized, is highly biased for this application due to the use of a logistic function in our ISD. (This behavior is discussed in more detail in Appendix~\ref{section:SNIS_behavior}.) Therefore, in this work, we use simple MC Integration to estimate $ C_S $ (the ISD normalizing constant). When comparing results between CVIS and MFIS/E-ACV for the same computational budget, the budget assigned to estimating $ \Tilde{P}_{F_L} $ for CVIS is split between estimating $ \Tilde{P}_{F_L} $ and estimating $ C_S $ for MFIS and E-ACV.
    \item The original E-ACV algorithm uses an IS estimator for $ \hat{P}_{F_L} $ in addition to $ \hat{Q} $ and $ \hat{Q}_L $. In this paper, we instead use SuS to estimate $ \hat{P}_{F_L} $ for the E-ACV algorithm. This ensures that the performance of the LF failure estimator is the same for both CVIS and E-ACV.
    \item Following from the previous point, since $ \hat{P}_{F_L} $ is now independent of both $ \hat{Q} $ and $ \hat{Q}_L $ for the E-ACV method as well, the $ \alpha $ parameter estimation is simplified in the following way. The procedure described in \cite{PhamGorodetsky2022} uses an ensemble of realizations for each of the estimators $ \hat{P}_{F_L} $, $ \hat{Q} $, and $ \hat{Q}_L $, which are used to compute the variances of each of these estimators as well as the covariances between them. Instead, when the E-ACV estimator is used in this manuscript, only a single realization of $ \hat{P}_{F_L} $ is needed; the variance of the estimate is calculated using the estimator presented in the original SuS paper~\cite{AuBeck2001}, and the covariances between $ \hat{P}_{F_L} $ and $ \hat{Q} $ and $ \hat{P}_{F_L} $ and $ \hat{Q}_L $ are set to zero, leveraging the independence between these estimators. Everything else remains unchanged.
\end{itemize}

\subsection{Example 1: Noisy and Biased Analytical Model}
\label{section:Example_1_biased_and_noise_LF_model}

\begin{figure}[!htbp]
\centering
\begin{subfigure}{.32\textwidth}
  \centering
  \includegraphics[width=\linewidth]{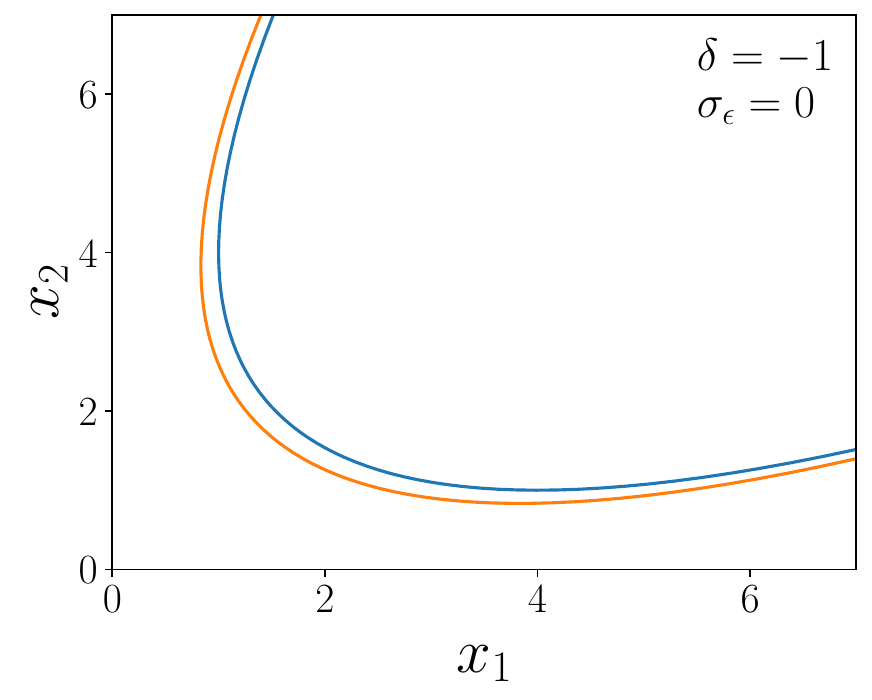}
  \caption{}
  \label{fig:Example_1_bias_-1_variance_0}
\end{subfigure}%
\begin{subfigure}{.32\textwidth}
  \centering
  \includegraphics[width=\linewidth]{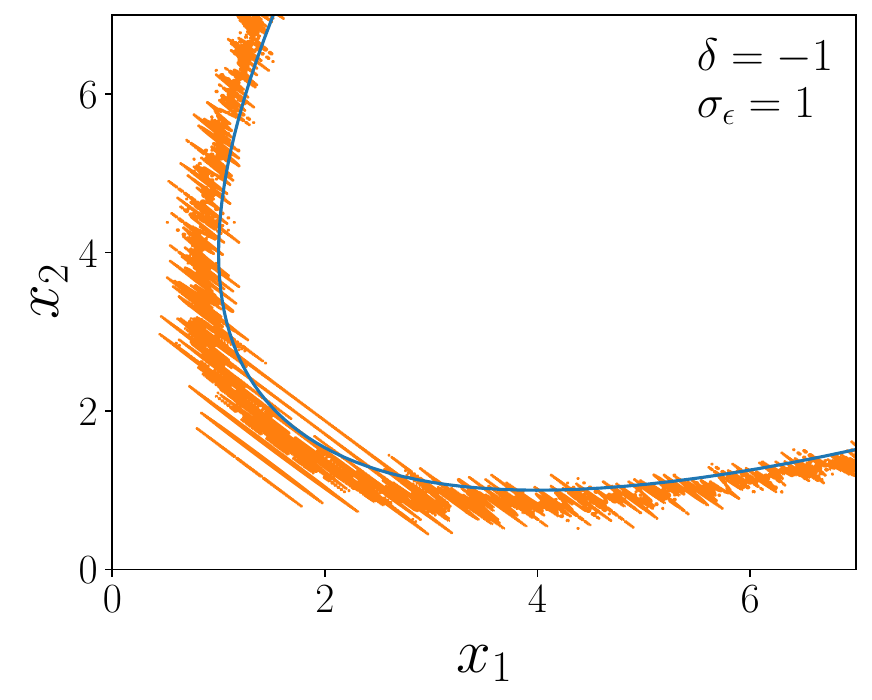}
  \caption{}
  \label{fig:Example_1_bias_-1_variance_1}
\end{subfigure}%
\begin{subfigure}{.32\textwidth}
  \centering
  \includegraphics[width=\linewidth]{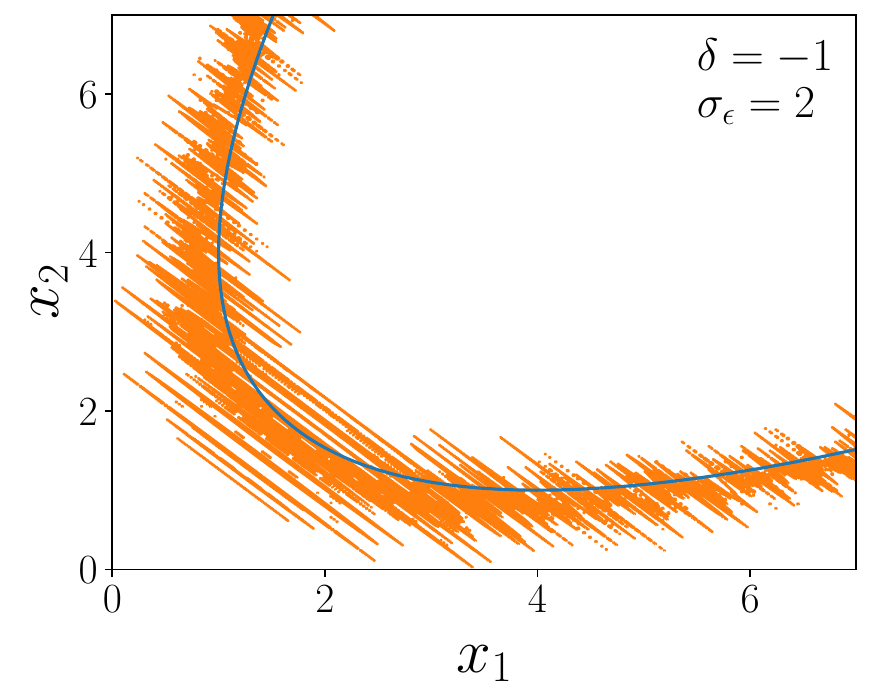}
  \caption{}
  \label{fig:Example_1_bias_-1_variance_2}
\end{subfigure}
\begin{subfigure}{.32\textwidth}
  \centering
  \includegraphics[width=\linewidth]{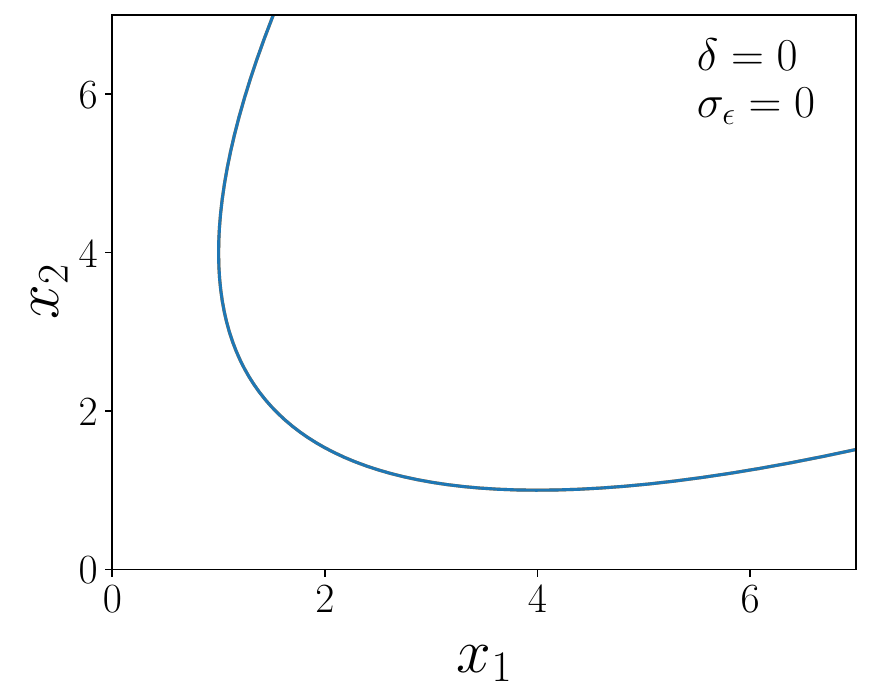}
  \caption{}
  \label{fig:Example_1_bias_0_variance_0}
\end{subfigure}%
\begin{subfigure}{.32\textwidth}
  \centering
  \includegraphics[width=\linewidth]{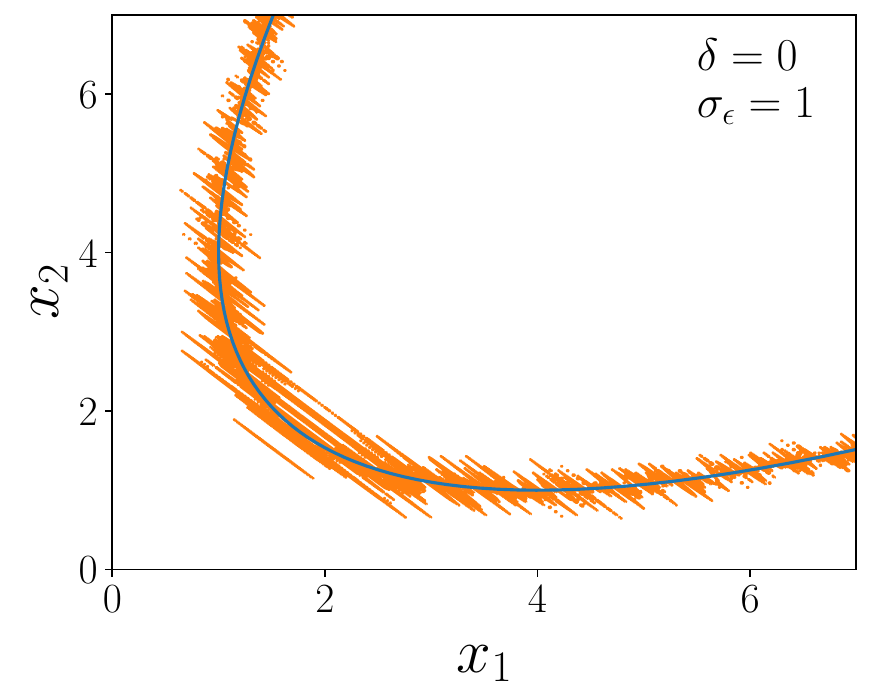}
  \caption{}
  \label{fig:Example_1_bias_0_variance_1}
\end{subfigure}%
\begin{subfigure}{.32\textwidth}
  \centering
  \includegraphics[width=\linewidth]{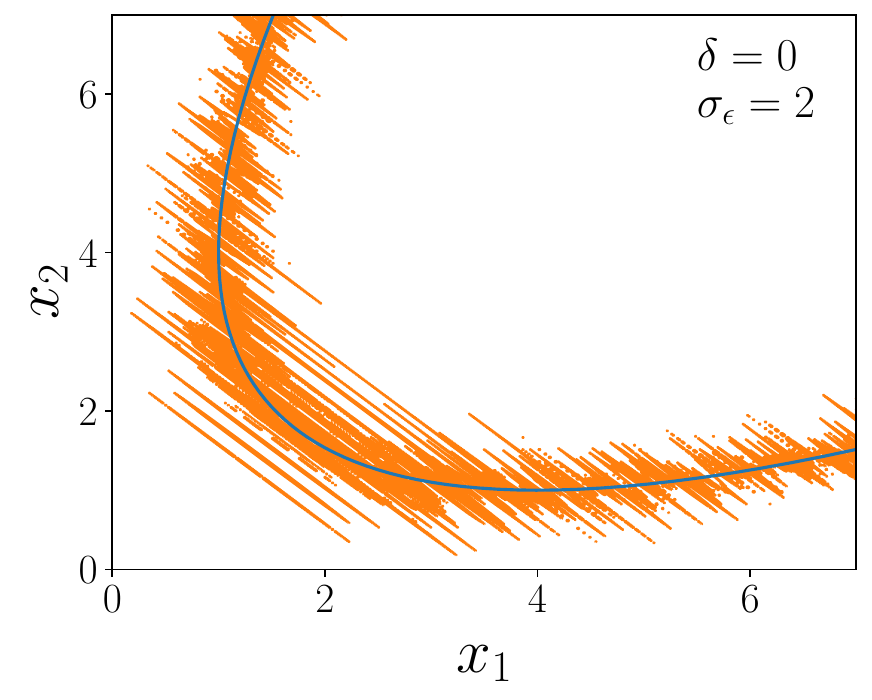}
  \caption{}
  \label{fig:Example_1_bias_0_variance_2}
\end{subfigure}
\begin{subfigure}{.32\textwidth}
  \centering
  \includegraphics[width=\linewidth]{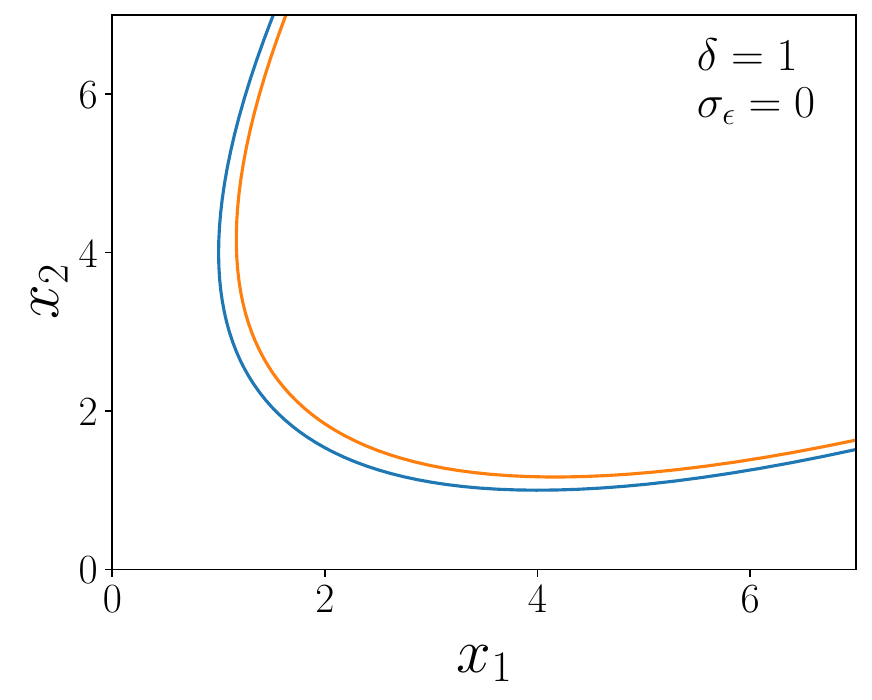}
  \caption{}
  \label{fig:Example_1_bias_1_variance_0}
\end{subfigure}%
\begin{subfigure}{.32\textwidth}
  \centering
  \includegraphics[width=\linewidth]{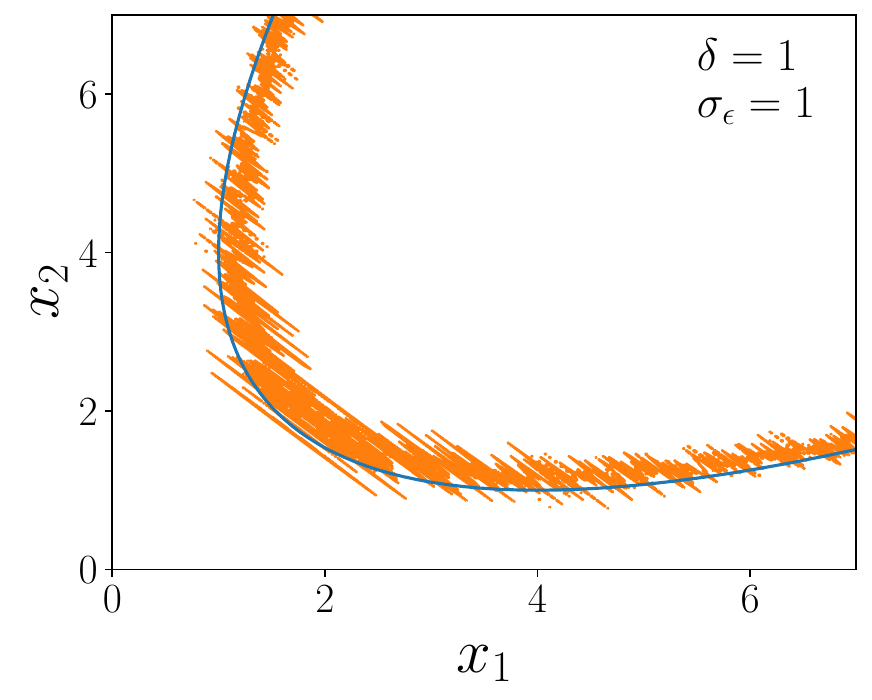}
  \caption{}
  \label{fig:Example_1_bias_1_variance_1}
\end{subfigure}%
\begin{subfigure}{.32\textwidth}
  \centering
  \includegraphics[width=\linewidth]{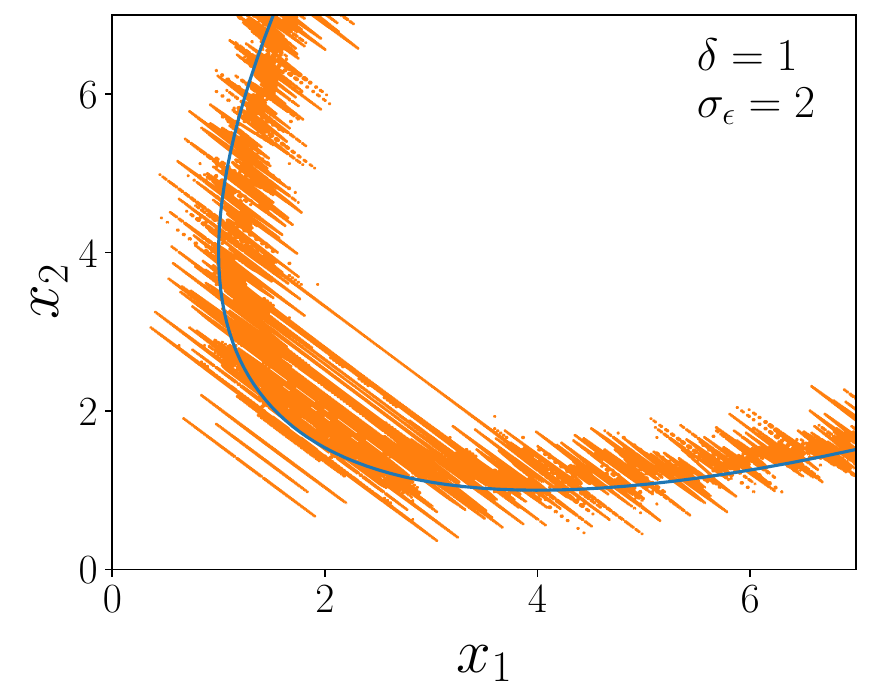}
  \caption{}
  \label{fig:Example_1_bias_1_variance_2}
\end{subfigure}
\begin{subfigure}{.32\textwidth}
  \centering
  \includegraphics[width=\linewidth]{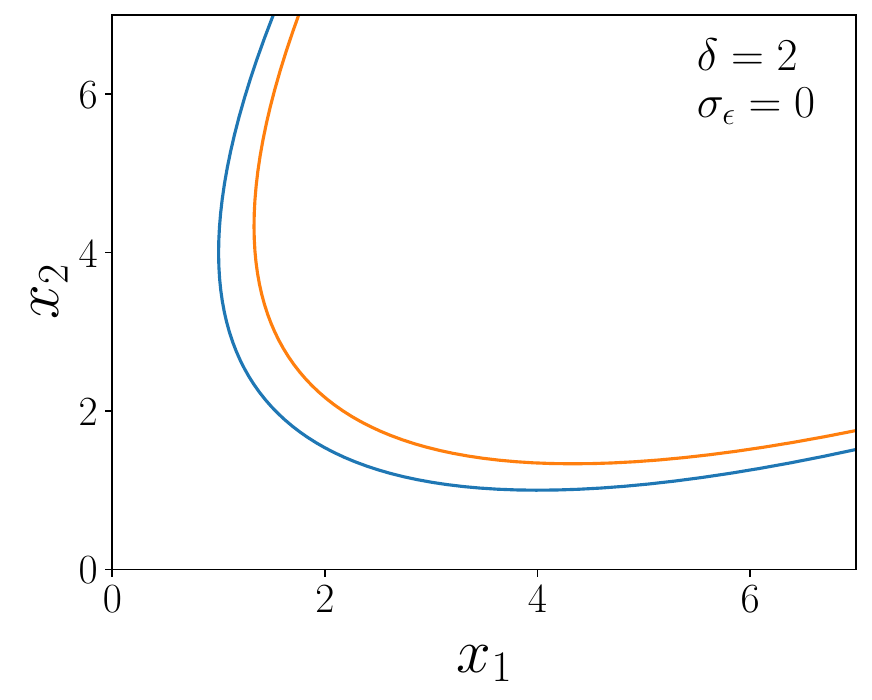}
  \caption{}
  \label{fig:Example_1_bias_2_variance_0}
\end{subfigure}%
\begin{subfigure}{.32\textwidth}
  \centering
  \includegraphics[width=\linewidth]{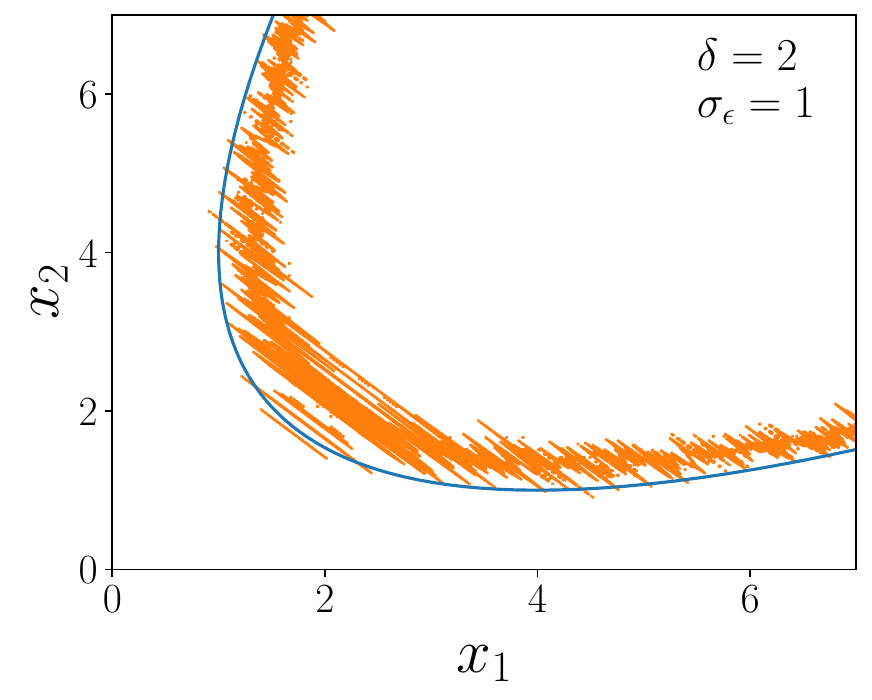}
  \caption{}
  \label{fig:Example_1_bias_2_variance_1}
\end{subfigure}%
\begin{subfigure}{.32\textwidth}
  \centering
  \includegraphics[width=\linewidth]{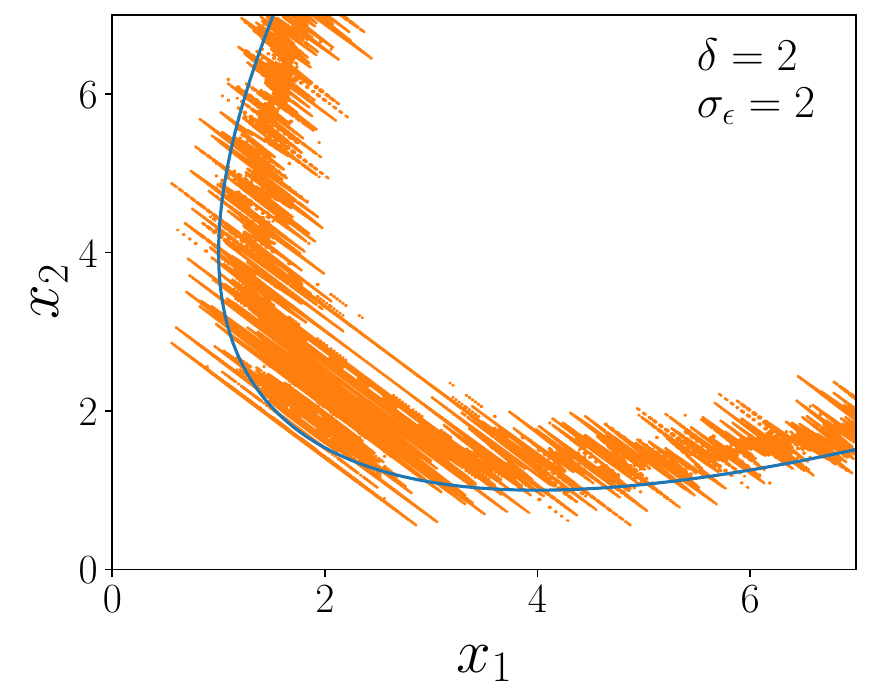}
  \caption{}
  \label{fig:Example_1_bias_2_variance_2}
\end{subfigure}
\caption{Example 1: Failure limit surface comparisons. In each figure, the solid blue line corresponds to $ H (\mathbf{x}) = 0 $, while the orange dashed line corresponds to $ L (\mathbf{x}) = 0 $. The title of each subfigure shows the values of $ \delta $ and $ \sigma_{\epsilon} $.}
\label{fig:Example_1_model_relationships}
\end{figure}

In this example, we study the performance of the proposed CVIS algorithm for varying levels of bias and noise between the HF and LF models. No difference in cost is considered between the HF and LF models, as the goal is only to characterize changes in performance for changing HF and LF model relationships. A simple analytical 2-dimensional system is considered, with the inputs following a standard Normal distribution. The HF model is given by
\begin{equation}
\label{eqn:example_1_HF_model}
    H(\mathbf{x}) = (x_1 - 2)^2 + (x_2 - 2)^2 - \frac{1}{2} \left(x_1 + x_2 - 1 \right)^2 + 3
\end{equation}
where $ \mathbf{x} = \begin{bmatrix} x_1 & x_2 \end{bmatrix}^T $. The LF model gives a biased and noisy prediction of the HF response as
\begin{equation}
\label{eqn:example_1_LF_model}
    L(\mathbf{x}) = H(\mathbf{x}) + \delta + \epsilon
\end{equation}
where $ \delta $ is a bias term and $ \epsilon \sim \mathcal{N} \left( 0, \sigma_{\epsilon}^2 \right) $ is a noise term. The existence of $ \epsilon $ makes $ L (\mathbf{x}) $ a stochastic field; however, for ease of analysis, only a single realization was used. Twelve cases were considered in all, corresponding to all combinations of $ \delta \in \left\{ -1, 0, 1, 2 \right\} $ and $ \sigma_{\epsilon} \in \left\{ 0, 1, 2 \right\} $. From Eqs.~\eqref{eqn:example_1_HF_model} and~\eqref{eqn:example_1_LF_model} it is clear to see that, for this example, the HF and LF models have comparable computational costs. The failure surfaces are compared for each case in Figure~\ref{fig:Example_1_model_relationships}. 

Table~\ref{tab:example_1_model_statistics} shows some statistics of the models, including the true failure probability as well as the correlation between the HF and LF models and the correlation between their respective indicator functions. In fact, this highlights one key reason it is difficult to apply CV for multi-fidelity reliability analysis. 
The standard CV estimator requires correlation between the indicators to achieve variance reduction. Even if two models are well correlated in terms of their response function predictions, the indicators might have a much weaker correlation, as illustrated in Table~\ref{tab:example_1_model_statistics}. Furthermore, even if model response correlation information is available through physical laws or engineering information, this usually does not translate into information about the correlation between the failure indicator functions constructed from the models. 
Therefore, a formulation like CVIS that does not require explicit information about the correlation between failure indicator functions for each model to achieve variance reduction is very convenient.
\begin{table}[!ht]
    \centering
    \begin{tabular}{c|c|c|c}
    Model & Failure Probability & $ \operatorname{\mathbb{\rho}} \left[ H(\mathbf{X}), L(\mathbf{X}) \right] $ & $ \operatorname{\mathbb{\rho}} \left[ I_H(\mathbf{X}), I_L(\mathbf{X}) \right] $ \\
    \hline \hline
    LF: $ \delta = -1 $, $ \sigma_{\epsilon} = 0 $ & 8.42 E-3 & 1.00 & 0.71 \\
    LF: $ \delta = -1 $, $ \sigma_{\epsilon} = 1 $ & 9.32 E-3 & 0.98 & 0.63 \\
    LF: $ \delta = -1 $, $ \sigma_{\epsilon} = 2 $ & 14.05 E-3 & 0.91 & 0.47 \\
    LF: $ \delta = 0 $, $ \sigma_{\epsilon} = 0 $ & 4.23 E-3 & 1.00 & 1.00 \\
    LF: $ \delta = 0 $, $ \sigma_{\epsilon} = 1 $ & 5.12 E-3 & 0.98 & 0.73 \\
    LF: $ \delta = 0 $, $ \sigma_{\epsilon} = 2 $ & 7.93 E-3 & 0.91 & 0.52 \\
    LF: $ \delta = 1 $, $ \sigma_{\epsilon} = 0 $ & 2.10 E-3 & 1.00 & 0.70 \\
    LF: $ \delta = 1 $, $ \sigma_{\epsilon} = 1 $ & 2.65 E-3 & 0.98 & 0.69 \\
    LF: $ \delta = 1 $, $ \sigma_{\epsilon} = 2 $ & 4.45 E-3 & 0.91 & 0.53 \\
    LF: $ \delta = 2 $, $ \sigma_{\epsilon} = 0 $ & 1.00 E-3 & 1.00 & 0.48 \\
    LF: $ \delta = 2 $, $ \sigma_{\epsilon} = 1 $ & 1.19 E-3 & 0.98 & 0.51 \\
    LF: $ \delta = 2 $, $ \sigma_{\epsilon} = 2 $ & 2.25 E-3 & 0.91 & 0.49 \\
    \hline
    HF & 4.23 E-3 & - & -
    \end{tabular}
    \caption{Example 1: HF and LF model statistics. $ \operatorname{\mathbb{\rho}} \left[ \cdot \right] $ denotes the correlation between two random variables. To compute these statistics, a set of $ 10^7 $ samples was generated from the 2D standard Normal input distribution. The HF model, as well as all 12 cases for the LF model, were evaluated on this sample set.}
    \label{tab:example_1_model_statistics}
\end{table}

For all twelve LF model cases, the sample allocations are summarised in Table~\ref{tab:example_1_sample_allocation_summary} for all methods. In all cases, the IS is conducted with 25 chains and 400 samples per chain using DE-MC for a total of 10,000 HF and LF model evaluations seeded from the SuS. The primary difference lies in the allocation of LF model evaluations prior to the initiation of IS. For CVIS, all samples are allocated to the SuS for estimation of $\tilde{P}_{F_L}$ using 10,000 samples per subset for a total of $10,000\times M_S$ LF model evaluations. 
However, for both MFIS and E-ACV 
the SuS budget from CVIS was split between SuS to estimate $\tilde{P}_{F_L}$ and Monte Carlo integration to estimate $\tilde{C}_{S}$. 
Two cases were considered: Case `A' -- 2000 samples per subset for SuS, and 8000$ \times M_{S} $ samples for the Monte Carlo integration; 
and Case `B' -- 5000 samples per subset for SuS, and 5000$ \times M_{S} $ samples for the Monte Carlo integration. Note that as the LF model changes, its failure probability changes; thus, $ M_{S} $ also changes. However, this sample allocation keeps the total number of HF and LF model calls equal among all three algorithms. 

\begin{table}[!ht]
    \centering
    \begin{tabular}{c|c|c|c|c|c|c}
    Sampling & \multicolumn{2}{c|}{IS (DE-MC)} & \multicolumn{2}{c|}{Subset Simulation} & \multicolumn{2}{c}{Monte Carlo Integration} \\
    \cline{2-7}
    Procedure & HF Calls & LF Calls & HF Calls & LF Calls & HF Calls & LF Calls \\
    \hline \hline
    CVIS & $ 10,000 $ & $ 10,000 $ & - & $ 10,000 \times M_{SS} $ & - & - \\
    MFIS-A & $ 10,000 $ & $ 10,000 $ & - & $ 2,000 \times M_{SS} $ & - & $ 8,000 \times M_{SS} $ \\
    E-ACV-A & $ 10,000 $ & $ 10,000 $ & - & $ 2,000 \times M_{SS} $ & - & $ 8,000 \times M_{SS} $ \\
    MFIS-B & $ 10,000 $ & $ 10,000 $ & - & $ 5,000 \times M_{SS} $ & - & $ 5,000 \times M_{SS} $ \\
    E-ACV-B & $ 10,000 $ & $ 10,000 $ & - & $ 5,000 \times M_{SS} $ & - & $ 5,000 \times M_{SS} $
    \end{tabular}
    \caption{Sample allocation summary for example 1. $ M_{SS} $ is the number of subsets required by the Subset Simulation procedure.}
    \label{tab:example_1_sample_allocation_summary}
\end{table}

\begin{figure}[!htbp]
\centering
\begin{subfigure}{.45\textwidth}
  \centering
  \includegraphics[width=\linewidth]{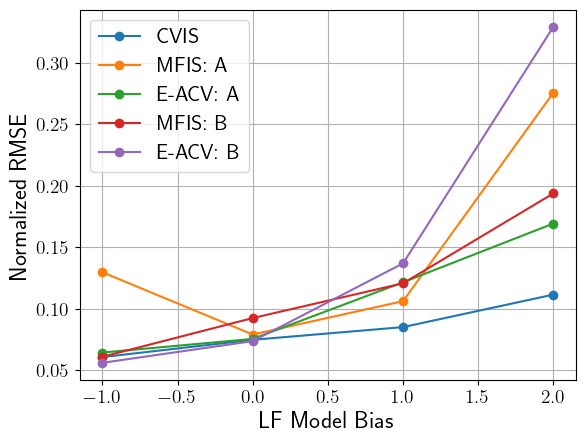}
  \caption{}
  \label{fig:RMSE_var_0}
\end{subfigure}%
\begin{subfigure}{.45\textwidth}
  \centering
  \includegraphics[width=\linewidth]{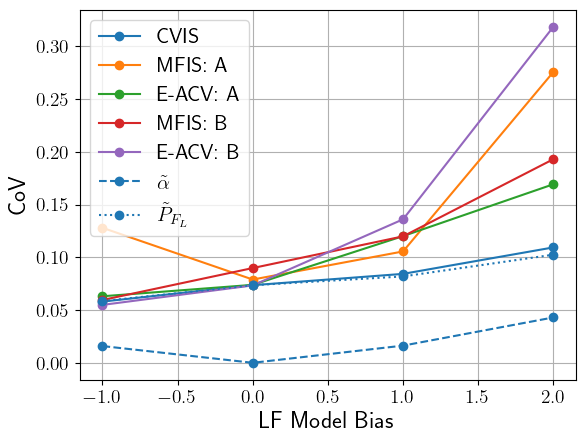}
  \caption{}
  \label{fig:COV_var_0}
\end{subfigure}
\begin{subfigure}{.45\textwidth}
  \centering
  \includegraphics[width=\linewidth]{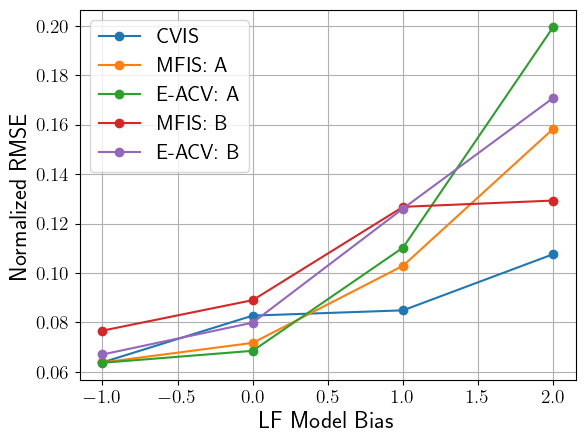}
  \caption{}
  \label{fig:RMSE_var_1}
\end{subfigure}%
\begin{subfigure}{.45\textwidth}
  \centering
  \includegraphics[width=\linewidth]{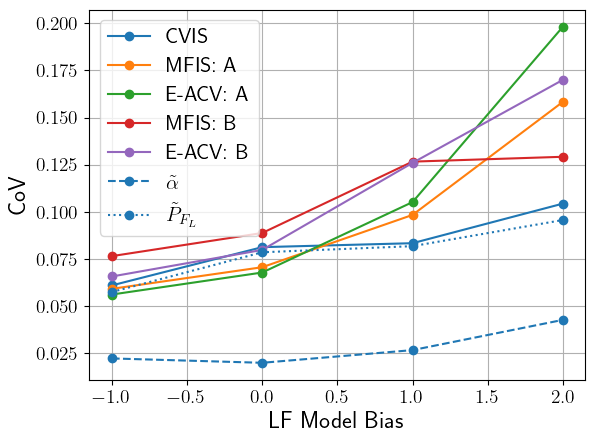}
  \caption{}
  \label{fig:COV_var_1}
\end{subfigure}
\begin{subfigure}{.45\textwidth}
  \centering
  \includegraphics[width=\linewidth]{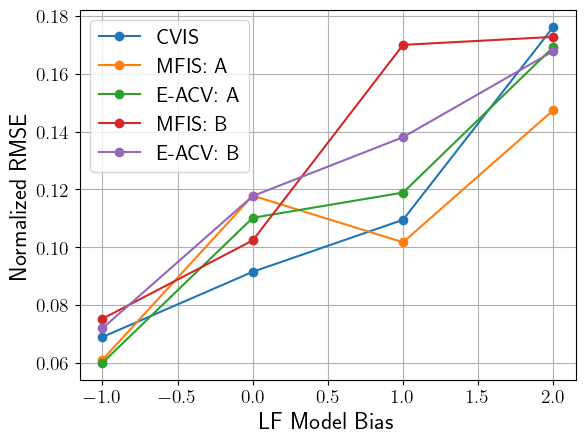}
  \caption{}
  \label{fig:RMSE_var_2}
\end{subfigure}%
\begin{subfigure}{.45\textwidth}
  \centering
  \includegraphics[width=\linewidth]{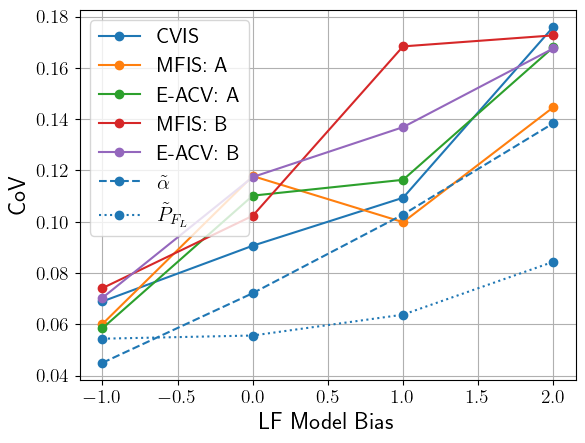}
  \caption{}
  \label{fig:COV_var_2}
\end{subfigure}
\caption{Example 1: (a), (c), and (e) show the root mean squared errors, and (b), (d), and (f) show the coefficient of variation for different magnitudes of noise and increasing model bias. Figures (a) and (b) correspond to $ \sigma_{\epsilon} = 0 $, figures (c) and (d) correspond to $ \sigma_{\epsilon} = 1 $, and figures (e) and (f) correspond to $ \sigma_{\epsilon} = 2 $. Statistics are computed as the mean from 100 independent trials of each algorithm.}
\label{fig:example_1_results}
\end{figure}

Figure~\ref{fig:example_1_results} compares statistics from CVIS against MFIS and E-ACV from one hundred independent trials for the sample allocations explained above.
These plots show the normalized root mean square error (RMSE) (Figures~\ref{fig:RMSE_var_0},~\ref{fig:RMSE_var_1}, and~\ref{fig:RMSE_var_2}) and coefficient of variation (CoV) (Figures~\ref{fig:COV_var_0},~\ref{fig:COV_var_1}, and~\ref{fig:COV_var_2}) of the estimators for different levels of noise and increasing model bias.
It is clear that the performance of CVIS is always comparable and usually slightly better than MFIS and E-ACV for the same computational budget, even as the quality of the LF model deteriorates due to increased bias or noise. Additionally, the strong similarity in the RMSE and CoV values for all cases indicates that there is no bias in the estimation.
The CoV plots in Figure~\ref{fig:example_1_results} also explicitly show the contributions of $ \Tilde{\alpha} $ (blue dashed lines) and $ \Tilde{P}_{F_{L}} $ (blue dotted lines) to the variance of $ \Tilde{P}_{F} $ (see Eq.~\eqref{eqn:cov_of_proposed_CVIS_estimator}). Such a breakdown is possible only due to the simple construction of the proposed CVIS estimate; for a more complex estimator like E-ACV, it is exceedingly difficult, if not impossible, to account for every source of uncertainty as well as the interactions between them. This ability of the CVIS estimator to cleanly separate sources of uncertainty is another strength of the formulation.

\subsection{Example 2: A Five-story Shear Building}
\label{section:Example_2_modal_analysis_shear_building}

\begin{figure}[!htbp]
\centering
\begin{subfigure}{.35\textwidth}
  \centering
  \includegraphics[width=\linewidth]{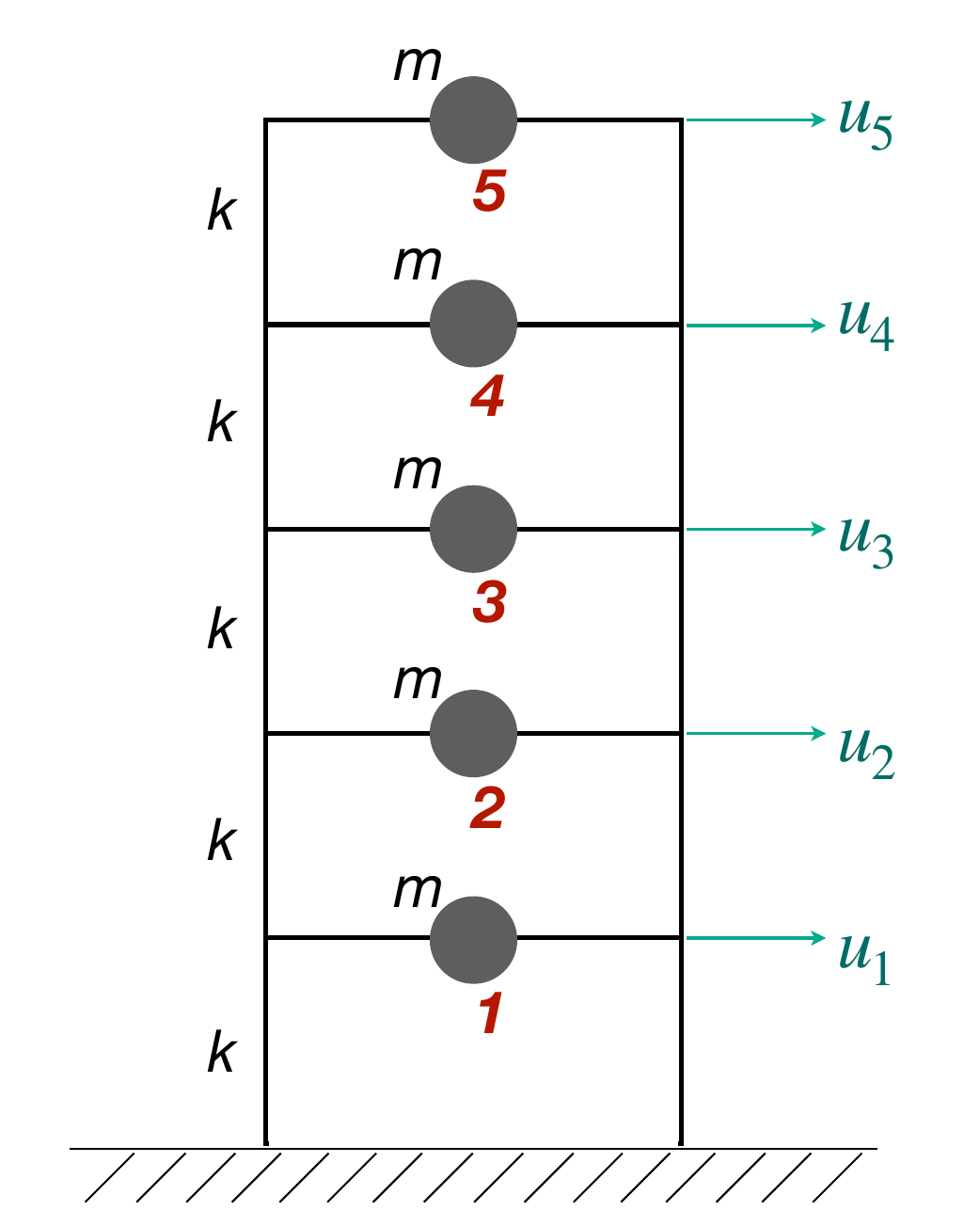}
  \caption{}
  \label{fig:example_2_frame}
\end{subfigure}%
\begin{subfigure}{.60\textwidth}
  \centering
  \includegraphics[width=\linewidth]{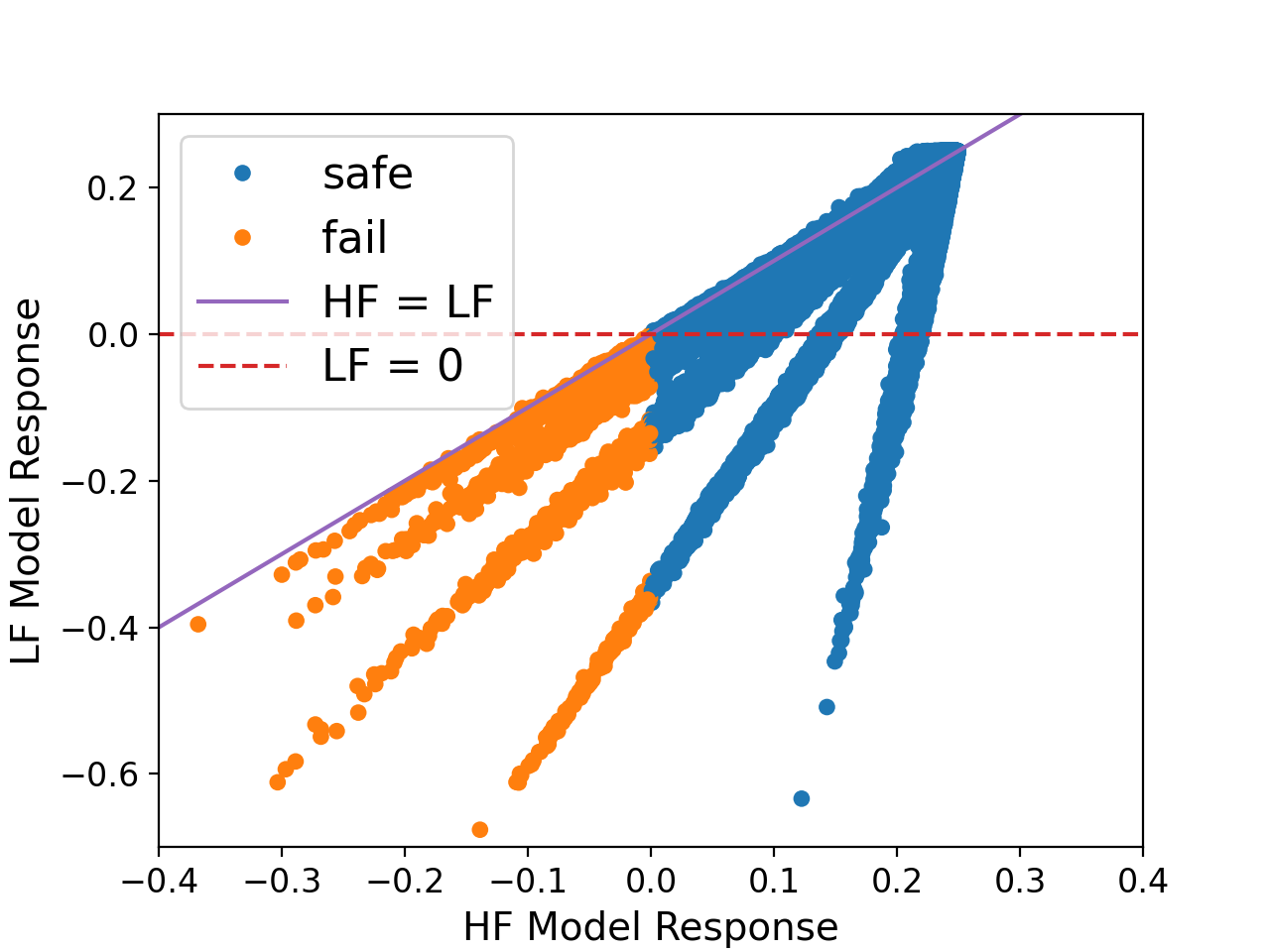}
  \caption{}
  \label{fig:example_2_model_relationship_scatterplot}
\end{subfigure}
\caption{Example 2: (a) A schematic of the five-story shear building. (b) Scatterplot of HF response predictions vs. LF response predictions using 1 million samples. The points are colored according to the HF model predictions; blue points correspond to safety, and orange points correspond to failure. On the other hand, the LF model predicts failure in all samples under the red dashed line.}
\label{fig:example_2_visualizations}
\end{figure}

We next consider the case of the five-story shear building adapted from \cite{chopra2017dynamics} and depicted in Figure~\ref{fig:example_2_frame}. 
Each floor is considered to have a lumped mass of $ m = 45,000 $ \si{\kilogram}, and the stories are all undamped with the same stiffness $ k = 20,000 $ \si{\kilo\newton\per\meter}.  A sinusoidal load of the form $ \mathbf{p}(t) = \mathbf{s} p_0 \sin{\left( \omega_p t \right)} $ is applied, where $ t $ denotes time, $ \omega_p $ is the frequency of the forcing function, $ p_0 = 100 $ \si{\kilo\newton} is the nominal force magnitude, and $ \mathbf{s} = \begin{bmatrix} s_1 & s_2 & s_3 & s_4 & s_5 \end{bmatrix} ^T $ is an unnormalized shape vector that controls the magnitude of the force applied to each story. The system is modeled by the set of linear differential equations given by:
\begin{equation}
    \mathbf{M}\ddot{\mathbf{u}}(t) + \mathbf{K}\mathbf{u}(t) = \mathbf{p}(t) 
    \label{eqn:eom}
\end{equation}
where $\mathbf{M}$ is the mass matrix, and $\mathbf{K}$ is the stiffness matrix. The displacements of each floor are denoted by the vector $ \mathbf{u}(t) = \begin{bmatrix} u_1(t) & u_2(t) & u_3(t) & u_4(t) & u_5(t) \end{bmatrix} ^T $.

Both $ \omega_p $ and the shape vector $ \mathbf{s} $ are considered to be stochastic;
therefore, the random vector of inputs for this problem is $ \mathbf{x} = \begin{bmatrix} s_1 & s_2 & s_3 & s_4 & s_5 & \omega_p \end{bmatrix} ^T $. Each $s_i$ is assumed to follow a standard normal distribution ($s_i\sim \mathcal{N} \left( 0, 1 \right)$), and $\omega_p$ is assumed uniform in the range [5, 50] Hz. ($\omega_p\sim\mathcal{U} \left( 5, 50 \right)$). Failure occurs if the relative displacement between any two stories or between the ground and the first story exceeds $ 25 $ \si{\centi\meter} at any time instant in the interval $ t \in \left[ 0, 1 \right]$ sec. The building is considered to be undeformed and at rest at $ t = 0 $.


The model is simplified by expressing the equations of motion (Eq.~\eqref{eqn:eom}) in modal coordinates as:
\begin{equation}
    \begin{split}
        \ddot{\mathbf{q}}(t) + \mathbf{\mathcal{K}}\mathbf{q}(t) = \mathbf{P}(t) \\
        \mathbf{u}(t) = \mathbf{\Phi}^T \mathbf{q}(t) \\
        \mathbf{\mathcal{K}} = \mathbf{\Phi}^T \mathbf{K} \mathbf{\Phi} \\
        \mathbf{P}(t) = \mathbf{\Phi}^T \mathbf{p}(t)
        \label{eqn:eom_modal}
    \end{split}
\end{equation}
where $ \mathbf{q}(t) = \begin{bmatrix} q_1 (t) & q_2 (t) & q_3 (t) & q_4 (t) & q_5 (t) \end{bmatrix} ^T $ is the modal coordinate vector (with $ q_i (t) $ being the $ i^{th}$ modal coordinate, $ i \in \left\{ 1, 2, 3, 4, 5 \right\} $), $ \mathbf{\mathcal{K}} $ is the diagonal modal stiffness matrix, and $ \mathbf{P}(t) $ is the modal force vector. $ \mathbf{\Phi} = \begin{bmatrix} \boldsymbol{\phi}_1 (t) & \boldsymbol{\phi}_2 (t) & \boldsymbol{\phi}_3 (t) & \boldsymbol{\phi}_4 (t) & \boldsymbol{\phi}_5 (t) \end{bmatrix} $ is the mode shape matrix where the column vector $ \boldsymbol{\phi}_i $ is the $ i^{th}$ mode shape. It is normalized such that $ \mathbf{\Phi}^T \mathbf{M} \mathbf{\Phi} = \mathbf{I} $, the identity matrix.

The HF model includes all five modes, therefore accurately modeling the full displacement history and thus accurately modeling structural failure. The LF model, on the other hand, only considers the one or two modes
whose natural frequencies are nearest to the forcing function frequency $ \omega_p $. 
Eq.~\eqref{eqn:example_2_lf_model_modal_summation_set} defines this criterion more rigorously. Moreover, the LF model only records the maximum story displacement that occurs at any time instant rather than recording all story displacements to compute the inter-story drift. Because this LF model sometimes underestimates the deformation that causes failure for the higher modes, a multiplicative safety factor of $ 2.0 $ is applied to the deflection predicted by the LF model to ensure that it is a conservative predictor of failure.


Let $ \omega_{n_i} $ be the $ i^{th}$ natural frequency of the structure, then using the modal coordinates (Eq.~\eqref{eqn:eom_modal}), the HF model is defined as 
\begin{gather}
    H (\mathbf{x}) = 0.25 - \arg \max_{0 \leq t \leq 1} \left( \lVert \mathbf{r} (t) \rVert_{\infty} \right) \label{eqn:example_2_HF_definition} \\
    \mathbf{r} (t) = \begin{bmatrix} \lvert u_1^{H}(t) \rvert & \lvert u_2^{H}(t) - u_1^{H}(t) \rvert & \lvert u_3^{H}(t)-u_2^{H}(t)\rvert & \lvert u_4^{H}(t)-u_3^{H}(t)\rvert & \lvert u_5^{H}(t) - u_4^{H}(t) \rvert \end{bmatrix} ^T \label{eqn:example_2_relative_displacement_flexural_deformation} \\
    \mathbf{u}^{H}(t) = \begin{bmatrix} u_1^{H}(t) & u_2^{H}(t) & u_3^{H}(t) & u_4^{H}(t) & u_5^{H}(t) \end{bmatrix} ^T = \sum_{i = 1}^5 \boldsymbol{\phi}_i q_i (t) \label{eqn:example_2_HF_displacement_history_modal_expansion}
\end{gather}
We similarly define the LF model as
\begin{gather}
    L (\mathbf{x}) = 0.25 - 2 \arg \max_{0 \leq t \leq 1} \left( \lVert \mathbf{u}^{L} (t) \rVert_{\infty} \right) \label{eqn:example_2_LF_definition} \\
    \mathbf{u}^{L}(t) = \begin{bmatrix} u_1^{L}(t) & u_2^{L}(t) & u_3^{L}(t) & u_4^{L}(t) & u_5^{L}(t) \end{bmatrix} ^T = \sum_{i \in \mathcal{I}} \boldsymbol{\phi}_i q_i (t) \label{eqn:example_2_LF_displacement_history_modal_expansion} \\
    \mathcal{I} = \begin{cases}
                    \left\{ 1 \right\} & \text{if } \omega_p < \omega_{n_1} \\
                    \left\{ j, j+1 \right\} & \text{if } \omega_{n_j} \leq \omega_p \leq \omega_{n_{j+1}} \text{ , } j = 1, 2, 3, 4 \\
                    \left\{ 5 \right\} & \text{if } \omega_p > \omega_{n_5}
                \end{cases} \label{eqn:example_2_lf_model_modal_summation_set}
\end{gather}
where the LF model is approximately $ 40 \% $ as expensive as the HF model since it only evaluates one or two of the modes instead of all five.

The scatterplot in Figure~\ref{fig:example_2_model_relationship_scatterplot} plots the HF response vs.\ the LF response for $ 10^6 $ random samples of the input $\mathbf{x}$, depicting the complicated relationship between the LF and HF models. Clearly, the covariance between these models is challenging to conceptualize, which begs the question of whether the optimal CV parameter can be meaningfully estimated. With this in mind, we once again compare CVIS against MFIS and E-ACV. Table~\ref{tab:example_2_sample_allocation_summary} lists the sample allocations for each method, which are again constructed to ensure that all three methods use the same computational budget.
\begin{table}[!ht]
    \centering
    \begin{tabular}{c|c|c|c|c|c|c}
    Sampling & \multicolumn{2}{c|}{Importance Sampling} & \multicolumn{2}{c|}{Subset Simulation} & \multicolumn{2}{c}{Monte Carlo Integration} \\
    \cline{2-7}
    Procedure & HF Calls & LF Calls & HF Calls & LF Calls & HF Calls & LF Calls \\
    \hline \hline
    CVIS & $ 40,000 $ & $ 40,000 $ & - & $ 5,000 \times M_{S} $ & - & - \\
    MFIS & $ 40,000 $ & $ 40,000 $ & - & $ 2,500 \times M_{S} $ & - & $ 2,500 \times M_{S} $ \\
    E-ACV & $ 40,000 $ & $ 40,000 $ & - & $ 2,500 \times M_{S} $ & - & $ 2,500 \times M_{S} $ \\
    \end{tabular}
    \caption{Example 2: Sample allocation summary. $ M_{S} $ is the number of subsets required by the Subset Simulation procedure.}
    \label{tab:example_2_sample_allocation_summary}
\end{table}

The results of this comparison are presented in Table~\ref{tab:example_2_results}, which shows the predicted failure probabilities for all three methods and the associated uncertainty (CoV) in the failure probability predictions. 
Note that for E-ACV we use the estimator presented in~\cite{PhamGorodetsky2022}, which treats the estimated $ \alpha $ value as a constant for variance estimation. For MFIS and CVIS, we use the Replicated Batch Means (RBM) estimator mentioned in Section~\ref{section:CVIS_algorithm_and_implementation}. Since a significant number of relatively short chains are used to ensure proper exploration of the ISD, the batch size $ b $ for the RBM method is taken to be the full length of each chain. For reference, Crude Monte Carlo using the HF model with $10^6$ samples is used to compute the true failure probability. The values for all three methods are very close to the true failure probability, indicating that all three are equally accurate for the same computational budget. The sample CoV, as well as the predicted CoV, further show that all three methods achieve comparable precision. Thus, we highlight the fact that CVIS is able to provide benefits above existing methods like MFIS or E-ACV without a drop in performance; these benefits include being able to circumvent estimation of the inter-model covariance calculations and ISD normalization constant, as well as the ability to better analyze the performance of the estimator through the $ \kappa $ diagnostic and the ability to account for the uncertainty in $ \Tilde{\alpha} $. (See Sections~\ref{section:goals_and_contributions} and~\ref{section:CV_vs_IS} for discussions of the benefits of CVIS.)

\begin{table}[!ht]
    \centering
    \begin{tabular}{c|c|c|c|c}
    \multirow{2}{*}{Algorithm} & \multicolumn{2}{c|}{Estimated $ \hat{P}_F $} & \multicolumn{2}{c}{Estimated $ \operatorname{CoV} \left[ \hat{P}_F \right] $} \\
    \cline{2-5}
     & Sample Mean & Sample CoV & Sample Mean & Sample CoV \\
    \hline \hline
    CVIS & 4.26 E-3 & 12.73\% & 13.46\% & 5.35\% \\
    MFIS & 4.33 E-3 & 13.75\% & 11.36\% & 7.62\% \\
    E-ACV & 4.30 E-3 & 13.50\% & 11.59\% & 8.16\% \\
    \hline \hline
    Target $ P_F $ & \multicolumn{4}{c}{4.27 E-3 (Computed using Crude Monte Carlo with $10^6$ samples)}
    \end{tabular}
    \caption{Example 2: The mean and coefficient of variation (CoV) of the predicted failure probability for the CVIS, MFIS, and E-ACV methods, computed from 100 independent trials. For each trial the CoV of the failure probability was also estimated. 
    The sample mean and CoV of these predictions from the 100 trials are also listed.}
    \label{tab:example_2_results}
\end{table}

\subsection{Example 3: Lid-driven Cavity Problem}
\label{section:Example_3_navier_stokes_lid_driven_cavity}

\begin{figure}[!htbp]
\centering
\begin{subfigure}{.35\textwidth}
  \centering
  \includegraphics[width=\linewidth]{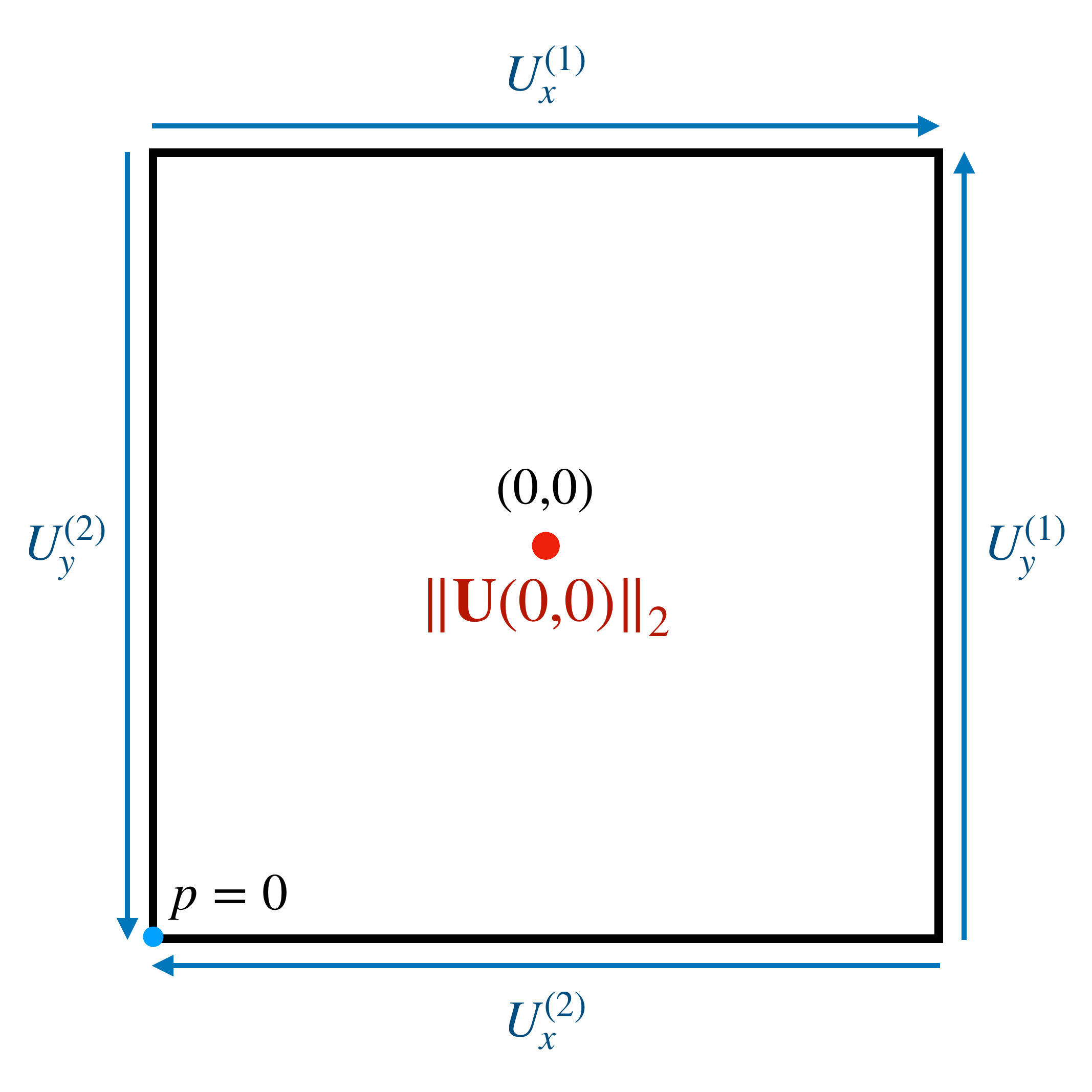}
  \caption{}
  \label{fig:example_3_schematic}
\end{subfigure}%
\begin{subfigure}{.48\textwidth}
  \centering
  \includegraphics[width=\linewidth]{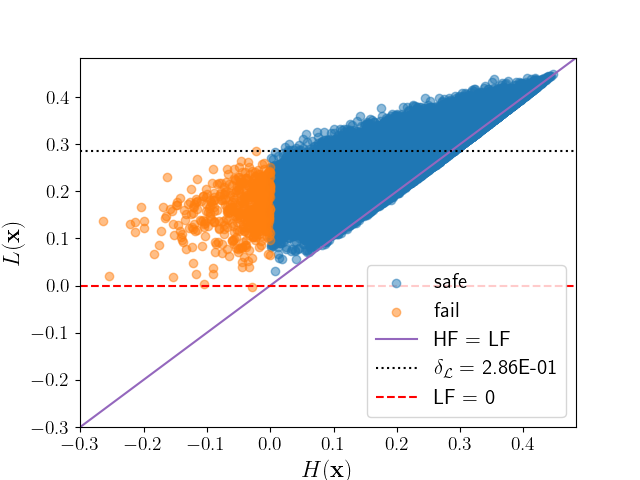}
  \caption{}
  \label{fig:example_3_scatterplot_safety_factor_1.0}
\end{subfigure}
\begin{subfigure}{.48\textwidth}
  \centering
  \includegraphics[width=\linewidth]{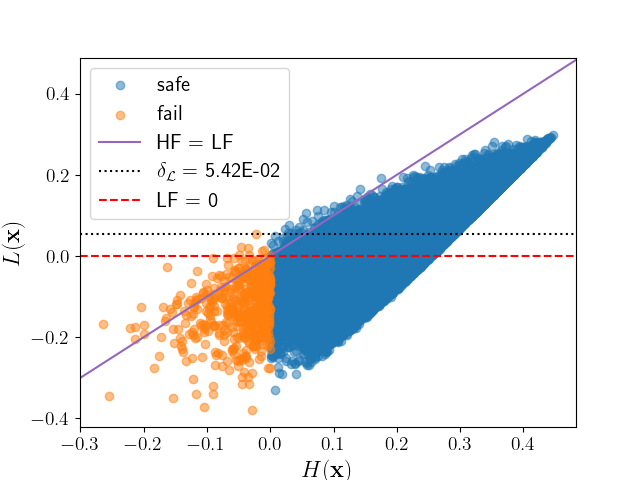}
  \caption{}
  \label{fig:fig:example_3_scatterplot_safety_factor_1.5}
\end{subfigure}%
\begin{subfigure}{.48\textwidth}
  \centering
  \includegraphics[width=\linewidth]{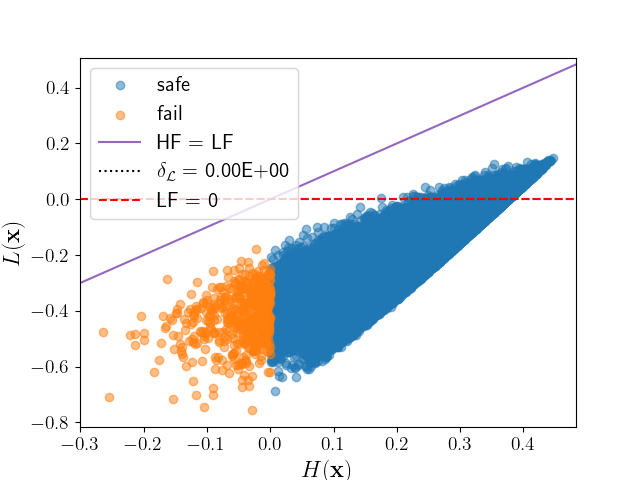}
  \caption{}
  \label{fig:fig:example_3_scatterplot_safety_factor_2.0}
\end{subfigure}
\caption{Visualizations for example 3. Figure (a) shows the flow domain for the four-sided lid-driven cavity problem, while figures (b), (c), and (d) show scatterplots of HF response predictions vs. LF response predictions for $ \zeta_L = 1.0 $, $ 1.5 $, and $ 2.0 $, using 200,000 samples. The points are colored according to the HF model predictions; blue points correspond to safety, and orange points correspond to failure. On the other hand, the LF model predicts failure in all samples under the red dashed line.}
\label{fig:example_3_schematic_and_scatterplots}
\end{figure}

Finally, we consider a numerical model of steady-state incompressible fluid flow: the four-sided lid-driven cavity problem. Here, fluid flow occurs in a 2-dimensional square-shaped domain as shown in Figure~\ref{fig:example_3_schematic}. The kinematic viscosity $ \nu $ and the density $ \rho $ of the fluid are both considered random variables, and the velocity at any point in the cavity is denoted by $ \mathbf{U} (x, y) \equiv \mathbf{U} = \begin{bmatrix} U_x & U_y \end{bmatrix} ^T $. Stochastic velocities $ U_x^{(1)} $,  $ U_y^{(1)} $, $ U_x^{(2)} $, and $ U_y^{(2)} $ are applied at the four boundaries of the square domain resulting in the following six-dimensional random vector of inputs: $ \mathbf{X} = \begin{bmatrix} U_x^{(1)} & U_x^{(2)} & U_y^{(1)} & U_y^{(2)} & \rho & \nu \end{bmatrix} ^T $. Table~\ref{tab:example_3_list_of_distributions} lists the distributions for each random variable.
\begin{table}[!ht]
    \centering
    \begin{tabular}{c|c|c}
    Random Variable & Distribution & Parameters \\
    \hline \hline
    $ \ln \left( U_x^{(1)} \right) $,  $ \ln \left(U_y^{(1)} \right) $, & \multirow{2}{*}{Truncated Normal} & Mean = $ \ln (0.75) $ , Variance = $ 0.0625 $ \\
    $ \ln \left( U_x^{(2)} \right) $, and $ \ln \left(U_y^{(2)} \right) $ &  & Domain = $ \left[ 0.5, 1.5 \right] $ \\
    \hline
    $ \rho $ & Uniform & Domain = $ \left[ 0.5, 1.5 \right] $ \\
    \hline
    \multirow{2}{*}{$ \ln (\nu) $} & \multirow{2}{*}{Truncated Normal} & Mean = $ \ln (0.025) $ , Variance = $ 0.25 $ \\
     &  & Domain = $ \left[ \ln (0.005), \ln (0.05) \right] $
    \end{tabular}
    \caption{Example 3: Distributions of the random variables.}
    \label{tab:example_3_list_of_distributions}
\end{table}

At any point in the interior of the domain, the fluid velocity $ \mathbf{U} $ can be solved using either the incompressible Navier-Stokes equations
\begin{align}
\begin{split}
    \frac{1}{\rho} \nabla p + \nabla \cdot \left( \nu \nabla \mathbf{U} \right) &= \left( \mathbf{U} \cdot \nabla \right) \mathbf{U} \label{eqn:navier_stokes_eq_1} \\
    \nabla \cdot \mathbf{U} &= 0 
\end{split}
\end{align}
(where $ p $ is the pressure) or using the Stokes approximation, which ignores the non-linear convective term
\begin{align}
\begin{split}
    \frac{1}{\rho} \nabla p + \nabla \cdot \left( \nu \nabla \mathbf{U} \right) &= 0 \label{eqn:stokes_eq_1} \\
    \nabla \cdot \mathbf{U} &= 0 
\end{split}
\end{align}

For our example, we solve for the velocity magnitude at the center of the domain (point $ (0, 0) $ in Figure~\ref{fig:example_3_schematic}) and define failure as velocity magnitude exceeding $ 0.75 $ units. The HF model (Eq.~\eqref{eqn:example_3_HF_model}) uses the Navier-Stokes equations, while the LF model (Eq.~\eqref{eqn:example_3_LF_model}) uses the Stokes approximations, such that the models can be expressed as
\begin{align}
    H (\mathbf{x}) &= 0.75 - \lVert \mathbf{U}_H (0, 0) \rVert _2 \label{eqn:example_3_HF_model} \\
    L (\mathbf{x}) &= 0.75 - \zeta_L \lVert \mathbf{U}_L (0, 0) \rVert _2 \label{eqn:example_3_LF_model}
\end{align}
where $ \mathbf{U}_H $ is the solution to Eq.~\eqref{eqn:navier_stokes_eq_1} and $ \mathbf{U}_L $ is the solution for Eq.~\eqref{eqn:stokes_eq_1}. Velocities are solved using the Navier-Stokes module~\cite{PETERSON201868} in the Multi-physics Object-Oriented Simulation Environment (MOOSE)~\cite{PERMANN2020100430,LINDSAY2022101202}, which solves for $ \mathbf{U}_H (0, 0) $ in $ \approx 4 $ CPU-seconds and for $ \mathbf{U}_L (0, 0) $ in $ \approx 1 $ CPU-second. $ \zeta_L $ is a factor that allows us to non-intrusively modify the model relationship between the HF and LF models, effectively adjusting $\delta_{\mathcal{L}}$. The effect of changing $ \zeta_L $ is shown in Figure~\ref{fig:example_3_schematic_and_scatterplots}, showing that it shifts the LF model output relative to the limit surface $L(\mathbf{x})=0$.

Table~\ref{tab:example_3_model_relationship_statistics} shows a set of diagnostics for the CVIS method for different values of $ \zeta_L $. First, we see that the Stokes approximation loses its predictive accuracy for larger flow velocities, as is clear from the deteriorating correlation between $ H (\mathbf{x}) $ and $ L (\mathbf{x}) $ for values in the failure domain $\mathbf{x} \in \Omega_{\mathcal{F}}$. 
Not only does this complicate the idea of inter-model covariance for use in a standard CV framework, but it also means that $ L (\mathbf{x}) $ vastly underestimates failure for $ \zeta_L = 1.0 $.
The large value of $ \delta_{\mathcal{L}} = 0.286 $ for this case (see Remark~\ref{remark:smallest_contour_of_LF_that_covers_HF_failure_region}) suggests that Stipulation~\ref{stipulation:fundamental_model_quality_assumption} is not satisfied. This is confirmed by computing the diagnostic $\kappa$ value (last column in Table~\ref{tab:example_3_model_relationship_statistics}), which indicates quite strongly ($\kappa$ is much less than 0.5) that the CVIS algorithm will not yield a variance reduction in this case. However, by changing the $ \zeta_L $ value, we can improve the value of $ \kappa $. (It is important to remember that $ \kappa $ is a classifier - larger values of $ \kappa $ do not imply more variance reduction. By ``improving'' $ \kappa $, we refer only to changing the relationship between the HF and LF models such that $ \kappa \geq 0.5 $, which ensures variance reduction according to Theorem~\ref{thm:CVIS_alpha_variance_reduction_diagnostic}. In particular, the $ \kappa $ criterion cannot be used to distinguish between the performance of $ \zeta_L = 1.5 $ and $ \zeta_L = 2.0 $.) 
\begin{table}[!ht]
    \centering
    \begin{tabular}{c|c|c|c|c|c}
    \multirow{2}{*}{$ \zeta_L $ Value} & \multicolumn{2}{c|}{$ \rho \left[ H (\mathbf{X}), L (\mathbf{X}) \right] $} & \multirow{2}{*}{$ \rho \left[ I_H (\mathbf{X}), I_L (\mathbf{X}) \right] $} & \multirow{2}{*}{$ \kappa $ (Eq.~\eqref{eqn:kappa_definition_and_variance_reduction_diagnostic})} & \multirow{2}{*}{$ \delta_{\mathcal{L}} $ (Remark~\ref{remark:smallest_contour_of_LF_that_covers_HF_failure_region})} \\
    \cline{2-3}
     & $ \mathbf{x} \in \Omega $ & $ \mathbf{x} \in \Omega_{\mathcal{F}} $ &  &  &  \\
    \hline \hline
    $ 1.0 $ & 0.89 & 0.26 & 0.045 & 1.99 E-3 & 2.86 E-1 \\
    $ 1.5 $ & 0.89 & 0.26 & 0.100 & 9.84 E-1 & 5.42 E-2 \\
    $ 2.0 $ & 0.89 & 0.26 & 0.012 & 1.0 & 0.00 \\
    \end{tabular}
    \caption{Example 3: Model Relationship diagnostics. All reported values were calculated by Crude Monte Carlo estimation using 200,000 samples.}
    \label{tab:example_3_model_relationship_statistics}
\end{table}

Due to the computational expense of the models, only one CVIS run was conducted with $ \zeta_L = 1.5 $. This run used 100 chains with 400 samples each for the DE-MC procedure (i.e., the IS estimators) and 5,000 samples per subset for the LF SuS procedure to estimate $\tilde{P}_{F_L}$. Table~\ref{tab:example_3_results} presents the CVIS prediction and the corresponding estimated CoV.
We see that CVIS very accurately predicts the failure probability with reasonable precision.
\begin{table}[!ht]
    \centering
    \begin{tabular}{c|c|c|c}
    Algorithm & $ \Tilde{P}_F $ & $ \operatorname{CoV} \left[ \Tilde{P}_F \right] $ & Total Model Calls \\
    \hline \hline 
     CVIS & 2.95 E-3 & 15.81 \% & 40,000 HF + 45,000 LF \\
    \hline
    Target $ P_F $ & \multicolumn{3}{c}{2.51 E-3 (Computed using Crude Monte Carlo with $2\times 10^5$ samples)}
    \end{tabular}
    \caption{Example 3: CVIS Prediction}
    \label{tab:example_3_results}
\end{table}



\section{Conclusions}
\label{seciton:conclusions}

In this work, we present the CVIS estimator, a statistical estimator for rare-event reliability analysis that couples control variates and importance sampling under a bifidelity modeling paradigm.
Instead of focusing purely on optimal variance reduction, emphasis is placed on robustness, simplicity, and various quality-of-life improvements, including (a) an ISD constructed to fully represent system response information from the Low Fidelity model and explicitly consider false-positive safety predictions, (b) a CV formulation that circumvents the need for knowing or estimating any model correlation information, (c) a variance-reduction diagnostic at no additional cost, (d) an estimator construction that does not require complicated optimization routines to determine operating parameters and allows for plug-and-play style incorporation of multiple sampling strategies by functioning with samples from unnormalized distributions, and (e) a closed-form variance estimator that incorporates all sources of uncertainty in an easily separable form. The proposed estimator is backed up by rigorous mathematical proofs and nuanced discussions of its strengths and weaknesses. Finally, it is tested against established methods such as \textit{Multifidelity Importance Sampling} (Peherstorfer et al.~\cite{PEHERSTORFER_MFIS}) and \textit{Ensemble Approximate Control Variates} (Pham \& Gorodetsky~\cite{PhamGorodetsky2022}) on a number of analytical and numerical examples. Through these case studies, the CVIS estimator was found to perform as well as or better than its counterparts while providing the aforementioned practical benefits (that are not all available from the existing methods in the literature).

\appendix

\section{Abbreviations and Notations Used}
\label{appendix:abbreviations_and_notations}

The following Tables~\ref{tab:list_of_abbreviations} and \ref{tab:list_of_notations} collect all the important abbreviations and symbols, respectively, used throughout this document.

\begin{table}[h!t]
    \centering
    \begin{tabular}{c|c}
    Abbreviation & Full Form \\
    \hline \hline 
    HF & High-Fidelity model \\
    LF & Low-Fidelity model \\
    MC & Crude/Simple Monte Carlo \\
    IS & Importance Sampling \\
    CV & Control Variates \\
    ACV & Approximate Control Variates \\
    SACV & Simple Approximate Control Variates \\
    CVMC & Control Variates Monte Carlo \\
    CVIS & Control Variates - Importance Sampling \\
    MFIS & Multifidelity Importance Sampling \\
    E-ACV & Ensemble Approximate Control Variates \\
    ISD & Importance Sampling Density \\
    \end{tabular}
    \caption{List of Abbreviations}
    \label{tab:list_of_abbreviations}
\end{table}

\begin{table}[h!t]
    \centering
    \begin{tabular}{c|c}
    Symbol & Explanation \\
    \hline \hline 
    $ \cdot^* $ & Optimal value of a quantity \\
    $ \hat{\cdot} $ & A general estimator of a quantity \\
    $ \Tilde{\cdot} $ & A specific estimator of a quantity chosen for use in our CVIS formulation \\
    $P_F$ & Failure probability of HF model \\
    $P_{F_L}$ & Failure probability of LF model \\
    \multirow{2}{*}{$ Q $ and $ Q_L $} & Component failure estimators of the HF and LF models, respectively, \\
     & using the same sample set, required for any CV formulation \\
    $ \alpha $ & Ratio of the estimators of $ Q $ and $ Q_L $ \\
    $ \alpha^{\dagger} $ & $ P_F / P_{F_L} $ \\
    $ f_{\mathbf{X}} (\mathbf{x}) $ & Joint density function of input distribution \\
    $ q_{\mathbf{X}} (\mathbf{x}) $ & Importance Sampling Density \\
    $ \mathfrak{q}_{\mathbf{X}} (\mathbf{x}, \beta) $ & Importance Sampling Density used for our CVIS formulation \\
    $ S_L (\mathbf{x}, \beta) $ & Logistic Function used to construct $ \mathfrak{q} (\mathbf{x}, \beta) $ \\
    $ \beta $ & Tuning parameter for $ S_L (\mathbf{x}, \beta) $ \\
    $ I_H (\mathbf{x}) $ & HF model failure indicator function \\
    $ I_L (\mathbf{x}) $ & LF model failure indicator function \\
    $ \mathbb{E}_k \left[ \cdot \right] $ & Expected value under density function $ k (\mathbf{x}) $ \\
    $ \mathbb{V}ar_k \left[ \cdot \right] $ & Variance under density function $ k (\mathbf{x}) $ \\
    $ \operatorname{\mathbb{C}ov}_k \left[ \cdot, \cdot \right] $ & Covariance under density function $ k (\mathbf{x}) $ \\
    $ \operatorname{CoV} \left[ \cdot \right] $ & Coefficient of variation of an estimator \\
    $ \Omega $ & Full input domain \\
    $ \Omega_{\mathcal{F}} $ & Failure domain \\
    $ \Omega_{\mathcal{F}} $ & Smallest contour of the LF model that contains the failure domain \\
    $ A^C $ & Complement of some set $ A $.
    \end{tabular}
    \caption{List of Select Notations}
    \label{tab:list_of_notations}
\end{table}

\section{Considerations for Practical Implementation}
\label{appendix:practical_implementation_considerations}

This appendix contains short discussions on a variety of practical considerations that arise when applying the proposed CVIS algorithm.

\subsection{Tuning the ISD}
\label{appendix:ISD_tuning_insights}

Here, we first present an optimization problem (Remark~\ref{remark:optimally_tuned_ISD_CVIS}) that defines the ``optimally-tuned ISD'' mentioned in Section~\ref{section:model_relationship_requirements}. We then provide a heuristic to compute $ \beta^* $ in practice that avoids solving an optimization problem.

\subsubsection{Defining the ``Optimally-tuned'' ISD}
\label{appendix:optimally_tuned_ISD_explanation}

To set up the optimization problem that defines the ``optimally-tuned'' ISD, we make use of the following.

\begin{theorem}
\label{thm:logistic_function_covers_space}
    Let there be two constants $ 0 < \delta < \infty $ and $ 0 < \varepsilon < 0.5 $, and a system state $ \mathbf{x} \in \Omega $ such that $ L (\mathbf{x}) = \delta $. Then:
    \begin{enumerate}
        \item There exists a finite positive $ \beta $ such that $ S_L (\mathbf{x}, \beta) = \varepsilon $.
        \item If any two of the variables in the tuple $ \left( \beta, \delta, \varepsilon \right) $ are specified, then the third variable is determined uniquely.
    \end{enumerate}
\end{theorem}

The proof follows directly from the definition of the Logistic function.
From this statement, we can assert the following

\begin{corollary}
\label{corollary:good_ISD_exists}
    It is always possible to find a $ \beta $ (which uniquely defines the curve $ S_L (\mathbf{x}, \beta) $) such that the resultant $ \mathfrak{q}_{\mathbf{X}} (\mathbf{x}, \beta) \geq \xi f_{\mathbf{X}} (\mathbf{x})/C_S ~ \forall ~ \mathbf{x} \in \Omega_{\mathcal{L}} $, where $ \xi $ is a constant that can be chosen arbitrarily in the range $ 0 < \xi < 0.5 $.
\end{corollary}

In practice, the constant $ \xi $ is chosen to ensure that the ISD has enough weight in all regions of $ \Omega_{\mathcal{L}} $ (and by extension $ \Omega_{\mathcal{F}} $) for the resultant IS estimators to be unbiased and well-defined when implemented.


So far, we have made no comments about what happens if $ \mathfrak{q}_{\mathbf{X}} (\mathbf{x}, \beta) $ has significant weight outside of $ \Omega_{\mathcal{F}} $, i.e. $ \mathfrak{q}_{\mathbf{X}} (\mathbf{x}, \beta) > 0 ~ \forall ~ \mathbf{x} \in \Omega_{\varrho} $, where $ \Omega_{\varrho} \subseteq \Omega_{\mathcal{F}}^C $. Clearly, from Eq.~\eqref{eqn:optimal_ISD_form}, this is undesirable. The proposed ISD is formed by approximating $ I_H (\mathbf{x}) $ using $ S_L (\mathbf{x}, \beta) $ as per the limit-state approximation concept defined in Section~\ref{section:Importance_Sampling}, which clearly places weight outside $ \Omega_{\mathcal{F}} $. However, this is not a fatal issue. The efficiency of IS decreases as more and more weight is spread in the safe-predicted regions of the HF model, but the estimator remains unbiased as long as $ \mathfrak{q}_{\mathbf{X}} (\mathbf{x}, \beta) > 0 ~ \forall ~ \mathbf{x} \in \Omega_{\mathcal{F}} $. Moreover, based on our definitions and stipulations, we can ensure that our ISD remains as tight around the HF failure region as possible (i.e., Eq.~\eqref{eqn:proposed_ISD}). To this end, we define the ``optimally-tuned'' ISD as follows.

\begin{remark}
\label{remark:optimally_tuned_ISD_CVIS}
    Recall the definitions of $\delta_{\mathcal{L}}$ and $ \Omega_{\mathcal{L}}$ from Section~\ref{section:model_relationship_requirements}. For $ \delta_{\mathcal{L}} $, which defines $ \Omega_{\mathcal{L}} $, first define $ \xi $ such that $ S_L (\mathbf{x}, \beta) \geq \xi ~ \forall ~ \mathbf{x} \in \Omega_{\mathcal{L}} $ (where $ 0 < \xi < 0.5 $). ($ S_L (\mathbf{x}, \beta) $ is responsible for the change in shape between $ f_\mathbf{X} (\mathbf{x}) $ and $ \mathfrak{q}_\mathbf{X} (\mathbf{x}) $.) Then, the optimal choices of $ \beta $ and $ \xi $ are defined as:
    \begin{displaymath}
        \left( \beta^*, \xi^* \right) = \arg \min_{\beta, \xi} \left( \int_{\Omega_{\mathcal{L}}^C} \mathfrak{q}_{\mathbf{X}} (\mathbf{x}, \beta) d\mathbf{x} - \lambda \xi \right)
    \end{displaymath}
    for some fixed value of $ \lambda $ that must be chosen a priori.


\end{remark}

\begin{figure}[!htbp]
\centering
\includegraphics[width=0.6\linewidth]{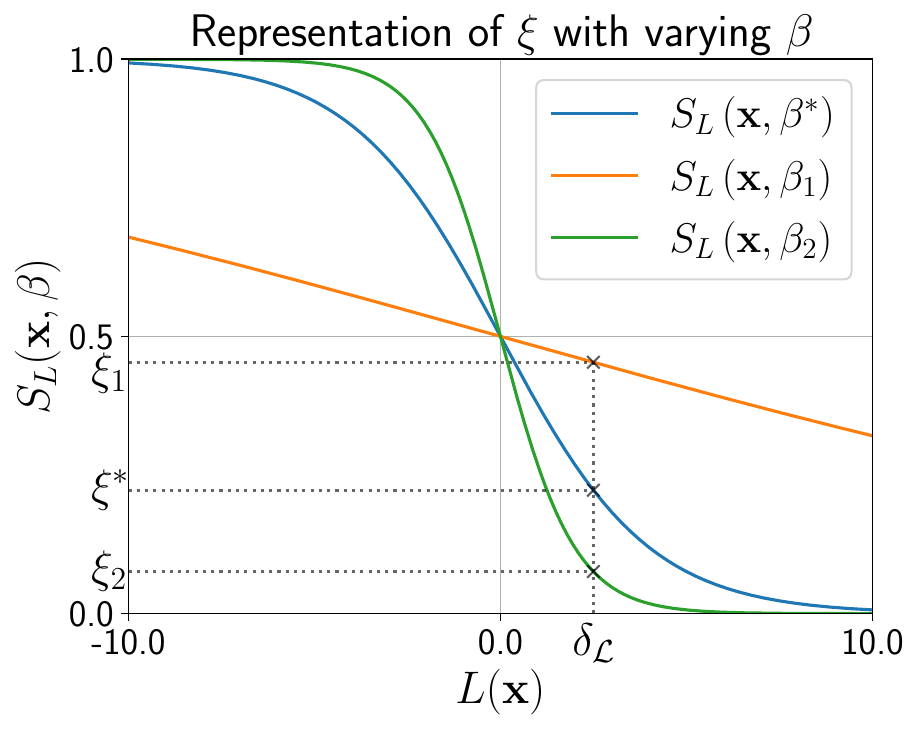}
\caption{The shape of $ S_L \left( \mathbf{x}, \beta \right) $ vs.\ $L(\mathbf{x})$ for three different values of $ \beta $. The blue curve represents the optimally tuned ISD (i.e., $ \beta = \beta^* $), while the orange curve represents a logistic that is too flat ($ \beta = \beta_1 < \beta^* $) and the green curve represents a logistic that is too sharp ($ \beta = \beta_2 > \beta^* $). The values of the three curves at $ L(\mathbf{x}) = \delta_{\mathcal{L}} $ are also marked on the graph as $ \xi^* $, $ \xi_1 $, and $ \xi_2 $ for $ \beta = \beta^* $, $ \beta = \beta_1 $, and $ \beta = \beta_2 $, respectively.}
\label{fig:logistic_behavior}
\end{figure}

To understand the reason for the above optimization problem, consider Figure~\ref{fig:logistic_behavior}, which illustrates the behavior of $ S_L \left( \mathbf{x}, \beta \right) $ for different values of $ \beta $ across a wide range of values $L(\mathbf{x})$ (x-axis). 

By definition, $ \Omega_{\mathcal{L}} = \left\{ \mathbf{x} \in \Omega : L(\mathbf{x}) < \delta_{\mathcal{L}} \right\} $ is the region on the horizontal axis of Figure~\ref{fig:logistic_behavior} having $ L(\mathbf{x}) < \delta_{\mathcal{L}} $. 
Consequently, $ \Omega_{\mathcal{L}}^C $ is the region on the horizontal axis having 
$ L(\mathbf{x}) > \delta_{\mathcal{L}} $. 
Since $ S_L (\mathbf{x}, \beta) $ is a monotonically decreasing function, the condition $ S_L (\mathbf{x}, \beta) \geq \xi ~ \forall \; \mathbf{x} \in \Omega_{\mathcal{L}} $ is satisfied if $ S_L (\mathbf{x}, \beta) = \xi $ when $ L (\mathbf{x}) = \delta_{\mathcal{L}} $.

We then rewrite the integral in the objective function from Remark~\ref{remark:optimally_tuned_ISD_CVIS} as
\begin{equation}
\label{eqn:one-dimensional_version_of_optimal_ISD_integral}
    \int_{\Omega_{\mathcal{L}}^C} \mathfrak{q}_{\mathbf{X}} (\mathbf{x}, \beta) d\mathbf{x} = \int^{\infty}_{\delta_{\mathcal{L}}} C_S^{-1} S_L (\mathbf{x}, \beta) f_L (L (\mathbf{x}) ) d L(\mathbf{x})
\end{equation}
where we use the definition of $ \mathfrak{q}_{\mathbf{X}} (\mathbf{x}, \beta) $ to express the integral over the input space as a one-dimensional integral over possible values of the LF response function $ L (\mathbf{x}) $. To do so, we define $ f_L (L (\mathbf{x}) ) $ through the line integral 
of $ f_{\mathbf{X}} (\mathbf{x}) $ along any given contour of $ L (\mathbf{x}) $. This also leads to the following relations
\begin{align}
    P_{F_L} &= \int_{-\infty}^{0} f_L (L (\mathbf{x}) ) d L(\mathbf{x}) \label{eqn:one_dimensional_version_of_LF_failure} \\
    P_F &\leq \int_{-\infty}^{\delta_{\mathcal{L}}} f_L (L (\mathbf{x}) ) d L(\mathbf{x}) \label{eqn:one_dimensional_inequality_of_HF_failure}
\end{align}
where Eq.~\eqref{eqn:one_dimensional_inequality_of_HF_failure} results as a consequence of the fact that $ \Omega_{\mathcal{F}} \subseteq \Omega_{\mathcal{L}} $.

The optimization problem in Remark~\ref{remark:optimally_tuned_ISD_CVIS} represents a balance between two terms: the integral in Eq.~\eqref{eqn:one-dimensional_version_of_optimal_ISD_integral} and the product $\lambda\xi$. According to the second term, we want $ \xi $ to be as large as possible ($\beta$ as small as possible) such that the ISD has a significant density in regions of $ \Omega_{\mathcal{L}} $, and therefore $ \Omega_{\mathcal{F}} $ by extension (as $ \Omega_{\mathcal{F}} \subseteq \Omega_{\mathcal{L}} $), that are not a part of the LF failure region. However, when $\beta$ is too small
the logistic is too flat, and the ISD has too much density in $ \Omega_{\mathcal{L}}^C $ as illustrated by the orange curve in Figure~\ref{fig:logistic_behavior}.
Conversely, simply minimizing the integral in Eq.~\eqref{eqn:one-dimensional_version_of_optimal_ISD_integral} leads to large $\beta$ values (and therefore small $\xi$ values) such that the ISD does not have sufficient density in the region $ 0 \leq L (\mathbf{x}) \leq \delta_{\mathcal{L}} $, as illustrated by the green curve in Figure~\ref{fig:logistic_behavior}. This makes it difficult to generate samples from HF failure regions that the LF model misclassifies as safe. 

This conflict occurs because the logistic curve pivots around the point $ L(\mathbf{x}) = 0 $, where $ S_L (\mathbf{x}, \beta) = 0.5 $ regardless of $ \beta $. Since our region of interest extends from $ L(\mathbf{x}) = -\infty $ up to $ L(\mathbf{x}) = \delta_{\mathcal{L}} $, we must balance the inclusion of favorable density in the region $ 0 \leq L (\mathbf{x}) \leq \delta_{\mathcal{L}} $ against the resulting unfavorable density in the region $ L(\mathbf{x}) > \delta_{\mathcal{L}} $. Such a balance is illustrated by the blue curve in Figure~\ref{fig:logistic_behavior}.

\subsubsection{Intuition behind the ISD Parameter Selection}
\label{appendix:LF_Pf_estimator_and_ISD_construction}

When Subset Simulation is applied in practice, each response function threshold value $ b_i $ (see Section~\ref{section:CVIS_algorithm_and_implementation}) is defined adaptively such that the probability $ P \left( \mathbf{x} \in \mathcal{S}_i | \mathbf{x} \in \mathcal{S}_{i-1} \right) = P_{i|i-1} = \pi_0 $ ($ \forall i = 1, 2, \dots, M_{S}-1 $), where $ \pi_0 $ is some pre-defined probability threshold, and the stopping criterion is such that $ P_{M_{S}|(M_{S}-1)} \geq \pi_0 $. Here, we aim to use these response function threshold values and the associated probabilities to estimate a reasonable value of $ \beta^* $ directly, without having to compute $ \delta_{\mathcal{L}} $ (see Section~\ref{section:model_relationship_requirements}).

Remember that Stipulation~\ref{stipulation:fundamental_model_quality_assumption} is interpreted as the LF model failure surface (defined by the curve $ L (\mathbf{x}) = 0 = b_{M_{S}} $) lying ``close to or before'' the HF model failure surface. Therefore, we argue that the curve defined by $ L (\mathbf{x}) = b_{M_{S}-1} $ is likely to be a conservative approximation of the HF model failure surface. Additionally, $ P_{M_{S}|(M_{S}-1)} $ gives an estimate of the volume of probability space between the curves $ L (\mathbf{x}) = b_{M_{S}-1} $ and $ L (\mathbf{x}) = 0 = b_{M_{S}} $. If this probability is small, then the two curves are ``far'' apart. Conversely, if the probability is large, then the two curves are close to one another. We use this information to qualitatively judge the likelihood that $ b_{M_{S}-1} \geq \delta_{\mathcal{L}} $, and hence guide our selection of $ \beta^* $. 

According to Theorem~\ref{thm:logistic_function_covers_space}, if we assign a desirable value to $ S_L (\mathbf{x_t}, \beta^*) $ at some point $ \mathbf{x_t} $ where $ L \left( \mathbf{x_t} \right) $ is known, then we can identify a unique value of $ \beta^* $. Consequently, using the argument above, we suggest the following approach to select $ \beta^* $:
\begin{gather}
    S_L \left( \mathbf{x_t}, \beta^* \right) = \frac{P_{M_{S}|(M_{S}-1)}}{2} \; \text{ , } \; L \left( \mathbf{x_t} \right) = \frac{b_{M_{S}-1} + b_{M_{S}}}{2} \\
    \Rightarrow \beta^* = \frac{2}{b_{M_{S}-1}} \ln \left( \frac{2}{P_{M_{S}|(M_{S}-1)}} - 1 \right)
\end{gather}


Also, the following theorem can be readily shown from the estimators presented in~\cite{AuBeck2001}:

\begin{theorem}
\label{thm:Subset_simulation_estimator_variance}
If a total of $ N_s $ samples are generated as part of the Subset Simulation procedure, then
\begin{equation}
\label{eqn:Subset_simulation_estimator_variance}
    \operatorname{\mathbb{V}ar} \left[ \Tilde{P}_{F_L} \right] = \frac{M_{S}}{N_s} O \left( P_{F_L}^2 \ln \left( \nicefrac{1}{P_{F_L}} \right) \right)
\end{equation}
\end{theorem}

Therefore, a Subset Simulation estimator follows the conditions required in Theorems~\ref{thm:model_limit_CVIS_alpha_optimal} and~\ref{thm:CVIS_alpha_variance_reduction_diagnostic} when $ N_s \approx M_{S} N $ (where $ N $ is the number of samples generated to compute the MC estimate $ \hat{Q}_L $, as required in the aforementioned theorems).

\subsection{Initializing the DE-MC Sampler}
\label{appendix:Sampling_from_k}

To seed the MCMC sampler, use the samples generated during Subset Simulation to compute $ \Tilde{P}_{F_{L}} $
that lie in the LF failure region. Therefore, these samples also lie within the main support of $ \mathfrak{q}_\mathbf{X} (\mathbf{x}, \beta^*) $. 
A weighted selection scheme can be used to select the seeds for the DE-MC procedure in such a way that the seeds also follow the target distribution, and therefore no burn-in is necessary to achieve convergence for the MCMC scheme. Although this introduces some correlation between $ \Tilde{P}_{F_{L}} $ and $ \Tilde{Q} $ or $ \Tilde{Q}_{L} $, the number of samples shared between the SuS and IS procedures are very small compared to the total number of samples generated in each method, which suggests that any correlation will be negligible. Additionally, a small burn-in length can be easily introduced for a few additional LF model calls (which is considerably cheaper than the HF model), which can effectively remove any correlations between the estimators for practical purposes.

\subsection{Statistics of MCMC Sampling-based Estimators}
\label{appendix:estimating_Q_or_alpha_stats}

The Markov Chain Central Limit Theorem~\cite{MCMCHandbookChapter1} allows us to derive statistics for estimators that use samples generated as states of one or multiple parallel independent Markov Chains. In general, let $ \pi_{\chi} $ be a probability distribution defined in a measurable space $ \Omega_\chi $ and $ \Psi : \Omega_\chi \to \mathbb{R}^d $ be a function, and say we are interested in estimating $ \mu_{\Psi} = \int_{\Omega_\chi} \Psi (\omega) \pi_{\chi} (\omega) d\omega $. Then, if $ \left\{ \mathcal{X}_{k, t} \right\} $ ($ t = 1, 2, \dots $ and $ k = 1, 2, \dots, \mathcal{C} $) denote $ \mathcal{C} $ independent ergodic Markov Chains each following the target distribution $ \pi_{\chi} $, we can construct an estimator $ \hat{\mu}_{\Psi} $ as
\begin{equation}
\label{eqn:generic_MCMC_mean_estimator}
    \hat{\mu}_{\Psi} = \sum_{k=1}^{\mathcal{C}} \sum_{t=1}^{\mathcal{T}} \frac{\Psi (\mathcal{X}_{k, t})}{\mathcal{C} \mathcal{T}}
\end{equation}
where $ \mathcal{T} $ is the number of states considered. The Markov Chain Central Limit Theorem (CLT) states that, as $ \mathcal{T} \to \infty $,
\begin{equation}
\label{eqn:Markov_chain_CLT}
    \sqrt{\mathcal{C} \mathcal{T}} \left( \hat{\mu}_{\Psi} - \mu_{\Psi} \right) \xrightarrow{dist.} \mathcal{N} \left( 0, \Sigma_d \right)
\end{equation}
where $ \Sigma_d $ is a $ d $-dimensional covariance matrix called the asymptotic variance for the Markov Chain, defined as
\begin{equation}
\label{eqn:asymptotic_variance_estimator}
    \begin{split}
        \Sigma_d = \operatorname{\mathbb{C}ov} \left[ \Psi (\mathcal{X}_{k, t}) \right] + 2 \sum_{\tau=1}^{\infty} \operatorname{\mathbb{C}ov} \left[ \Psi (\mathcal{X}_{k, t}), \Psi (\mathcal{X}_{k, t+\tau}) \right] \\
        \text{i.e., } \; \Sigma_d = \Gamma (0) + 2 \sum_{\tau=1}^{\infty} \Gamma (\tau)
    \end{split}
\end{equation}
Note that since all the chains are ergodic (therefore also stationary) and follow the target distribution, $ \Sigma_d $ is the same regardless of the value of $ k $ and $ t $. Thus, for ease of notation, we represent $ \operatorname{\mathbb{C}ov} \left[ \Psi (\mathcal{X}_{k, t}), \Psi (\mathcal{X}_{k, t+\tau}) \right] $ by  $ \Gamma (\tau) $. By extension, $ \Gamma (0) = \operatorname{\mathbb{C}ov} \left[ \Psi (\mathcal{X}_{k, t}), \Psi (\mathcal{X}_{k, t}) \right] = \operatorname{\mathbb{C}ov} \left[ \Psi (\mathcal{X}_{k, t}) \right] $.

The Markov Chain CLT is used to derive the statistical properties of our estimate $ \hat{\mu}_{\Psi} $; for any fixed and sufficiently large value of $ \mathcal{C} $ and $ \mathcal{T} $, Eq.~\eqref{eqn:Markov_chain_CLT} can be rewritten as
\begin{equation}
\label{eqn:MC_CLT_estimator_properties}
    \hat{\mu}_{\Psi} \sim \mathcal{N} \left( \mu_{\Psi} , \frac{1}{\mathcal{C} \mathcal{T}} \Sigma_d \right)
\end{equation}

Clearly, to assign uncertainty bounds on $ \hat{\mu}_{\Psi} $, we need to estimate $ \Sigma_d $. A naive estimate of $ \Sigma_d $ can be constructed by individually estimating each of the terms in the right-hand-side sum of Eq.~\eqref{eqn:asymptotic_variance_estimator} from finite length chains (i.e., $ \mathcal{T} < \infty $) and then summing them up.

We can estimate $ \Gamma (0) $ and $ \Gamma (\tau) $ for each individual chain (i.e. $ k = 1, \dots, \mathcal{C} $) as $ \Gamma_k (0) $ and $ \Gamma_k (\tau) $ in the following way
\begin{equation}
\label{eqn:individual_cov_estimators_MC_CLT}
    \begin{split}
        \hat{\Gamma}_k (0) = \frac{1}{\mathcal{T}-1} \sum_{t=1}^\mathcal{T} \left[ \Psi (\mathcal{X}_{k, t}) - \overline{\Psi}_k \right] \left[ \Psi (\mathcal{X}_{k, t}) - \overline{\Psi}_k \right]^T \\
        \hat{\Gamma}_k (\tau) = \frac{1}{\mathcal{T} - \tau} \sum_{t=1}^{\mathcal{T} - \tau} \left[ \Psi (\mathcal{X}_{k, t}) - \overline{\Psi}_k \right] \left[ \Psi (\mathcal{X}_{k, t+\tau}) - \overline{\Psi}_k \right]^T
    \end{split}
\end{equation}
where $ \overline{\Psi}_k = \mathcal{T}^{-1} \sum_{t=1}^\mathcal{T} \Psi (\mathcal{X}_{k, t}) $. Thus, the naive estimator of $ \Sigma_d $ for each chain can be written as
\begin{equation}
    \label{eqn:naive_asymptotic_variance_estimator}
    {}_k \hat{\Sigma}_{d, \text{naive}} = \hat{\Gamma}_k (0) +  2 \sum_{\tau=1}^{\mathcal{T}-1} \hat{\Gamma} (\tau)
\end{equation}
and for all chains taken together as
\begin{equation}
    \hat{\Sigma}_{d, \text{naive}} = \frac{1}{\mathcal{C}} \sum_{k=1}^\mathcal{C} {}_k \hat{\Sigma}_{d, \text{naive}}
\end{equation}
Note that the maximum value possible for $ \tau $ when each chain has a finite length $ \mathcal{T} $ is $ \mathcal{T} - 1 $.

Unfortunately, it is well known that $ {}_k \hat{\Sigma}_{d, \text{naive}} $ is not a consistent estimator of the asymptotic variance, i.e., $ \operatorname{\mathbb{V}ar} \left[ {}_k \hat{\Sigma}_{d, \text{naive}} \right] \nrightarrow 0 $ as $ \mathcal{T} \to \infty $~\cite{MCMCHandbookChapter1,GeyerPracticalMCMC,Priestley1981}. Since $ \mathcal{C} $ is usually fixed a priori by external considerations, this implies that the variance in $ \hat{\Sigma}_{d, \text{naive}} $ does not decrease as data is increased. Therefore, the problem of estimating the asymptotic variance for Markov Chains (for both single chains and multiple parallel chains) has received significant attention in the literature. There are various types of estimators to do so, such as Initial Sequence Estimators~\cite{GeyerPracticalMCMC}, Spectral Window Methods~\cite{AgarwalVatsGloballyCenteredAutocovariance}, and Batch Means Methods~\cite{ArgonAndradottirRBM,gupta2020globallycenteredbatchmeans}. Further, work has also been done in exploring the optimal parameters for these estimators~\cite{FlegalJonesBatchSize,LiuVatsOptimalBatchSize} as well as improving the overall quality of the estimates~\cite{VatsLugsail}.

\section{Analyzing the Behavior of Self-normalized Importance Sampling}
\label{section:SNIS_behavior}

As mentioned in Section~\ref{section:Implementation_details}, Self-normalized Importance Sampling (SNIS) produces highly biased estimates when used for rare-event failure estimation using the proposed $ \mathfrak{q} (\mathbf{x}, \beta) $ as the ISD. To explain this, we first analyze the SNIS formulation. From section~\ref{section:Importance_Sampling}, we know that 
\begin{equation}
\label{eqn:IS_general}
    \hat{P}_{F_{IS}} = \frac{1}{N} \sum_{i=1}^N \left[ I_H (\mathbf{x}_i) \frac{f_\mathbf{X} (\mathbf{x}_i)}{q_\mathbf{X} (\mathbf{x}_i)} \right]
\end{equation}
where the notation is as per Table~\ref{tab:list_of_notations}, and each $ \mathbf{x}_i $ is drawn from $ q_\mathbf{X} (\mathbf{x}) $. 
If $ (\cdot)^u $ indicates the unnormalized form of a density function, then the SNIS estimator can be given by
\begin{equation}
\label{eqn:IS_self-normalized_general}
    \hat{P}_{F_{SNIS}} = \frac{ \frac{1}{N} \sum_{i=1}^N \left[ I_H (\mathbf{x}_i) \frac{f^u_\mathbf{X} (\mathbf{x}_i)}{q^u_\mathbf{X} (\mathbf{x}_i)} \right] }{ \frac{1}{N} \sum_{i=1}^N \left[ \frac{f^u_\mathbf{X} (\mathbf{x}_i)}{q^u_\mathbf{X} (\mathbf{x}_i)} \right] } = \frac{ \sum_{i=1}^N \left[ I_H (\mathbf{x}_i) \frac{f^u_\mathbf{X} (\mathbf{x}_i)}{q^u_\mathbf{X} (\mathbf{x}_i)} \right] }{ \sum_{i=1}^N \left[ \frac{f^u_\mathbf{X} (\mathbf{x}_i)}{q^u_\mathbf{X} (\mathbf{x}_i)} \right] }
\end{equation}
Note that Eq.~\eqref{eqn:IS_self-normalized_general} can be used with the normalized density in place of its unnormalized form for either $ f^u_\mathbf{X} (\mathbf{x}) $ or $ q^u_\mathbf{X} (\mathbf{x}) $. The particular case of interest for us is when $ f_\mathbf{X} (\mathbf{x}) $ and $ q^u_\mathbf{X} (\mathbf{x}) $ are involved. If we define $ C_q $ to be the normalizing constant of $ q^u_\mathbf{X} (\mathbf{x}) $ (i.e., $ q_\mathbf{X} (\mathbf{x}) = q^u_\mathbf{X} (\mathbf{x}) / C_q $), then we can write
\begin{align}
    1 &= \int_{\Omega} f_\mathbf{X} (\mathbf{x}) d \mathbf{x} = \int_{\Omega} \frac{f_\mathbf{X} (\mathbf{x})}{q_\mathbf{X} (\mathbf{x})} q_\mathbf{X} (\mathbf{x}) d \mathbf{x} \\
     &= C_q \int_{\Omega} \frac{f_\mathbf{X} (\mathbf{x})}{q^u_\mathbf{X} (\mathbf{x})} q_\mathbf{X} (\mathbf{x}) d \mathbf{x} \\
    \Rightarrow C_q &= \frac{1}{\int_{\Omega} \frac{f_\mathbf{X} (\mathbf{x})}{q^u_\mathbf{X} (\mathbf{x})} q_\mathbf{X} (\mathbf{x}) d \mathbf{x}} \\
    \Rightarrow \hat{C}_{q_{IS}} &= \frac{1}{\frac{1}{N} \sum_{i=1}^N \left[ \frac{f_\mathbf{X} (\mathbf{x}_i)}{q^u_\mathbf{X} (\mathbf{x}_i)} \right]} \label{eqn:normalizing_constant_estimate}
\end{align}
where each $ \mathbf{x}_i $ is drawn from $ q_\mathbf{X} (\mathbf{x}) $. We can also see that Eq.~\eqref{eqn:IS_general} can be rewritten as
\begin{equation}
\label{eqn:IS_general_norm_constant}
    \hat{P}_{F_{IS}} = C_q \frac{1}{N} \sum_{i=1}^N \left[ I_H (\mathbf{x}_i) \frac{f_\mathbf{X} (\mathbf{x}_i)}{q^u_\mathbf{X} (\mathbf{x}_i)} \right]
\end{equation}

Comparing Eq.~\eqref{eqn:IS_general_norm_constant} with Eqs.~\eqref{eqn:IS_self-normalized_general} and~\eqref{eqn:normalizing_constant_estimate}, it is apparent that when the input distribution is normalized but the ISD is not, the SNIS estimator is the same as the usual IS estimator but with the ISD normalizing constant estimated using IS as well.

For our choice of ISD, if we define $ \mathfrak{q}^u_{\mathbf{X}} (\mathbf{x}, \beta) = S_L (\mathbf{x}, \beta) f_\mathbf{X} (\mathbf{x}) $ (adapted from Eq.~\ref{eqn:proposed_ISD}), then we can use Eq.~\eqref{eqn:normalizing_constant_estimate} to write an IS estimate of our normalizing constant $ C_S $ as
\begin{equation}
    \label{eqn:C_S_estimate_IS}
    \hat{C}_{S_{IS}} = \frac{1}{\frac{1}{N} \sum_{i=1}^N \left[ \frac{1}{S_L (\mathbf{x}_i, \beta)} \right]}
\end{equation}
where each $ \mathbf{x}_i $ is drawn from $ \mathfrak{q}_\mathbf{X} (\mathbf{x}, \beta) $. Eq.~\eqref{eqn:C_S_estimate_IS} provides the key insight into why using SNIS for MFIS and E-ACV produces such biased estimates. In particular, we will compare it to the MC estimator
\begin{equation}
    \label{eqn:C_S_estimate_MC}
    \hat{C}_{S_{MC}} = \frac{1}{N} \sum_{i=1}^N S_L (\mathbf{x}_i, \beta) 
\end{equation}
where each $ \mathbf{x}_i $ is drawn from $ f_\mathbf{X} (\mathbf{x}) $.

First, notice that for Eq.~\eqref{eqn:C_S_estimate_MC}, most of the samples are concentrated near the peak of $ f_\mathbf{X} (\mathbf{x}) $. Assuming that failure occurs in the tail of the input distribution and that the LF model reflects this to some degree, we see that $ S_L (\mathbf{x}, \beta) $ takes very small values for most samples. On the other hand, for Eq.~\eqref{eqn:C_S_estimate_IS}, most of the samples are concentrated near the peak of $ \mathfrak{q}_\mathbf{X} (\mathbf{x}, \beta) $, which occurs near the LF model failure surface. In this region, $ S_L (\mathbf{x}, \beta) $ takes significantly larger values.

Next, observe that $ S_L (\mathbf{x}, \beta) $, the logistic function, takes values between $ 0 $ and $ 1 $ in its entire domain, implying that $ 1 / S_L (\mathbf{x}, \beta) $ takes values between $ 1 $ and $ \infty $. Thus, singular outliers do not impact estimators using $ S_L (\mathbf{x}, \beta) $ by much, but can skew an estimator dependent on $ 1 / S_L (\mathbf{x}, \beta) $ by massive amounts. For example, consider a sample $ \mathbf{x} $ that lies in the region where $ S_L (\mathbf{x}, \beta) \approx 0.001 $; a little change in its location (say making $ S_L (\mathbf{x}, \beta) \approx 0.002 $) impacts $ \hat{C}_{S_{MC}} $ very minimally while it can potentially impact $ \hat{C}_{S_{IS}} $ a lot (the reciprocal of the logistic goes from $ 1000 $ to $ 2000 $). This significantly heightens any bias in the underlying sample set between $ \hat{C}_{S_{MC}} $ and $ \hat{C}_{S_{IS}} $.

Putting these observations together, we see that the sample set used in $ \hat{C}_{S_{IS}} $ is concentrated in regions having larger values of $ S_L (\mathbf{x}, \beta) $  and therefore smaller values of $ 1 / S_L (\mathbf{x}, \beta) $. Thus, we expect to see any imperfections in the sampling scheme (associated with small values of $ S_L (\mathbf{x}, \beta) $) show up as significant positive bias in the $ \hat{C}_{S_{IS}} $ estimator. Considering that due to computational budget restrictions, the MCMC scheme used to sample from the ISD uses short chains, one can expect to see some bias in the MCMC-generated samples from $ \mathfrak{q}_\mathbf{X} (\mathbf{x}, \beta) $. Thus, we expect SNIS-based estimators to give positively biased estimates when used for MFIS and E-ACV.

\begin{figure}[!htbp]
\centering
\begin{subfigure}{.45\textwidth}
  \centering
  \includegraphics[width=\linewidth]{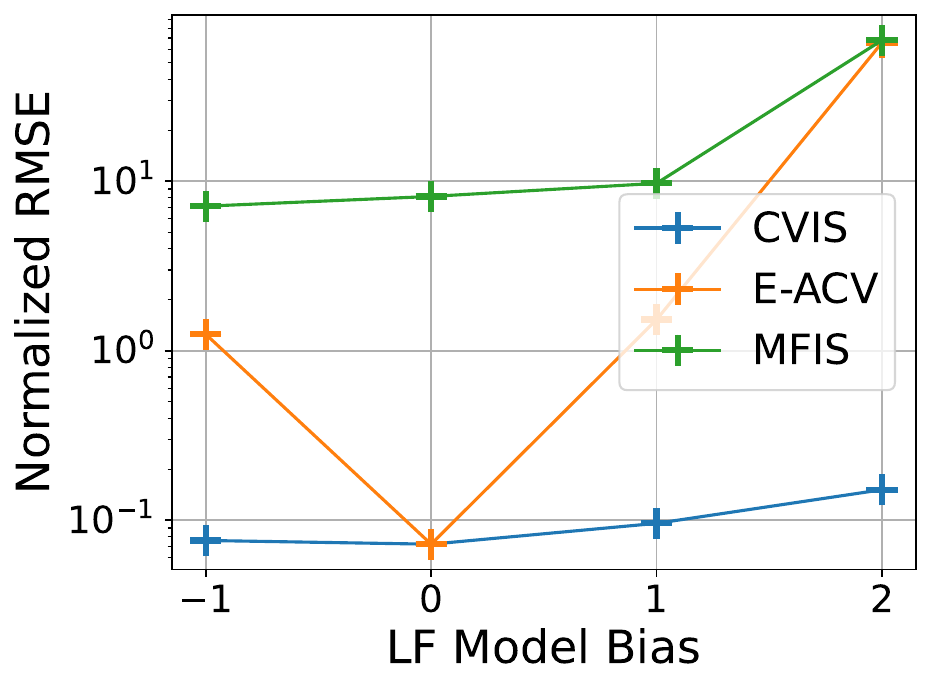}
  \caption{}
  \label{fig:RMSE_var_0_SNIS}
\end{subfigure}%
\begin{subfigure}{.45\textwidth}
  \centering
  \includegraphics[width=\linewidth]{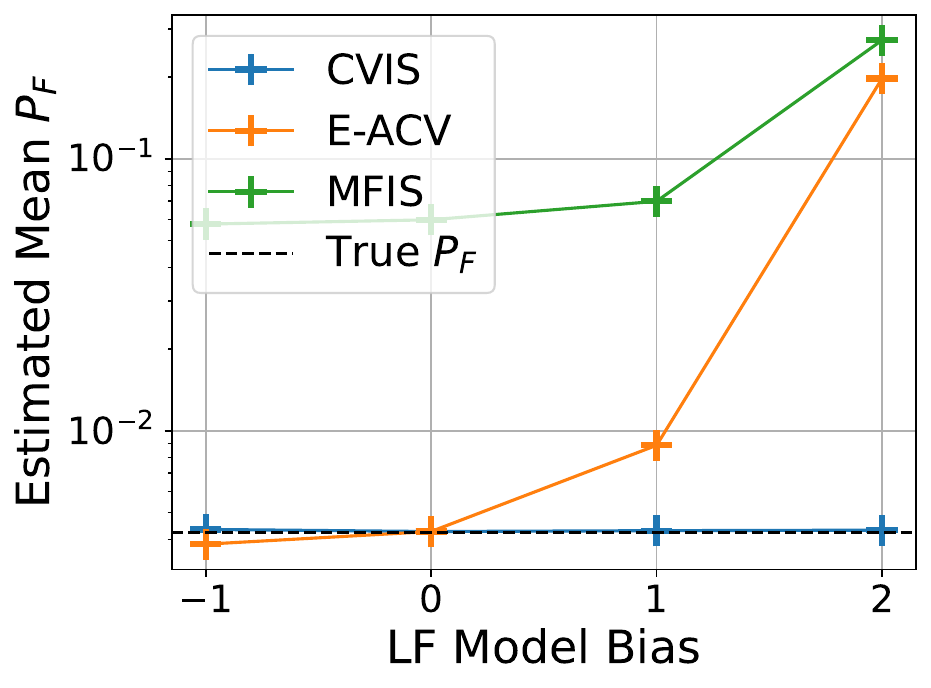}
  \caption{}
  \label{fig:mean_var_0_SNIS}
\end{subfigure}
\begin{subfigure}{.45\textwidth}
  \centering
  \includegraphics[width=\linewidth]{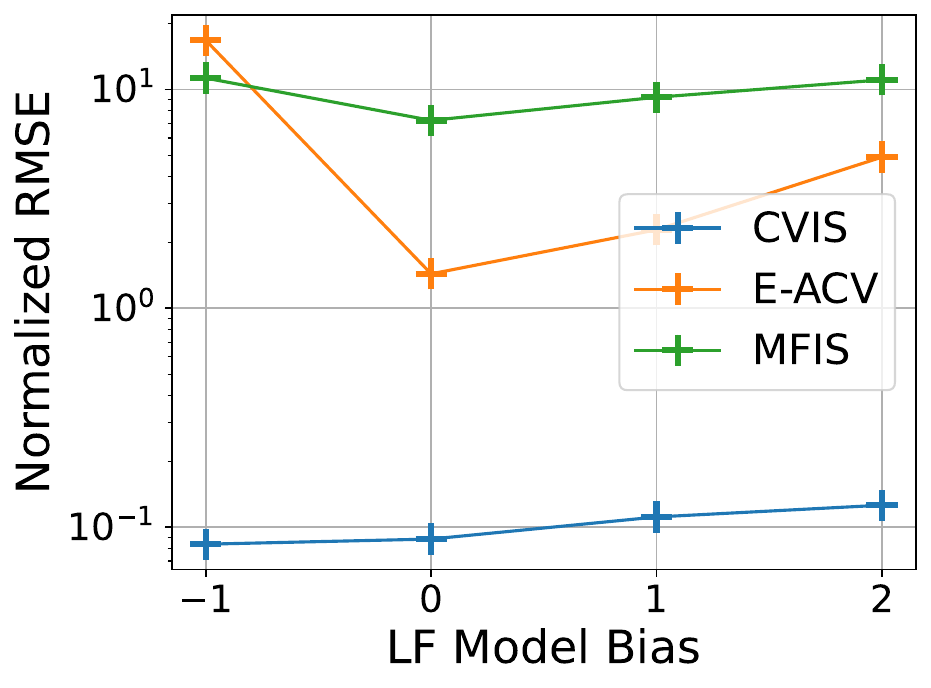}
  \caption{}
  \label{fig:RMSE_var_1_SNIS}
\end{subfigure}%
\begin{subfigure}{.45\textwidth}
  \centering
  \includegraphics[width=\linewidth]{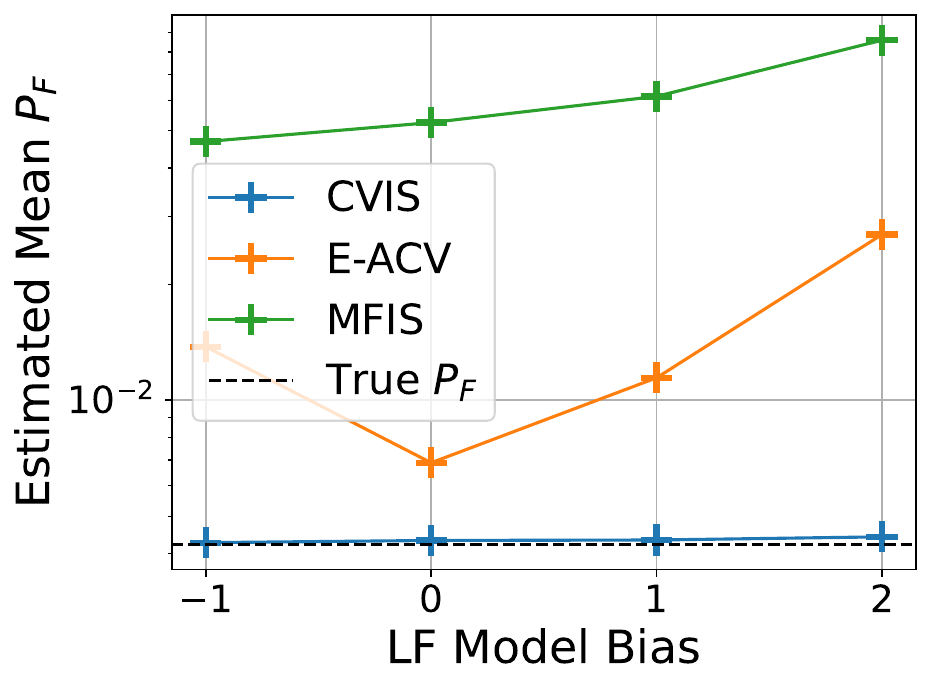}
  \caption{}
  \label{fig:mean_var_1_SNIS}
\end{subfigure}
\begin{subfigure}{.45\textwidth}
  \centering
  \includegraphics[width=\linewidth]{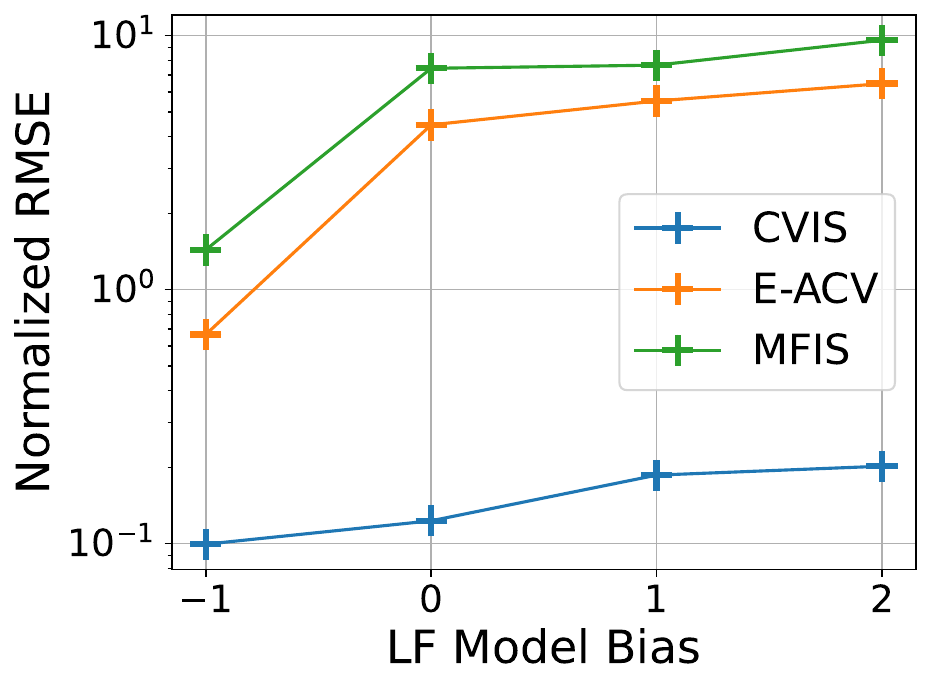}
  \caption{}
  \label{fig:RMSE_var_2_SNIS}
\end{subfigure}%
\begin{subfigure}{.45\textwidth}
  \centering
  \includegraphics[width=\linewidth]{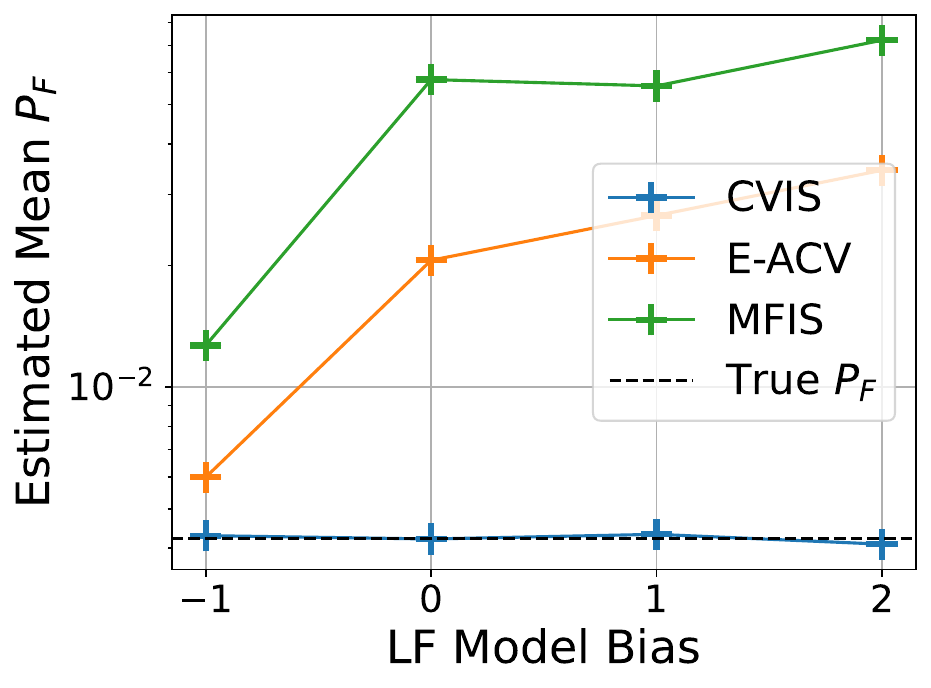}
  \caption{}
  \label{fig:mean_var_2_SNIS}
\end{subfigure}
\caption{Results for Example 1 using SNIS for MFIS and E-ACV. (a), (c), and (e) show the root mean squared errors, and (b), (d), and (f) show the estimated failure probability for different magnitudes of noise and increasing model bias. The dotted black lines in (b), (d), and (f) represent the true failure probability. Figures (a) and (b) correspond to $ \sigma_{\epsilon} = 0 $, figures (c) and (d) correspond to $ \sigma_{\epsilon} = 1 $, and figures (e) and (f) correspond to $ \sigma_{\epsilon} = 2 $. Statistics are computed as the mean from 100 independent trials.}
\label{fig:SNIS_results}
\end{figure}

This is indeed what we observe 
when we apply SNIS in Example 1 (Section~\ref{section:Example_1_biased_and_noise_LF_model}) using identical sample allocation for CVIS, MFIS, and E-ACV (as defined in row 1 in Table~\ref{tab:example_1_sample_allocation_summary}).
Figure~\ref{fig:SNIS_results} plots the RMSE of each estimator normalized by the true failure probability from 100 independent trials. It also plots the failure probability predicted by each estimator, with the true failure probability shown using the dashed black line. As expected, we see high positive bias in the estimates of MFIS and E-ACV, sometimes of more than an order of magnitude.

Based on the above argument, one might question the apparently greater stability of $ \Tilde{\mathcal{Q}} = \sum_{i=1}^N \left[ I_H \left( \mathbf{x}_i \right) / S_L \left( \mathbf{x}_i , \beta \right) \right] / N $ and $ \Tilde{\mathcal{Q}}_L = \sum_{i=1}^N \left[ I_L \left( \mathbf{x}_i \right) / S_L \left( \mathbf{x}_i , \beta \right) \right] / N $, which both also have $ S_L \left( \mathbf{x}_i , \beta \right) $ in the denominator. For $ \Tilde{\mathcal{Q}}_L $ this explanation is straightforward; by definition $ S_L \left( \mathbf{x}_i , \beta \right) \geq 0.5 $ if $ I_L \left( \mathbf{x}_i \right) = 1 $, and thus $ S_L \left( \mathbf{x}_i , \beta \right) \in \left[ 1, 2 \right] $ wherever $ I_L \left( \mathbf{x}_i \right) $ is not zero, which makes $ \Tilde{\mathcal{Q}}_L $ much less sensitive to biases in the sample set used. Similarly, under the conditions on the relative behavior of the HF and LF models that make CVIS useful in practice, $ I_H \left( \mathbf{x} \right) = 0 $ in most regions where $ S_L \left( \mathbf{x}, \beta \right) $ takes the smallest values. This makes $ \Tilde{\mathcal{Q}} $ fairly stable as well.

In light of these results, we do not recommend using SNIS with our ISD formulation unless significantly longer MCMC chains can be run, thus improving sample quality from the ISD. For CVIS, however, the normalization constant is absent from the formulation without introducing significant bias to the estimator.

\section{Proofs, Additional Theorems, and Mathematical Derviations}
\label{appendix:proofs_theorems_derivations}

This Appendix presents proofs for the theorems presented in the manuscript and provides some additional theorems and derivations to facilitate the understanding of the theory supporting our framework.

\subsection{Proof of Theorem~\ref{thm:CVIS_variance_less_than_CV_variance}}
\label{appendix:proof_CVIS_variance_less_than_CV_variance}

\begin{proof}[Proof of Theorem~\ref{thm:CVIS_variance_less_than_CV_variance}]

Considering that $ \hat{P}_{F_L} $ is independent from from $ \hat{Q} $ and $ \hat{Q}_L $, the variance of $ \hat{P}_{F_{CVMC}} $ can be written from Eq.~\eqref{eqn:acv_with_cmc} as follows
\begin{equation}
\label{eqn:variance_CVMC_general}
    \operatorname{\mathbb{V}ar} \left[ \hat{P}_{F_{CVMC}} \right] = \operatorname{\mathbb{V}ar} \left[ \hat{Q}_{MC} \right] + \alpha^2 \operatorname{\mathbb{V}ar} \left[ \hat{Q}_{L_{MC}} \right] + \alpha^2 \operatorname{\mathbb{V}ar} \left[ \hat{P}_{F_L} \right] - 2 \alpha \operatorname{\mathbb{C}ov} \left[ \hat{Q}_{MC}, \hat{Q}_{L_{MC}} \right]
\end{equation}
Applying Monte Carlo estimators for $ \hat{Q}_{MC} $ and $ \hat{Q}_{L_{MC}} $, 
we get the following simplification
\begin{equation}
\label{eqn:variance_CVMC_simplified}
    \operatorname{\mathbb{V}ar} \left[ \hat{P}_{F_{CVMC}} \right] = \frac{\operatorname{\mathbb{V}ar}_f \left[ I_H \right]}{N} + \alpha^2 \frac{\operatorname{\mathbb{V}ar}_f \left[ I_L \right]}{N} + \alpha^2 \operatorname{\mathbb{V}ar} \left[ \hat{P}_{F_L} \right] - 2 \alpha \frac{\operatorname{\mathbb{C}ov}_f \left[ I_H , I_L \right]}{N}
\end{equation}
Here, $ \operatorname{\mathbb{V}ar}_f \left[ \cdot \right] $ and $ \operatorname{\mathbb{C}ov}_f \left[ \cdot \right] $ represent the variance and covariance operators with respect to the distribution $ f_{\mathbf{X}} (\mathbf{x}) $.

Similarly, the variance of $ \hat{P}_{F_{CVIS}} $ can be written from Eq.~\eqref{eqn:acv_with_is} as
\begin{align}
    \operatorname{\mathbb{V}ar} \left[ \hat{P}_{F_{CVIS}} \right] & = \operatorname{\mathbb{V}ar} \left[ \hat{Q}_{IS} \right] + \alpha^2 \operatorname{\mathbb{V}ar} \left[ \hat{Q}_{L_{IS}} \right] + \alpha^2 \operatorname{\mathbb{V}ar} \left[ \hat{P}_{F_L} \right] - 2 \alpha \operatorname{\mathbb{C}ov} \left[ \hat{Q}_{IS}, \hat{Q}_{L_{IS}} \right] \label{eqn:variance_CVIS_general} \\
    & = C_S^2 \frac{\operatorname{\mathbb{V}ar}_q \left[ \frac{I_H}{S_L} \right]}{N} + \alpha^2 C_S^2 \frac{\operatorname{\mathbb{V}ar}_q \left[ \frac{I_L}{S_L} \right]}{N} + \alpha^2 \operatorname{\mathbb{V}ar} \left[ \hat{P}_{F_L} \right] - 2 \alpha C_S^2 \frac{\operatorname{\mathbb{C}ov}_q \left[ \frac{I_H}{S_L} , \frac{I_L}{S_L} \right]}{N} \label{eqn:variance_CVIS_simplified}
\end{align}
Note that dependence on $\mathbf{x}$ has been dropped for simplicity in notation.

Using the definition of the variance and covariance operators, we can expand Eq.~\eqref{eqn:variance_CVMC_simplified} as
\begin{gather}
    \operatorname{\mathbb{V}ar} \left[ \hat{P}_{F_{CVMC}} \right] = T_1^{(CVMC)} + T_2^{(CVMC)} \label{eqn:variance_CVMC_expansion} \\
    T_1^{(CVMC)} = \frac{\mathbb{E}_f \left[ I_H^2 \right]}{N} + \alpha^2 \frac{\mathbb{E}_f \left[ I_L^2 \right]}{N} - 2 \alpha \frac{\mathbb{E}_f \left[ I_H I_L \right]}{N} \label{eqn:variance_CVMC_expansion_t1} \\
    T_2^{(CVMC)} = \alpha^2 \operatorname{\mathbb{V}ar} \left[ \hat{P}_{F_L} \right] - \left\{\frac{ \left(\mathbb{E}_f \left[ I_H \right] \right)^2}{N} + \alpha^2 \frac{\left(\mathbb{E}_f \left[ I_L \right] \right)^2}{N} - 2 \alpha \frac{\mathbb{E}_f \left[ I_H \right] \mathbb{E}_f \left[ I_L \right]}{N} \right\} \label{eqn:variance_CVMC_expansion_t2}
\end{gather}

Similarly, the expansion of Eq.~\eqref{eqn:variance_CVIS_simplified} is
\begin{gather}
    \operatorname{\mathbb{V}ar} \left[ \hat{P}_{F_{CVIS}} \right] = T_1^{(CVIS)} + T_2^{(CVIS)} \label{eqn:variance_CVIS_expansion} \\
    T_1^{(CVIS)} = C_S^2 \left\{ \frac{\mathbb{E}_q \left[ \left( \frac{I_H}{S_L} \right)^2 \right]}{N} + \alpha^2 \frac{\mathbb{E}_q \left[ \left( \frac{I_L}{S_L} \right)^2 \right]}{N} - 2 \alpha \frac{\mathbb{E}_q \left[ \left( \frac{I_H}{S_L} \right) \left( \frac{I_L}{S_L} \right) \right]}{N} \right\} \label{eqn:variance_CVIS_expansion_t1} \\
    T_2^{(CVIS)} = \alpha^2 \operatorname{\mathbb{V}ar} \left[ \hat{P}_{F_L} \right] - C_S^2 \left\{\frac{ \left( \mathbb{E}_q \left[ \frac{I_H}{S_L} \right] \right)^2}{N} + \alpha^2 \frac{\left( \mathbb{E}_q \left[ \frac{I_L}{S_L} \right] \right)^2}{N} - 2 \alpha \frac{\mathbb{E}_q \left[ \frac{I_H}{S_L} \right] \mathbb{E}_q \left[ \frac{I_L}{S_L} \right]}{N} \right\} \label{eqn:variance_CVIS_expansion_t2}
\end{gather}

Using lemma~\ref{lemma:relationship_expectation_wrt_f_expectation_wrt_k}, we can say that $ T_2^{(CVIS)} = T_2^{(CVMC)} $ for all values of $ \alpha $. Therefore, to prove the theorem, we simply need to show that $ T_1^{(CVIS)} \leq T_1^{(CVMC)} $.

Next, recognizing that $ I_H^2 = I_H $ and $ I_L^2 = I_L $, we can use lemma~\ref{lemma:intersection_failure_smaller_than_individual_failure} to expand $ T_1^{(CVMC)}$ (Eq.~\eqref{eqn:variance_CVMC_expansion_t1}) in the following manner
\begin{equation}
    \begin{aligned}
        T_1^{(CVMC)} &= \frac{\mathbb{E}_f \left[ I_H^2 \right]}{N} + \alpha^2 \frac{\mathbb{E}_f \left[ I_L^2 \right]}{N} - 2 \alpha \frac{\mathbb{E}_f \left[ I_H I_L \right]}{N} \\
        &= \frac{1}{N} \int_{\Omega} I_H^2 (\mathbf{x}) f (\mathbf{x}) d\mathbf{x} + \frac{\alpha^2}{N} \int_{\Omega} I_L^2 (\mathbf{x}) f (\mathbf{x}) d\mathbf{x} - \frac{2 \alpha}{N} \int_{\Omega} I_H (\mathbf{x}) I_L (\mathbf{x}) f (\mathbf{x}) d\mathbf{x} \\
        &= \frac{\left(1 - \alpha\right)^2}{N} \int_{\Omega_{HL}} I_H (\mathbf{x}) I_L (\mathbf{x}) f_{\mathbf{X}} (\mathbf{x}) d\mathbf{x} + \frac{1}{N} \int_{\Omega_{HL}^C} \left( I_H (\mathbf{x}) + \alpha^2 I_L (\mathbf{x}) \right) f_{\mathbf{X}} (\mathbf{x}) d\mathbf{x}
    \end{aligned}
\end{equation}
where $ \Omega_{HL} = \left\{ \mathbf{x} \in \Omega : I_H (\mathbf{x}) I_L (\mathbf{x}) = 1 \right\} $ and $ \Omega_{HL}^C = \Omega \setminus \Omega_{HL} $. Thus,
\begin{gather}
    T_1^{(CVMC)} = T_3^{(CVMC)} + T_4^{(CVMC)} \label{eqn:t1_CVMC_expansion} \\
    T_3^{(CVMC)} = \frac{\left(1 - \alpha\right)^2}{N} \int_{\Omega_{HL}} I_H (\mathbf{x}) I_L (\mathbf{x}) f_{\mathbf{X}} (\mathbf{x}) d\mathbf{x} \label{eqn:t1_CVMC_expansion_t3} \\
    T_4^{(CVMC)} = \frac{1}{N} \int_{\Omega_{HL}^C} \left( I_H (\mathbf{x}) + \alpha^2 I_L (\mathbf{x}) \right) f_{\mathbf{X}} (\mathbf{x}) d\mathbf{x} \label{eqn:t1_CVMC_expansion_t4}
\end{gather}

We can similarly expand $ T_1^{(CVIS)} $ (Eq.~\eqref{eqn:variance_CVIS_expansion_t1}) as
\begin{gather}
    T_1^{(CVIS)} = T_3^{(CVIS)} + T_4^{(CVIS)} \label{eqn:t1_CVIS_expansion} \\
    T_3^{(CVIS)} = \frac{\left( 1 - \alpha \right)^2 C_S^2}{N} \int_{\Omega_{HL}} \frac{ I_H (\mathbf{x}) I_L (\mathbf{x})}{ \left( S_L (\mathbf{x}, \beta^*) \right)^2 } \mathfrak{q}_{\mathbf{X}} (\mathbf{x}, \beta^*) d\mathbf{x} \label{eqn:t1_CVIS_expansion_t3} \\
    T_4^{(CVIS)} = \frac{C_S^2}{N} \int_{\Omega_{HL}^C} \frac{ \left( I_H (\mathbf{x}) + \alpha^2 I_L (\mathbf{x}) \right) }{ \left( S_L (\mathbf{x}, \beta^*) \right)^2 } \mathfrak{q}_{\mathbf{X}} (\mathbf{x}, \beta^*) d\mathbf{x} \label{eqn:t1_CVIS_expansion_t4}
\end{gather}

For equations~\ref{eqn:t1_CVMC_expansion} -~\ref{eqn:t1_CVMC_expansion_t4} we use lemma~\ref{lemma:intersection_failure_smaller_than_individual_failure} with $ r (\mathbf{x}) \coloneqq 1 $ and $ \pi (\mathbf{x}) \coloneqq f_{\mathbf{X}} (\mathbf{x}) $, while for equations~\ref{eqn:t1_CVIS_expansion} -~\ref{eqn:t1_CVIS_expansion_t4} we use lemma~\ref{lemma:intersection_failure_smaller_than_individual_failure} with $ r (\mathbf{x}) \coloneqq \left( S_L (\mathbf{x}, \beta^*) \right)^2 $ and $ \pi (\mathbf{x}) \coloneqq \mathfrak{q}_{\mathbf{X}} (\mathbf{x}, \beta^*) $.

Finally, we apply lemma~\ref{lemma:division_by_logistic_inequality} to obtain
\begin{align}
    C_S^2 \int_{\Omega_{HL}} \frac{ I_H (\mathbf{x}) I_L (\mathbf{x})}{ \left( S_L (\mathbf{x}, \beta^*) \right)^2 } \mathfrak{q}_{\mathbf{X}} (\mathbf{x}, \beta^*) d\mathbf{x} &\leq \int_{\Omega_{HL}} I_H (\mathbf{x}) I_L (\mathbf{x}) f_{\mathbf{X}} (\mathbf{x}) d\mathbf{x} \label{eqn:lemma_division_by_logistic_eq1} \\
    C_S^2 \int_{\Omega_{HL}^C} \frac{ I_H (\mathbf{x}) }{ \left( S_L (\mathbf{x}, \beta^*) \right)^2 } \mathfrak{q}_{\mathbf{X}} (\mathbf{x}, \beta^*) d\mathbf{x}  &\leq \int_{\Omega_{HL}^C}  I_H (\mathbf{x}) f_{\mathbf{X}} (\mathbf{x}) d\mathbf{x} \label{eqn:lemma_division_by_logistic_eq2} \\
    C_S^2 \int_{\Omega_{HL}^C} \frac{ I_L (\mathbf{x}) }{ \left( S_L (\mathbf{x}, \beta^*) \right)^2 } \mathfrak{q}_{\mathbf{X}} (\mathbf{x}, \beta^*) d\mathbf{x}  &\leq \int_{\Omega_{HL}^C} I_L (\mathbf{x}) f_{\mathbf{X}} (\mathbf{x}) d\mathbf{x} \label{eqn:lemma_division_by_logistic_eq3}
\end{align}

From Eq.~\eqref{eqn:lemma_division_by_logistic_eq1} we see that $ T_3^{(CVIS)} \leq T_3^{(CVMC)} $, and from Eqs.~\eqref{eqn:lemma_division_by_logistic_eq2} and~\eqref{eqn:lemma_division_by_logistic_eq3} we see that $ T_4^{(CVIS)} \leq T_4^{(CVMC)} $ for arbitrary $ \alpha $.

Therefore, $ T_1^{(CVIS)} \leq T_1^{(CVMC)} $ for all values of $ \alpha $, which completes the proof. 
\end{proof}

\begin{lemma}
\label{lemma:relationship_expectation_wrt_f_expectation_wrt_k}
For $ \lambda \in \left\{ H, L, HL \right\} $, where $ I_{HL} (\mathbf{x}) = I_H (\mathbf{x}) I_L (\mathbf{x}) $ and $ I_H (\mathbf{x}) $ and $ I_L (\mathbf{x}) $ are as previously defined
\begin{equation}
    C_S \mathbb{E}_q \left[ \frac{I_{\lambda}}{S_L} \right] = \mathbb{E}_f \left[ I_{\lambda} \right]
\end{equation}
\end{lemma}

\begin{proof}[Proof of Lemma~\ref{lemma:relationship_expectation_wrt_f_expectation_wrt_k}]
By definition
\begin{equation}
    C_S \mathbb{E}_q \left[ \frac{I_{\lambda}}{S_L} \right] = C_S \int_{\Omega} \frac{I_{\lambda} \left( \mathbf{x} \right)}{S_L \left( \mathbf{x}, \beta^* \right)} \mathfrak{q}_{\mathbf{X}} \left( \mathbf{x}, \beta^* \right) d\mathbf{x}
\end{equation}
Substituting Eq.~\eqref{eqn:proposed_ISD}, and recognizing the definition of $ \mathbb{E}_f \left[ I_{\lambda} \right] $ , we get
\begin{align}
    C_S \mathbb{E}_q \left[ \frac{I_{\lambda}}{S_L} \right] &= C_S \int_{\Omega} \frac{I_{\lambda} \left( \mathbf{x} \right)}{C_S} f_{\mathbf{X}} \left( \mathbf{x} \right) d\mathbf{x}
    = \int_{\Omega} I_{\lambda} \left( \mathbf{x} \right) f_{\mathbf{X}} \left( \mathbf{x} \right) d\mathbf{x} \\
    &= \mathbb{E}_f \left[ I_{\lambda} \right]
\end{align}
\end{proof}

\begin{lemma}
\label{lemma:intersection_failure_smaller_than_individual_failure}
Let $ r (\mathbf{x}) : \Omega \to \mathbb{R} $ be some function of $ \mathbf{x} $ such that $ r (\mathbf{x}) > 0 $ $ \forall \mathbf{x} \in \Omega $, and $ \pi(\mathbf{x}) $ be some probability density function on $ \Omega $ (i.e. $ \operatorname{supp} \left( \pi(\mathbf{x}) \right) = \Omega $). Then, for $ \lambda \in \left\{ H, L \right\} $,
\begin{equation}
    \int_{\Omega} \frac{I_{\lambda} (\mathbf{x})}{r(\mathbf{x})} \pi(\mathbf{x}) d\mathbf{x} = \int_{\Omega_{HL}} \frac{I_{HL} (\mathbf{x})}{r(\mathbf{x})} \pi(\mathbf{x}) d\mathbf{x} + \int_{\Omega_{HL}^C} \frac{I_{\lambda} (\mathbf{x})}{r(\mathbf{x})} \pi(\mathbf{x}) d\mathbf{x}
\end{equation}
where $ \Omega_{HL} = \left\{ \mathbf{x} \in \Omega : I_{HL} (\mathbf{x}) = 1 \right\} $, $ \Omega_{HL}^C = \Omega \setminus \Omega_{HL} $, $ I_{HL} (\mathbf{x}) $ is as defined in lemma~\ref{lemma:relationship_expectation_wrt_f_expectation_wrt_k}, and $ I_H (\mathbf{x}) $ and $ I_L (\mathbf{x}) $ are as previously defined.
\end{lemma}

\begin{proof}[Proof of Lemma~\ref{lemma:intersection_failure_smaller_than_individual_failure}]
Notice that by definition, $ I_{HL} (\mathbf{x}) = 1 \Rightarrow I_{\lambda} (\mathbf{x}) = 1 $. Therefore, $ I_{\lambda} (\mathbf{x}) = I_{HL} (\mathbf{x}) ~ \forall \mathbf{x} \in \Omega_{HL} $

Now, by decomposing the integral over $ \Omega $ into two integrals over the non-overlapping regions $ \Omega_{HL} $ and $ \Omega_{HL}^C $ (by definition $ \Omega_{HL} \cup \Omega_{HL}^C = \Omega $), we can write the following
\begin{align}
    \int_{\Omega} \frac{I_{\lambda} (\mathbf{x})}{r(\mathbf{x})} \pi(\mathbf{x}) d\mathbf{x} &= \int_{\Omega_{HL}} \frac{I_{\lambda} (\mathbf{x})}{r(\mathbf{x})} \pi(\mathbf{x}) d\mathbf{x} + \int_{\Omega_{HL}^C} \frac{I_{\lambda} (\mathbf{x})}{r(\mathbf{x})} \pi(\mathbf{x}) d\mathbf{x} \\
    &= \int_{\Omega_{HL}} \frac{I_{HL} (\mathbf{x})}{r(\mathbf{x})} \pi(\mathbf{x}) d\mathbf{x} + \int_{\Omega_{HL}^C} \frac{I_{\lambda} (\mathbf{x})}{r(\mathbf{x})} \pi(\mathbf{x}) d\mathbf{x}
\end{align}
\end{proof}



\begin{lemma}
\label{lemma:division_by_logistic_inequality}
Given a set $ \Omega_s \subseteq \Omega $, if $ \lambda \in \left\{ H, L, HL \right\} $, where $ I_{HL} (\mathbf{x}) $ is as defined in lemma~\ref{lemma:relationship_expectation_wrt_f_expectation_wrt_k} and $ I_H (\mathbf{x}) $ and $ I_L (\mathbf{x}) $ are as previously defined, then
\begin{equation}
    C_S^2 \int_{\Omega_s} \frac{I_{\lambda} (\mathbf{x})}{\left( S_L (\mathbf{x}, \beta^*) \right)^2} \mathfrak{q}_{\mathbf{X}} \left( \mathbf{x}, \beta^* \right) d\mathbf{x} \leq \int_{\Omega_s} I_{\lambda} (\mathbf{x}) f_{\mathbf{X}} \left( \mathbf{x} \right) d\mathbf{x}
\end{equation}
\end{lemma}

\begin{proof}[Proof of Lemma~\ref{lemma:division_by_logistic_inequality}]
Substituting Eq.~\eqref{eqn:proposed_ISD}, we can simplify the left-hand-side
\begin{equation}
\label{eqn:division_by_logistic_LHS}
    C_S^2 \int_{\Omega_s} \frac{I_{\lambda} (\mathbf{x})}{\left( S_L (\mathbf{x}, \beta^*) \right)^2} \mathfrak{q}_{\mathbf{X}} \left( \mathbf{x}, \beta^* \right) d\mathbf{x} = C_S \int_{\Omega_s} \frac{I_{\lambda} (\mathbf{x})}{S_L (\mathbf{x}, \beta^*)} f_{\mathbf{X}} \left( \mathbf{x} \right) d\mathbf{x}
\end{equation}
Next, from remark~\ref{remark:optimally_tuned_ISD_CVIS}, we know that $ S_L \left( \mathbf{x}, \beta^* \right) \geq \xi^*  $ $ \forall \mathbf{x} \in \Omega_{\mathcal{L}} $. Recognizing that $ \Omega_{\lambda} \subseteq \Omega_{\mathcal{L}} $, we infer that $ \forall \mathbf{x} \in \Omega_{\lambda} $
\begin{equation}
    \frac{1}{S_L (\mathbf{x}, \beta^*)} \leq \frac{1}{\xi^*}
\end{equation}
Recollecting the definition of $ \Omega_{\lambda} $ and the fact that $ I_{\lambda} (\mathbf{x}) $ only takes values $ 0 $ and $ 1 $, we can extend Eq.~\eqref{eqn:division_by_logistic_LHS} into the following inequality
\begin{equation}
\label{eqn:division_by_logistic_final_step}
    C_S \int_{\Omega_s} \frac{I_{\lambda} (\mathbf{x})}{S_L (\mathbf{x}, \beta^*)} f_{\mathbf{X}} \left( \mathbf{x} \right) d\mathbf{x} \leq \frac{C_S}{\xi^*} \int_{\Omega_s} I_{\lambda} (\mathbf{x}) f_{\mathbf{X}} \left( \mathbf{x} \right) d\mathbf{x}
\end{equation}

Since $ C_S \sim O \left( P_{F_L} \right) $, and for rare events (and using Stipulation~\ref{stipulation:fundamental_model_quality_assumption}) $ P_{F_L} \ll 0.5 $, $ C_S < \xi^* $ for rare events ($ \xi^* $ is bounded between $ 0 $ and $ 0.5 $). This fact and Eq.~\eqref{eqn:division_by_logistic_final_step} taken together prove the lemma.
\end{proof}

\subsection{Proof of Theorem~\ref{thm:model_limit_CVIS_alpha_optimal}}
\label{appendix:proof_optimal_alpha_in_model_limit}

\begin{proof}[Proof of Theorem~\ref{thm:model_limit_CVIS_alpha_optimal}]
Consider $ \hat{P}_F^{(CVMC)} $ as defined in Eq.~\eqref{eqn:acv_with_cmc} and apply control variates theory to find the value of $ \alpha $ that optimizes variance reduction (Eq.~\eqref{eqn:ACV_optimal_alpha}). Let this value of $ \alpha $ be called $ \alpha^* $. Since $ \hat{P}_{F_L} $ is independent of $ \hat{Q}_{MC} $ and $ \hat{Q}_{L_{MC}} $,
\begin{equation}
\label{eqn:CVMC_optimal_alpha}
    \alpha^* = \frac{\operatorname{\mathbb{C}ov} \left[ \hat{Q}_{MC}, \hat{Q}_{L_{MC}} \right]}{\operatorname{\mathbb{V}ar} \left[ \hat{Q}_{L_{MC}} \right] + \operatorname{\mathbb{V}ar} \left[ \hat{P}_{F_L} \right]}
\end{equation}

From condition 3 of the theorem
, we know that $ \Bigl( \operatorname{\mathbb{V}ar} \left[ \hat{Q}_{L_{MC}} \right] + \operatorname{\mathbb{V}ar} \left[ \hat{P}_{F_L} \right] \Bigr) \approx \operatorname{\mathbb{V}ar} \left[ \hat{Q}_{L_{MC}} \right] $.
Further, by plugging in the definitions for the estimators and recalling that they use independent samples, we get
\begin{equation}
\label{eqn:CVMC_optimal_alpha_expansion}
    \alpha^* = \frac{\frac{1}{N} \operatorname{\mathbb{C}ov}_f \left[ I_H (\mathbf{x}), I_L (\mathbf{x}) \right]}{\frac{1}{N} \operatorname{\mathbb{V}ar}_f \left[ I_L (\mathbf{x}) \right]}
\end{equation}

Next, we expand the variance and covariance operators according to their definitions, and use condition 2 from the statement of the theorem to argue that $ \mathbb{E}_f \left[ I_H (\mathbf{x}) \right] \mathbb{E}_f \left[ I_L (\mathbf{x}) \right] \ll \mathbb{E}_f \left[ I_H (\mathbf{x}) I_L (\mathbf{x}) \right] $ and $ \left( \mathbb{E}_f \left[ I_L (\mathbf{x}) \right] \right)^2 \ll \mathbb{E}_f \left[ \left( I_L (\mathbf{x}) \right)^2 \right] $. (To do so we use stipulation~\ref{stipulation:fundamental_model_quality_assumption} to assert that if $ P_F \ll 1 $, then $ P_{F_L} \ll 1 $.) Therefore, Eq.~\eqref{eqn:CVMC_optimal_alpha_expansion} further simplifies to
\begin{equation}
    \alpha^* = \frac{\mathbb{E}_f \left[ I_H (\mathbf{x}) I_L (\mathbf{x}) \right]}{\mathbb{E}_f \left[ \left( I_L (\mathbf{x}) \right)^2 \right]}
\end{equation} 

Notice that
\begin{equation}
    \begin{aligned}
        \mathbb{E}_f \left[ I_H (\mathbf{x}) I_L (\mathbf{x}) \right] &= \int_{\Omega} I_H (\mathbf{x}) I_L (\mathbf{x}) f_{\mathbf{X}} (\mathbf{x}) d\mathbf{x} \\
        &= \int_{\Omega} I_H (\mathbf{x}) f_{\mathbf{X}} (\mathbf{x}) d\mathbf{x} - \int_{\Omega_{\Delta}} I_H (\mathbf{x}) f_{\mathbf{X}} (\mathbf{x}) d\mathbf{x} \\
        &= \mathbb{E}_f \left[ I_H (\mathbf{x}) \right] - \int_{\Omega_{\Delta}} I_H (\mathbf{x}) f_{\mathbf{X}} (\mathbf{x}) d\mathbf{x}
    \end{aligned}
\end{equation}

Also, $ \mathbb{E}_f \left[ \left( I_L (\mathbf{x}) \right)^2 \right] = \mathbb{E}_f \left[ I_L (\mathbf{x}) \right] = P_{F_L} $ and $ \mathbb{E}_f \left[ I_H (\mathbf{x}) \right] = P_F $. Therefore,
\begin{equation}
    \alpha^* = \frac{P_F}{P_{F_L}} - \frac{1}{P_{F_L}} \int_{\Omega_{\Delta}} I_H (\mathbf{x}) f(\mathbf{x}) d\mathbf{x}
\end{equation}

As $ \Omega_{\Delta} \to \varnothing $, $ \alpha^* \to \nicefrac{P_F}{P_{F_L}} $. Finally, note from Eq.~\eqref{eqn:CVIS_alpha} and Eq.~\eqref{eqn:limiting_or_true_value_of_cvis_alpha} that $ \Tilde{\alpha} $ is an estimate of $ \nicefrac{P_F}{P_{F_L}} $.
\end{proof}

\subsection{Proof of Theorem~\ref{thm:CVIS_alpha_variance_reduction_diagnostic} and Corollary~\ref{corollary:CVIS_alpha_diagnostic_terms_of_correlation}}
\label{appendix:proof_variance_reduction_diagnostic}

\begin{proof}[Proof of Theorem~\ref{thm:CVIS_alpha_variance_reduction_diagnostic}]
For variance reduction to occur, $ \operatorname{\mathbb{V}ar} \left[ \hat{P}_{F_{CVMC}} \right] - \operatorname{\mathbb{V}ar} \left[ \hat{Q}_{MC} \right] \leq 0 $. Expanding $ \operatorname{\mathbb{V}ar} \left[ \hat{P}_{F_{CVMC}} \right] $ (see Eq.~\eqref{eqn:variance_CVMC_general}), and replacing $ \alpha $ with $ \alpha^{\dagger} $, this can be expanded into the following. (Recall that $ \alpha^{\dagger} = \nicefrac{P_F}{P_{F_L}} \neq 0 $.)
\begin{gather}
    \alpha^{\dagger^2} \operatorname{\mathbb{V}ar} \left[ \hat{Q}_{L_{MC}} \right] + \alpha^{\dagger^2} \operatorname{\mathbb{V}ar} \left[ \hat{P}_{F_L} \right] - 2 \alpha^{\dagger} \operatorname{\mathbb{C}ov} \left[ \hat{Q}_{MC}, \hat{Q}_{L_{MC}} \right] \leq 0 \\
    \Rightarrow \alpha^{\dagger} \leq \frac{2 \operatorname{\mathbb{C}ov} \left[ \hat{Q}_{MC}, \hat{Q}_{L_{MC}} \right]}{ \operatorname{\mathbb{V}ar} \left[ \hat{Q}_{L_{MC}} \right] + \operatorname{\mathbb{V}ar} \left[ \hat{P}_{F_L} \right]} \label{eqn:alpha_for_CV_variance_reduction}
\end{gather}

Next, applying the conditions stated in the theorem to simplify Eq.~\eqref{eqn:alpha_for_CV_variance_reduction} in the same manner as in the proof of theorem~\ref{thm:model_limit_CVIS_alpha_optimal}, we get
\begin{equation}
    \alpha^{\dagger} \leq \frac{2 \mathbb{E}_f \left[ I_H (\mathbf{x}) I_L (\mathbf{x}) \right]}{\mathbb{E}_f \left[ \left( I_L (\mathbf{x}) \right)^2 \right]}
\end{equation}

Using the definition of $ P_{HL} $ and recognizing that $ \mathbb{E}_f \left[ \left( I_L (\mathbf{x}) \right)^2 \right] = \mathbb{E}_f \left[ I_L (\mathbf{x}) \right] = P_{F_L} $, and substituting the definition of $ \alpha^{\dagger} $, this becomes
\begin{align}
\label{eqn:CVIS_alpha_variance_reduction_diagnostic_final_condition}
    \frac{P_F}{P_{F_L}} &\leq \frac{2 P_{HL}}{P_{F_L}} \\
    \Rightarrow \kappa = \frac{P_{HL}}{P_F} &\geq \frac{1}{2}
\end{align}
\end{proof}

\begin{proof}[Proof of Corollary~\ref{corollary:CVIS_alpha_diagnostic_terms_of_correlation}]
Starting from Eq.~\eqref{eqn:alpha_for_CV_variance_reduction}, we apply the definitions for the estimators as well as condition 3 from theorem~\ref{thm:CVIS_alpha_variance_reduction_diagnostic} (since the corollary requires the same conditions as that theorem), to get
\begin{equation}
    \alpha^{\dagger} \leq \frac{2 \operatorname{\mathbb{C}ov}_f \left[ I_H (\mathbf{x}), I_L (\mathbf{x}) \right]}{ \operatorname{\mathbb{V}ar}_f \left[ I_L (\mathbf{x}) \right]}
\end{equation}

By definition, $ \operatorname{\mathbb{C}ov}_f \left[ I_H (\mathbf{x}), I_L (\mathbf{x}) \right] = \rho_{HL} \sqrt{ \operatorname{\mathbb{V}ar}_f \left[ I_H (\mathbf{x}) \right] \operatorname{\mathbb{V}ar}_f \left[ I_L (\mathbf{x}) \right] } $. Substituting,
\begin{equation}
    \alpha^{\dagger} \leq 2 \rho_{HL} \sqrt{ \frac{\operatorname{\mathbb{V}ar}_f \left[ I_H (\mathbf{x}) \right]}{ \operatorname{\mathbb{V}ar}_f \left[ I_L (\mathbf{x}) \right]} }
\end{equation}

Since $ I_H (\mathbf{x}) $ is a Bernoulli random variable, $ \operatorname{\mathbb{V}ar}_f \left[ I_H (\mathbf{x}) \right] = P_F - P_F^2 $, and similarly, $ \operatorname{\mathbb{V}ar}_f \left[ I_L (\mathbf{x}) \right] = P_{F_L} - P_{F_L}^2 $. Further, since the target probability is small, we argue that $ P_F - P_F^2 \approx P_F $ and $ P_{F_L} - P_{F_L}^2 \approx P_{F_L} $. (This is similar to the argument made in the proof of theorem~\ref{thm:model_limit_CVIS_alpha_optimal}.)
\begin{equation}
    \Rightarrow \alpha^{\dagger} \leq 2 \rho_{HL} \sqrt{ \frac{P_F}{P_{F_L}} }
\end{equation}

Using the definition of $ \alpha^{\dagger} $, we get,
\begin{align}
    \alpha^{\dagger} &\leq 2 \rho_{HL} \sqrt{\alpha^{\dagger}} \\
    \Rightarrow \alpha^{\dagger} &\leq 4 \rho_{HL}^2
\end{align}
\end{proof}

\subsection{Proof of Theorem~\ref{thm:alpha_variance}}
\label{appendix:proof_alpha_variance}

\begin{proof}[Proof of Theorem~\ref{thm:alpha_variance}]

The procedure used here was first described by Katz et al. \cite{Katz1978}.

Define the following
\begin{gather}
    \mu_H = \mathbb{E} \left[ \Tilde{\mathcal{Q}} \right] \; \text{ and } \; \sigma_H^2 = \operatorname{\mathbb{V}ar} \left[ \Tilde{\mathcal{Q}} \right] \label{eqn:stats_of_HF_IS_estimator} \\
    \mu_L = \mathbb{E} \left[ \Tilde{\mathcal{Q}}_L \right] \; \text{ and } \; \sigma_L^2 = \operatorname{\mathbb{V}ar} \left[ \Tilde{\mathcal{Q}}_L \right] \label{eqn:stats_of_LF_IS_estimator}
\end{gather}


If $ N $, i.e., the total number of samples used to evaluate $ \Tilde{\mathcal{Q}} $ and $ \Tilde{\mathcal{Q}}_L $ is large enough for the estimators to be practically useful, then the Central Limit Theorem also applies, and we can write
\begin{gather}
    \Tilde{\mathcal{Q}} = \mu_H + Z_H \; \text{ where } \; Z_H \sim \mathcal{N} \left( 0, \sigma_H^2 \right)  \label{eqn:rv_expansion_of_HF_IS_estimator} \\
    \Tilde{\mathcal{Q}}_L = \mu_L + Z_L \; \text{ where } \; Z_L \sim \mathcal{N} \left( 0, \sigma_L^2 \right) \label{eqn:rv_expansion_of_LF_IS_estimator}
\end{gather}

Therefore, substituting equations~\ref{eqn:rv_expansion_of_HF_IS_estimator} and~\ref{eqn:rv_expansion_of_LF_IS_estimator} into Eq.~\eqref{eqn:proposed_alpha_definition_alternate}, $ \Tilde{\alpha} $ becomes
\begin{equation}
\label{eqn:proposed_alpha_CLT_rv_expression}
    \Tilde{\alpha} = \frac{\mu_H + Z_H}{\mu_L + Z_L} = \frac{\mu_H}{\mu_L} \frac{\left( 1 + \frac{Z_H}{\mu_H} \right)}{\left( 1 + \frac{Z_L}{\mu_L} \right)}
\end{equation}
\begin{equation}
\label{eqn:log_of_proposed_alpha_CLT_rv_expression}
    \Rightarrow \ln{\left( \Tilde{\alpha} \right)} =  \ln{\left(\frac{\mu_H}{\mu_L}\right)} + \ln{\left( 1 + \frac{Z_H}{\mu_H} \right)} - \ln{\left( 1 + \frac{Z_L}{\mu_L} \right)}
\end{equation}

Notice also that since $ \Tilde{\mathcal{Q}} $ and $ \Tilde{\mathcal{Q}}_L $ are estimators, $ \sigma_H \to 0 $ and $ \sigma_L \to 0 $ as $ N \to \infty $. (In fact, in most practical cases $ \nicefrac{\sigma_H}{\mu_H} \leq 0.1 $, and analogously for $ \sigma_L $.) Therefore, again for sufficiently large $ N $,
\begin{equation}
\label{eqn:H_and_L_are_small}
    \frac{Z_H}{\mu_H} \ll 1 \; \text{ and } \; \frac{Z_L}{\mu_L} \ll 1
\end{equation}

We now make use of the following Taylor Series Expansion for small $ \delta $,
\begin{equation}
\label{eqn:small_delta_taylor_series_expansion_log_1_+_delta}
    \ln{(1 + \delta)} = \delta - \frac{\delta^2}{2} + \frac{\delta^3}{3} - \frac{\delta^4}{4} \dots
\end{equation}
which, in conjunction with equations~\ref{eqn:log_of_proposed_alpha_CLT_rv_expression} and~\ref{eqn:H_and_L_are_small}, allows us to write
\begin{gather}
\label{eqn:simplified_approximate_log_alpha_expansion}
    \ln{\left( \Tilde{\alpha} \right)} \approx  \ln{\left(\frac{\mu_H}{\mu_L}\right)} + R_{HL} \\
    \text{where } \; R_{HL} = \frac{Z_H}{\mu_H} - \frac{Z_L}{\mu_L}
\end{gather}

Since $ Z_H $ and $ Z_L $ are Gaussian random variables, $ R_{HL} $ is also a Gaussian random variable.
\begin{equation}
\label{eqn:distribution_of_R_HL}
    R_{HL} \sim \mathcal{N} \left( 0, \frac{\sigma_H^2}{\mu_H^2} + \frac{\sigma_L^2}{\mu_L^2} \right)
\end{equation}

This directly shows us that
\begin{gather}
    \ln{\left( \Tilde{\alpha} \right)} \sim \mathcal{N} \left( \ln{\left(\frac{\mu_H}{\mu_L}\right)} , \frac{\sigma_H^2}{\mu_H^2} + \frac{\sigma_L^2}{\mu_L^2} \right)
\end{gather}
\end{proof}

\section{Special Case: Nested Failure Regions}
\label{appendix:nested_failure}

In the special case when the failure region predicted by the LF model contains the failure region predicted by the HF model, the proposed CVIS estimator collapses into the Multifidelity Importance Sampling method proposed by Peherstorfer et. al~\cite{PEHERSTORFER_MFIS}.

In this special case, $ I_L (\mathbf{x}) = I_H (\mathbf{x}) = 1 $ $ \forall \mathbf{x} \in \Omega_{\mathcal{F}} $. Thus, the most efficient Importance Sampling Density of our selected form (Eq.~\eqref{eqn:proposed_ISD}) occurs when $ \beta^* = \infty $. Additionally, 
we know that for this value of $ \beta^* $, $ S_L \left( \mathbf{x}, \beta^* \right) = I_L (\mathbf{x}) $, and thus our proposed ISD converges to the ISD suggested in~\cite{PEHERSTORFER_MFIS}.

We can also recognize that in this case, $ \hat{Q}_L = \hat{C}_S $ (Eq.~\eqref{eqn:LF_IS_estimator_for_CVIS}) and $ \hat{Q} = \hat{\mathbb{E}}_q \left[ I_H \left( \mathbf{X} \right) \right] $ (Eq.~\eqref{eqn:HF_IS_estimator_for_CVIS}), where $ \hat{\mathbb{E}}_q \left[ \cdot \right] $ is an estimator of $ \mathbb{E}_q \left[ \cdot \right] $. Further, $ \hat{C}_S = \hat{P}_{F_L} $. We now see from equations~\ref{eqn:CVIS_alpha} and~\ref{eqn:CVIS_estimator}, that, for this special case

\begin{align}
    \Tilde{\alpha} &= \frac{\hat{\mathbb{E}}_q \left[ I_H \left( \mathbf{X} \right) \right]}{\hat{P}_{F_L}} \\
    \Rightarrow \Tilde{P}_F &= \hat{\mathbb{E}}_q \left[ I_H \left( \mathbf{X} \right) \right]
\end{align}

This also satisfies our intuition as there is no variance in the LF indicator function within the probability density used as the ISD by our framework in this special case. Therefore, there is no additional variance reduction possible by Control Variates as there is no correlation between the two models within the chosen probability distribution (and $ \alpha^* $ = 0 as per Eq.~\eqref{eqn:ACV_optimal_alpha}). Hence, all variance reduction is achieved by Importance Sampling, which is optimal under this framework.

\section*{Acknowledgments}
This research is supported through the INL Laboratory Directed Research and Development (LDRD) Program under DOE Idaho Operations Office Contract DE-AC07-05ID14517. This research made use of the resources of the High-Performance Computing Center at INL, which is supported by the Office of Nuclear Energy of the US DOE and the Nuclear Science User Facilities under Contract No. DE-AC07-05ID14517. Portions of this work were carried out at the Advanced Research Computing at Hopkins (ARCH) core facility (https://www.arch.jhu.edu/), which is supported by the National Science Foundation (NSF) Grant Number OAC1920103. The authors are grateful to Professor Dootika Vats for her suggestions on MCMC variance estimators.

\bibliographystyle{unsrt}  

\end{document}